\documentclass{aa}  

\usepackage{caption}
\usepackage{graphicx}
\usepackage{ulem}
\usepackage{txfonts}
\usepackage{color}

\usepackage{amsmath}
\usepackage{cases}
\definecolor{linkcolor}{rgb}{0,0,.7}
\definecolor{oldText}{rgb}{0.5,0,0}
\definecolor{newText}{rgb}{0,.5,0}

\usepackage[colorlinks=true]{hyperref}
\hypersetup{
  colorlinks   = true,
  urlcolor     = linkcolor,
  linkcolor    = linkcolor, 
  citecolor   = linkcolor 
}

\renewcommand{\H}{\text{H}}
\renewcommand{\O}{\text{O}}
\renewcommand{\doi}[1]{\textcolor{linkcolor}{\href{http://doi.org/#1}{\textbf{doi}}}}
\newcommand{\ads}[1]{\textcolor{linkcolor}{\href{#1}{\textbf{ads}}}}
\renewcommand{\arxiv}[1]{\textcolor{linkcolor}{\href{https://arxiv.org/abs/#1}{\textbf{arXiv}}}}

\renewcommand{\u}[1]{\underline{#1}}

\newcommand{\N}{\text{N}}
\newcommand{\C}{\text{C}}

\newcommand{\der}[2]{\frac{d{#1}}{d{#2}}}
\newcommand{\e}[1]{\cdot 10^{#1}}
\newcommand{\parder}[2]{\frac{\partial{#1}}{\partial{#2}}}

\newcommand{\unit}[2]{\,\text{#1}^{#2}}

\begin{document} 

   \title{SHAMPOO: A stochastic model for tracking dust particles under the influence of non-local disk processes}

   \author{M. Oosterloo
          \inst{1}
          \and
          I. Kamp
          \inst{1}
          \and
          W. van Westrenen
          \inst{2}
          \and
          C. Dominik
          \inst{3}
          }

   \institute{Kapteyn Astronomical Institute, University of Groningen, Landleven 12, 9747 AD Groningen, Netherlands
                  \and
               Department of Earth Sciences, Vrije Universiteit Amsterdam, De Boelelaan 1085,
1081 HV Amsterdam, Netherlands
\and 
 Anton Pannekoek Institute for Astronomy, University of Amsterdam, Science Park 904, 1098XH, Amsterdam, Netherlands
               }

   \date{Received: 23 November 2022; accepted: 25 April 2023}

    % \abstract{}{}{}{}{} 
% 5 {} token are mandatory
 
  \abstract
   {The abundances of carbon, hydrogen, nitrogen, oxygen, and sulfur (CHNOS) are crucial for understanding the initial composition of planetesimals and, by extension, planets. At the onset of planet formation, large amounts of these elements are stored in ices on dust grains in planet-forming disks. The evolution of the ice in dust, however, is affected by disk processes, including dynamical transport, collisional growth and fragmentation, and the formation and sublimation of ice. These processes can be highly coupled and non-local.}
   {In this work, we aim to constrain the disk regions where dynamical, collisional, and ice processing are fully coupled. Subsequently, we aim to develop a flexible modelling approach that is able to predict the effects of these processes acting simultaneously on the CHNOS budgets of planetesimal-forming material in these regions.}
   {We compared the timescales associated with these disk processes to constrain the disk regions where such an approach is necessary, and subsequently developed the SHAMPOO (\u{S}toc\u{ha}stic \u{M}onomer \u{P}r\u{o}cess\u{o}r) code, which tracks the CHNOS abundances in the ice mantle of a single 'monomer' dust particle of bare mass $m_\text{m}$, embedded in a larger 'home aggregate'. The monomer inside its home aggregate is affected by aerodynamic drag, turbulent stirring, collision processes, and ice adsorption and desorption simultaneously. The efficiency of adsorption onto and the photodesorption of the monomer here depends on the depth $z_\text{m}$ at which the monomer is embedded in the home aggregate. We used SHAMPOO to investigate the effect of the fragmentation velocity $v_\text{frag}$ and home aggregate filling factor $\phi$ on the amount of CHNOS-bearing ices for monomers residing at $r=10$ AU.}
   {The timescale analysis shows that the locations where disk processes are fully coupled depend on both grain size and ice species. We find that monomers released at 10 AU embedded in smaller, more fragile, aggregates with fragmentation velocities of \mbox{1 m/s} are able to undergo adsorption and photodesorption more often than monomers in aggregates with fragmentation velocities of \mbox{5 m/s} and 10 m/s. Furthermore, we find that at 10 AU in the midplane, aggregates with a filling factor of $\phi=10^{-3}$ are able to accumulate ice 22 times faster on average than aggregates with $\phi=1$ under the same conditions.}
   {Since different grain sizes are coupled through collisional processes and the grain ice mantle typically consists of multiple ice species, it is difficult to isolate the locations where disk processes are fully coupled, necessitating the development of the SHAMPOO code. Furthermore, the processing of ice may not be spatially limited to dust aggregate surfaces for either fragile or porous aggregates.}


    \maketitle

    \section{Introduction}
    
% Start off from planetary habitability, which depends on planet composition. 
The light elements carbon, hydrogen, nitrogen, oxygen, and sulfur (CHNOS) play an important role in the evolution of rocky planets. For example, their absolute and relative surface and atmospheric abundances in the form of volatile molecules such as H$_2$O, CO$_2$, or N$_2$ play a key role in the surface conditions and, in extension, habitability of planets \citep[e.g.][]{Kasting+1993, Kasting&Catling2003, Kopparapu+2013}. The interior properties and evolution of a planet are also affected by the planetary budgets of CHNOS. CHNOS abundances have profound effects on the physical state (solid versus liquid) of planetary cores \citep[e.g.][]{Tronnes+2019}, on the melting temperatures and mineralogy of their silicate mantles \citep[e.g.][]{Kushiro+1969, Dasgupta&Hirschmann2006, Hakim+2019}, and on volcanic outgassing speciation \citep[e.g.][]{Bower+2022}.\\
In order to understand the evolution of a planet, it is crucial to identify how much CHNOS a planet initially inherits from its building blocks. Planets are thought to form in a few megayears from micrometre-sized dust inferred to be present in planet-forming disks around young stars  \citep[e.g.][]{Andrews2020, Raymond&Morbidelli2020}. The first stage of planet formation involves the coagulation of these micron-sized dust grains into millimetre- to centimetre-sized particles through pairwise collisions \citep[][]{Dominik&Tielens1997, Birnstiel+2012, Krijt+2016}. These may subsequently grow into planetesimals either through continuous coagulation or through gravitational collapse triggered by, for example, streaming instabilities \citep[][]{Okuzumi+2012,Youdin&Goodman2005, Johansen+2007, Johansen+2014}.\\
In the colder regions of the disk, a considerable fraction of the solid-phase CHNOS mass budget is likely incorporated as ices onto dust grains in molecular species such as H$_2$O, CO, CO$_2$, CH$_4$, NH$_3$, and H$_2$S \citep[][]{Boogert+2015, Oberg&Bergin2020, Krijt+2022}. On the one hand, evidence from the Solar System  suggests that these ices present on dust grains are inherited from the interstellar medium at least to some degree \citep[e.g.][]{Altwegg+2017, Drozdovskaya+2019}. On the other hand, the amount of ice is also affected by the local balance of molecule adsorption and desorption, which are highly sensitive to the local temperature, radiation field, and composition of the gas phase \citep[e.g.][]{Cuppen+2017}. As dust grains grow into larger aggregates through collisions, dynamical processes including vertical settling, radial drift and turbulent diffusion result in significant displacement of dust throughout the disk \citep[][]{Weidenschilling1977, Armitage2010, Ciesla2010, Ciesla2011}. This dynamical transport exposes individual dust grains to a wide range of local conditions, which could have profound consequences on the amount and composition of ice \citep[][]{Ciesla2010, Ciesla2011}. \\
Many modelling efforts have attempted to constrain the effects of dust transport and collisional processes, and suggest large scale transport of volatiles throughout the disk \citep[][]{Cuzzi+2004, Krijt+2016b, Bosman+2018, Krijt+2018, Krijt+2020, Bergner+2021}. However, these studies usually focus on one molecular species, such as H$_2$O \citep[][]{Krijt+2016b,Schoonenberg+2018}, CO \citep[][]{Kama+2016, Krijt+2018, Krijt+2020, VanClepper+2022}, or CO$_2$ \citep{Bosman+2018}, or emphasize a limited subset of migration, collision and ice proccesses \citep[][]{Ciesla2010, Ciesla2011, Krijt&Ciesla2016, Bergner+2021}, or disk chemistry \citep[][]{Krijt+2020}. Alternatively, the models are sometimes limited to either the radial \citep[][]{Schoonenberg+2018, Booth&Ilee2019} or vertical disk dimension \citep[][]{Krijt+2016b, Krijt&Ciesla2016}.\\
In this work we investigate for the first time the full coupling of dynamical transport (vertical settling, radial drift, and turbulent diffusion), collisional processes (coagulation and fragmentation), and the adsorption and desorption of multiple ices. We develop a stochastic model where we track the behaviour of dust grains with an ice mantle containing multiple CHNOS-bearing molecules in response to these processes. All processes are treated in a fully coupled, 2D fashion. In our model, we follow the evolution of individual dust grains. This enables predictions for local solid-phase CHNOS budgets throughout the disk via statistical analysis of the behaviour of a large set of individual dust grain models.\\
We discuss our model and methods in Sect. \ref{sec:2}, and benchmark them against a few key earlier works. Subsequently, in Sect. \ref{sec:3}, we analyse the coupling behaviour of the various disk processes as a function of vertical and radial position throughout the disk, and we explore the behaviour of individual dust grains under coupled disk processing. We discuss the parameter sensitivities of our model in Sect. \ref{sec:4}, and summarize our key conclusions in Sect. \ref{sec:5}.

    \section{Modelling Approach}
    \label{sec:2}

\begin{figure}
    \centering
    \includegraphics[width=.48\textwidth]{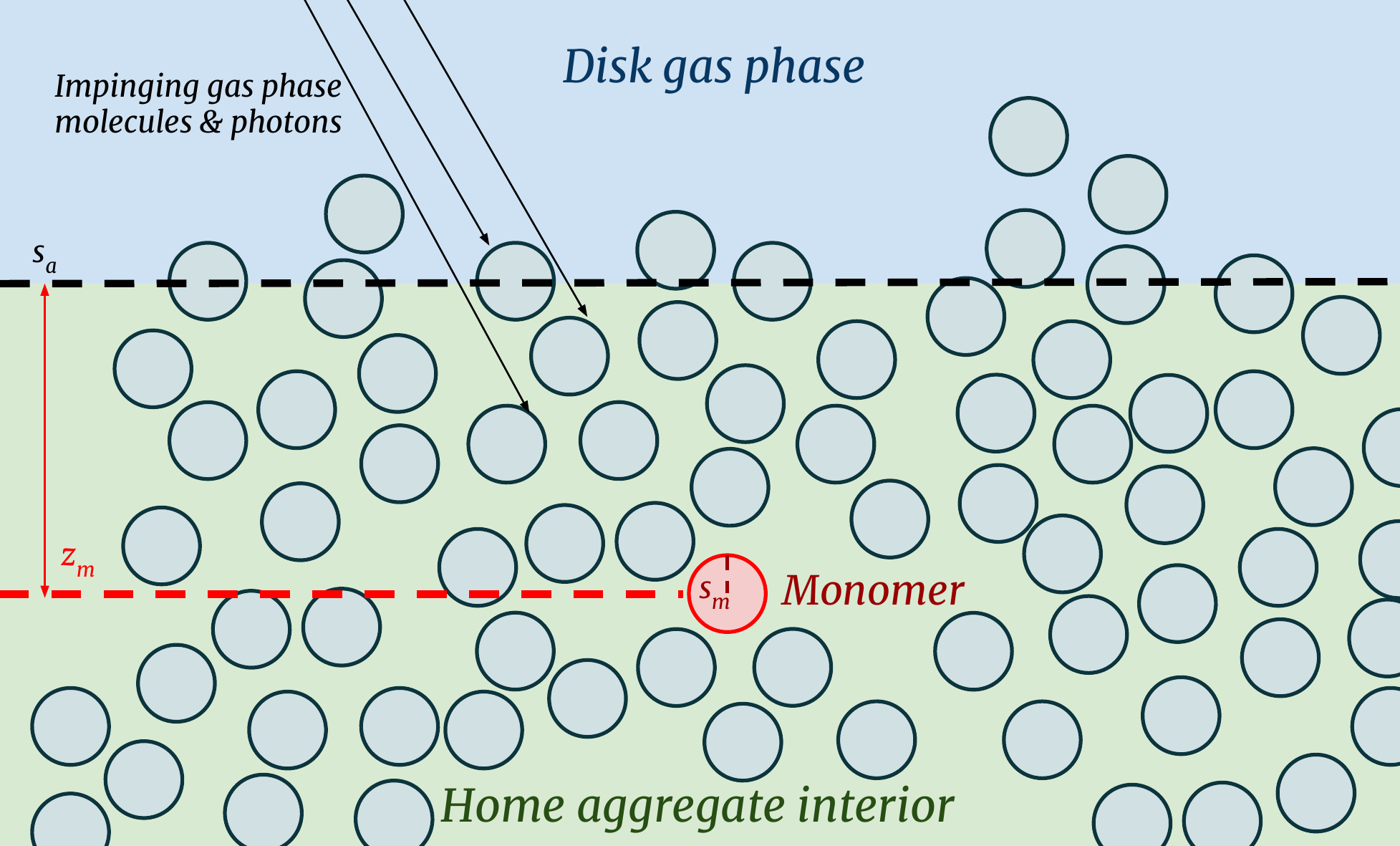}
    \caption{The key model idea: A monomer of radius $s_m$ embedded at some monomer depth $z_m$ inside a home aggregate of effective radius $s_a$.}
    \label{fig:2MonomerAndHomeAggregate}
\end{figure}

In SHAMPOO (\u{S}toc\u{ha}stic \u{M}onomer \u{P}r\u{o}cess\u{o}r), we follow a small tracer dust particle called a \textit{monomer}, which has size\footnote{Throughout this work, particle "size" signifies the (effective) radius of a dust particle.} $s_m$ (see Fig. \ref{fig:2MonomerAndHomeAggregate}). In practice, the monomer will usually be embedded at a certain depth $z_m$ in a larger dust aggregate of effective radius $s_a$ as a consequence of collisions with other monomers and dust aggregates. Therefore, we associate each monomer at any given time with a \textit{home aggregate}: the dust aggregate that hosts the monomer tracked by our model. Inferences at any given time about the properties of local dust populations can subsequently be made by tracing the evolution of a large number of monomers.\\
The monomer and its home aggregate interact with the disk environment through a number of processes. Aerodynamic drag and turbulent diffusion may displace the home aggregate including the monomer both radially and vertically throughout the disk \citep[e.g.][]{Weidenschilling1977, Armitage2010, Ciesla2010, Ciesla2011}. In addition, collisions with other dust aggregates may result in changes in the home aggregate size $s_a$ and monomer depth $z_m$ through coagulation and fragmentation \citep[e.g.][]{Dominik&Tielens1997, Blum&Munch1993, Birnstiel+2011}. Lastly, gas phase molecules impinging on the home aggregate can be adsorbed as ice on the monomer. Here, $z_m$ plays an important role as the monomer depth determines the probability for gas molecules impinging on the home aggregate to be able to reach the monomer. All these processes depend on the properties of the local disk environment in which the monomer is located. This local disk environment is fully described by the thermo-chemical disk model ProDiMo\footnote{Official webpage: \url{https://prodimo.iwf.oeaw.ac.at/}}  \citep{Woitke+2009, Kamp+2010, Thi+2011, Thi+2013}.\\
We discuss the background disk model in Sect. \ref{sec:2.1}, and elaborate on the dynamical model in Sect. \ref{sec:2.2}. Subsequently, we introduce our collision model in Sect. \ref{sec:2.3}, and our treatment of ice formation in Sect. \ref{sec:2.4}.

\subsection{Background disk model}
\label{sec:2.1}
\begin{table*}
    \begin{center}
    \begin{tabular}{cp{0.4\textwidth}rlc}
    \hline
         \textbf{Parameter}& \textbf{Name} &\textbf{Value} & \textbf{Unit} &\textbf{Reference} \\\hline\hline
$M_\star$ & Stellar mass                        & $0.7$     & $M_\odot$     & (1) \\
$L_\star$ & Total stellar luminosity            & $6$       & $L_\odot$     & (2,3) \\
$T_\text{eff}$ & Stellar effective temperature  & $4000$    & K              & (2,3) \\
$f_\text{UV}$   & Stellar UV excess             &0.01       & - & (1) \\ 
$p_\text{UV}$   & UV power law index            & 1.3       & - & (1) \\
$L_X$           & X-ray luminosity              & $1\e{23}$ &$\unit{J}{}\unit{s}{-1}$     & (1) \\
$T_X$ & X-ray emission temperature              & $2\e{7}$   & K & (1) \\\hline
$\rho_m$        & Monomer material density      & $2094$     & $\unit{kg}{}\unit{m}{-3}$ & (1)\\
$a_\text{min}$ & Minimum background model grain size & $5\e{-8}$ & m & (1)\\
$a_\text{pow}$ & Background grain size distribution power law slope & 3.5 & - & (1)\\
$N_\text{bins}$ & Number of grain size bins & $100$ & - & (1) \\\hline
$M_\text{disk}$ & Disk mass                     & 0.1 & $M_\odot$ & (3)                           \\
$\delta$    & Dust to gas mass ratio            & 0.01          & -     & (1)    \\
$r_\text{in}$ & Inner disk radius               & 0.07          & AU    & (1)    \\
$r_\text{out}$ & Outer disk radius              & 600           & AU    & (3)                     \\
$\epsilon$ & Column density exponent            & 1             & -     & (1)    \\
$r_\text{taper}$ & Tapering-off radius          & 100           & AU    & (1)    \\
$H_0$ & Reference scale height                  & 10            & AU    & (1)    \\
$r_0$ & Reference scale height radial distance  & 100           & AU    & (1)    \\
$\gamma$ & Disk tapering-off exponent power                         & 1          & -     & (1)                    \\
$\beta$ & Flaring power                         & 1.1           & -     & (3)                     \\
$\alpha$ & Turbulence strength parameter        & $10^{-3}$     & -     & (3)                     \\

\hline
\end{tabular}
%\captionsetup{width=.9\textwidth}
    \caption{Input parameters used in the background disk model. References: (1) \cite{Woitke+2016}, (2) \cite{Siess+2000}, (3) this work.}
    \label{tab:DiskTable}
    
    \end{center}
\end{table*}

\begin{figure}
    \centering
    \includegraphics[width=0.49\textwidth]{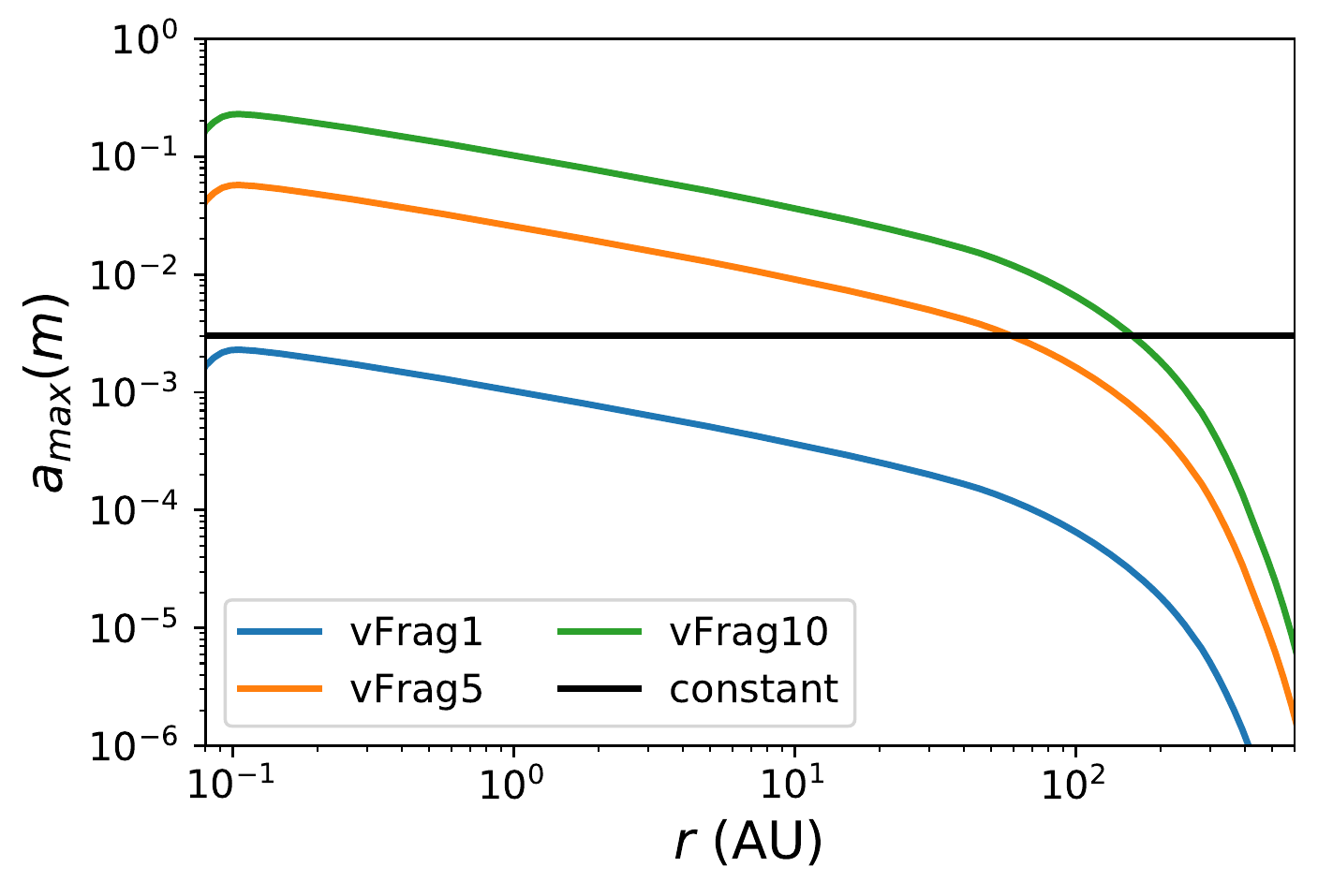}
    \caption{The radial behaviour of $a_\text{max}$ for the four different background models considered throughout this work.}
    \label{fig:21BackgroundAMAX}
\end{figure}

The processes which alter the ice mantle of the monomer all depend on the local disk environment as characterized by the spatial physical, thermal and chemical structure of the disk. For this purpose we utilize the thermochemical disk model ProDiMo \citep{Woitke+2009, Kamp+2010, Thi+2011, Thi+2013}. This code has been developed to calculate the local physical, thermal, and chemical structure in an azimuthally symmetric disk. We use ProDiMo to determine the local gas and dust density $\rho_\text{g}$, $\rho_\text{d}$, and temperature $T_\text{g}$, $T_\text{d}$, respectively. Time-dependent chemistry in ProDiMo also allows us to infer the local molecule number densities $n_x$ of all species $x$. We here select molecular species based on their importance for the solid-phase CHNOS mass budgets. Therefore, we will restrict ourselves to volatile H$_2$O, CO, CO$_2$, CH$_4$, NH$_3$, and H$_2$S throughout the rest of this work. Lastly, we use the local UV radiation field $\chi_\text{RT}$, which is calculated from the 2D radiative transfer model.\\
The gas density structure of the disk is calculated from a parametrized column density structure \citep{Woitke+2009, Woitke+2016}.
\begin{align}
    \Sigma_\text{g}(r) = \Sigma_0\cdot r^{-\epsilon}\cdot \exp\left[\left(\frac{r}{r_\text{taper}}\right)^{2-\gamma}\right],
\end{align}
where $\epsilon$ is the density profile power law index, and $r_\text{taper}$ denotes the tapering radius beyond which an exponential cutoff occurs. $\Sigma_0$ is derived from the normalization condition
\begin{align}
    M_\text{disk}=2\pi\int\limits_0^\infty \Sigma_\text{g}(r)\,dr.
\end{align}
The vertical gas density structure is given by a Gaussian
\begin{align}
\label{eq:GasStructure}
\rho_\text{g}(r,z)=\rho_0\exp\left(\frac{z^2}{2H_\text{g}^2(r)}\right),\qquad  H_\text{g}(r)=H_0\left(\frac{r}{r_0}\right)^\beta
\end{align}
Here, $H_\text{g}$ denotes the gas scale height, $H_0$ the reference scale height at a reference distance $r_0$, and $\beta$ is the flaring exponent. The normalization $\rho_0$ is determined via
\begin{align}
    \Sigma_\text{g}(r)=2\int\limits_0^{z_\text{max}(r)}\rho_\text{g}(r,z)\,dz,
\end{align}
where $z_\text{max}(r)=0.5r$ is the maximum vertical height considered in the model \citep{Woitke+2009}.\\
The initial dust density structure is treated in a similar fashion, where the total unsettled dust density $\rho_\text{d}^\star$ is related to the gas density as $\rho_\text{d}^\star=\delta\rho_\text{g}$. Here, $\delta$ denotes the dust-to-gas mass ratio. However, we do account for the settling of the largest dust grains via the procedure outlined in \cite{Woitke+2016}. We assume that the initial unsettled dust size distribution follows the same power law everywhere in the disk:
\begin{align}
\label{eq:dustSizeDistribution}
    f(a)\, &
    \begin{cases}
    \propto a^{-a_\text{pow}}\quad & \text{if } a\in [a_\text{min},a_\text{max}]\\
    = 0 & \text{elsewhere.}
    \end{cases}
\end{align}
Here, $a_\text{min}=5\e{-8}$ m denotes the minimum grain size. Similarly, $a_\text{max}$ denotes the maximum grain size, which is usually determined by the behaviour of local collisional growth processes. However, we note that grain growth in the outer regions of older disks may become drift-limited instead \citep{Birnstiel+2012}. Assuming dust fragmentation by turbulence-driven relative motion, \cite{Birnstiel+2012} expressed the maximum grain size as
\begin{align}
\label{eq:amax}
    a_\text{max}=f_\text{f}\frac{2}{3\pi}\frac{\Sigma_\text{g}}{\rho_\text{a}\alpha}\left(\frac{v_\text{frag}}{c_\text{s}}\right)^2.
\end{align}
Here, $f_\text{f}$ denotes an offset parameter of order unity, which is assumed $f_\text{f}=\frac{1}{2}$ throughout this work. $\rho_\text{a}$ denotes the density of dust aggregates, and is assumed equal to the home aggregate density. $\alpha$ represents the turbulence strength \citep{Shakura&Sunyaev1973}. $v_\text{frag}$ denotes the relative velocity above which aggregates undergo fragmentation, while $c_\text{s}$ denotes the local isothermal soundspeed. \\
The proportionality constant in Eq. \eqref{eq:dustSizeDistribution} is determined by requiring the total mass density resulting from integrating over all grain sizes $a$ to be equal to the total unsettled dust mass density $\rho_\text{d}^\star$ \citep{Woitke+2016}
\begin{equation}
    \rho_\text{d}^\star = \frac{4\pi}{3}\rho_\text{m}\int\limits_{a_\text{min}}^{a_\text{max}}f(a)a^3\,da.
\end{equation}
Here, $\rho_\text{m}=2094$ kg/m$^3$ denotes the dust grain material density, which is assumed equal to the monomer density. The monomer and dust aggregate density $\rho_\text{a}$ are related as $\rho_\text{a}=\phi \rho_\text{m}$, where $\phi$ denotes the dust aggregate mass filling factor.\\
The grain sizes are sampled log-uniformly over $N_\text{bins}=100$ size bins $i$ between $a_\text{min}$ and $a_{\text{max},D}$. Here $a_{\text{max},D}$ denotes the largest value for $a_\text{max}$ encountered anywhere throughout the disk, as calculated via Eq. \eqref{eq:amax}. Subsequently, the dust scale height for each size bin is calculated via \citep{Dubrulle+1995}
\begin{align}
\label{eq:DubrulleSettling}
H_\text{d}=H_\text{g}\sqrt{\frac{\alpha}{\text{St}\sqrt{3}+\alpha}}.
\end{align}
Here, St denotes the Stokes number, which is a function of grain size, soundspeed and the gas density in the disk midplane (see also Sect. \ref{sec:2.2}). We note that Eq. \eqref{eq:DubrulleSettling} is similar to the relation between $H_\text{d}$ and $H_\text{g}$ derived by \cite{Youdin&Lithwick2007}. Dust only redistributes vertically, such that the surface density associated with a grain of size $a$ at given radial distance $r$
\begin{align}
    \Sigma_\text{d}(a,r)=2\int_0^{z_\text{max}}\rho_\text{d}(a,r,z)\,dz,
\end{align}
remains constant before and after settling.\\
The thermal structure of the gas and dust, and the local UV radiation field $\chi_\text{RT}$ are derived from the local 2D radiation field $J_\nu$, which is calculated with ProDiMo's radiative transfer module. The gas temperature $T_\text{g}$ and dust temperature $T_\text{d}$ are treated separately. The dust temperature is found by assuming radiative equilibrium for the dust, while the gas temperature is derived from a detailed heating/cooling balance. For a full description of the radiative transfer and heating and cooling model we refer the reader to \cite{Woitke+2009, Woitke+2016, Thi+2011, Aresu+2011} and \cite{Oberg+2022}. \\
We use the time-dependent chemistry in ProDiMo to calculate the local gas phase molecule number densities $n_x$ of molecular species $x$. In the models used in this work, we considered the large DIANA chemical network containing 13 elements and 235 species. Within this chemical network, ions and ices of a particular molecular species are treated as different chemical species. Furthermore, we use the adsorption energies listed in the 2012 edition of the UMIST database \citep{McElroy+2013}. However, we note that it is not possible to define a single set of adsorption energies which is consistent throughout the entire disk due to the different binding energies associated with different grain surface compositions \citep{Kamp+2017}. For a more elaborate discussion of this specific chemical network, we refer the reader to \cite{Kamp+2017}, whereas the treatment of chemistry in ProDiMo has been discussed in \cite{Woitke+2009, Aresu+2011}.\\ We use a two-step approach for the chemistry. The first step involves solving the time-dependent chemistry from initially atomic chemical abundances under molecular cloud conditions. We here choose an integration time of $1.7\e{5}$ yr, where we follow \cite{Helling+2014}. This value has previously been derived as the best-fit lifetime for the Taurus Molecular Cloud 1 \citep[][]{McElroy+2013}. During the second step, we use the resulting abundances from this first step as initial chemical abundances for our disk model, and subsequently evolve this for another $2\e{5}$ yr under the local disk conditions to yield representative chemical abundances at the onset of planetesimal formation. Although time-dependent chemistry is used to inform the chemical structure of the background disk models during the simulations, the local chemical abundances in the background models are kept fixed during SHAMPOO simulations.\\
Throughout this work, we consider four different background disk models, distinguished by the radial behaviour of $a_\text{max}$ outlined in Fig. \ref{fig:21BackgroundAMAX}. For three of these models, the radial behaviour of $a_\text{max}$ is given by Eq. \eqref{eq:amax} for a fragmentation velocity $v_\text{frag}=1,5,10$ m/s. We will refer to these models as \texttt{vFrag1}, \texttt{vFrag5}, and \texttt{vFrag10}. Furthermore, we also consider a background model \texttt{const} where the maximum grain size is fixed as a function of $r$ to $a_\text{max}=3\e{-3}$ m. For all other disk model parameters, we consider the values listed in Table \ref{tab:DiskTable} for all four background models. These parameters are primarily set to values of the typical T Tauri model of the DIANA project  \citep{Woitke+2016, Kamp+2017, Dionatos+2019, Woitke+2019}. We also used this \mbox{DIANA T Tauri} model to inform our value for $a_\text{max}=3\e{-3}$ m for the \texttt{const}-model. However, it is thought that the onset of the formation of the first generation of planetesimals starts in the \mbox{class I} disk stage \citep[e.g.][]{Nixon+2018}. Therefore, we consider a more massive disk around a younger, more luminous pre-main sequence (PMS) star with respect to the standard T Tauri model parameters ($M_\text{disk}=0.1M_\odot$, $L_\star=6\,L_\odot$). This luminosity is consistent with a $0.7M_\odot$ PMS star of $2\e{5}$ yr and solar metallicity $Z_\oplus\approx0.02$ \citep{Siess+2000}.\\
The above approach gives rise to the disk structures depicted in Figs. \ref{fig:ABBackgroundModelconstStructure}-\ref{fig:ABBackgroundModelvFrag10Structure} shown in Appendix \ref{sec:ABP}, which show the density ($\rho_\text{g}$,$\rho_\text{d}$) and thermal ($T_\text{g}$,$T_\text{d}$) structures besides the UV radiation field $\chi_\text{RT}$ for the four different background models. Furthermore, Figs. \ref{fig:ABBackgroundModelconstAbundances}-\ref{fig:ABBackgroundModelvFrag10Abundances} show $n_x$ as a function of position in the disk for H$_2$O, CO, CO$_2$, CH$_4$, NH$_3$, and H$_2$S for the different background disk models, both in the gas and solid phase at the end of the evaluation of the time-dependent chemistry in ProDiMo. Qualitatively, the main differences between the background models are caused by the degree of settling, with background models with larger dust grains (\texttt{vFrag5} and \texttt{vFrag10}) having their dust settled in a thinner disk compared to the gas density structure, which is identical for all four background disk models. In particular for the \texttt{vFrag10}-model, Fig. \ref{fig:ABBackgroundModelvFrag10Structure} shows that $\rho_\text{d}$ is significantly higher in the disk midplane due to settling, at the cost of lower values at higher $z/r$. As a consequence of dust settling, the midplane region which is shielded from UV radiation is thinner in disk models where $a_\text{max}$ is larger. This results in higher temperatures $T_\text{g}$,$T_\text{d}$ down to lower $z/r$, which also limits the ice-forming region towards lower $z/r$.

\subsection{Dust dynamics}
\label{sec:2.2}
The key purpose of the dynamical model is to describe the evolution of the radial ($r$) and vertical ($z$) position of the monomer during every time step $\Delta t$. Key processes which can displace the monomer and its home aggregate are aerodynamic drag and turbulent diffusion. Our model is based on the work of \cite{Ciesla2010, Ciesla2011}.
The dynamical behaviour of the monomer is fully determined by the properties of its home aggregate, and by how well the home aggregate is coupled to the gas. An important measure which characterizes the nature of how the home aggregate interacts with the surrounding gas is the Stokes number \citep[e.g.][]{Armitage2010, Krijt+2018, Visser+2021}:
\begin{numcases}{\text{St} =}
\sqrt{\frac{\pi}{8}}\frac{
\rho_\text{a}s_\text{a}}{\rho_\text{g}c_\text{s}}\Omega\quad &if\quad $s_\text{a}< \frac{9\lambda_\text{mfp}}{4}$\quad Epstein drag \label{eq:StEpstein}\\
\sqrt{\frac{\pi}{8}}\frac{4}{9\lambda_\text{mfp}}\frac{\rho_\text{a}s_\text{a}^2}{\rho_\text{g}c_\text{s}}\Omega \quad &if\quad $s_\text{a}\geq\frac{9\lambda_\text{mfp}}{4}$\quad\text{Stokes drag} \label{eq:StStokes}
\end{numcases}
Here, $\rho_\text{a}$ and $s_\text{a}$ denote the material density and size of the home aggregate. In addition, $\rho_\text{g}$ and $c_\text{s}$ denote the local gas density and isothermal soundspeed. $\Omega$ denotes the Keplerian orbital frequency
\begin{align}
\Omega = \sqrt{\frac{GM_\star}{(r^2+z^2)^3}},
\end{align}
where $G$ is the gravitational constant and $M_\star$ the mass of the host star. In addition, $\lambda_\text{mfp}$ denotes the mean free path and is calculated as
\begin{align}
    \lambda_\text{mfp}=\frac{\rho_\text{g}}{\sqrt{2}\mu m_\text{p}\sigma_\text{mol}},
\end{align}
where $\mu=2.3$ denotes the mean molecular weight in atomic mass units, $m_\text{p}$ the proton mass, and $\sigma_\text{mol}=2\e{-19}$ m$^2$ the mean molecular cross section \citep[see e.g.][]{Okuzumi+2012, Krijt+2018}.
\\
As the Stokes number is a measure of how decoupled the motion of the home aggregate is from the gas, it can be used to characterize the dynamical behaviour of the home aggregate due to aerodynamic drag. The radial and vertical velocities due to aerodynamic drag $v_r$ and $v_z$ are given by \citep[e.g.][]{Armitage2010}
\begin{align}
    v_r&= -2\eta r \Omega \frac{\text{St}}{1+\text{St}^2}\label{eq:vr},\\
    v_z&=-\Omega z \text{St}\label{eq:vz}.
\end{align}
$\eta$ here denotes the dimensionless gas pressure gradient, given by
\begin{align}
    \eta = -\frac{1}{2}\left(\frac{c_\text{s}}{r\Omega}\right)^2\parder{\ln \rho_\text{g}}{\ln r}.
\end{align}
Altogether, if a monomer is embedded in a home aggregate with significant Stokes number (St $\gtrsim \alpha$), the monomer will move towards the disk midplane and radially inward as a consequence of the vertical settling and radial drift of the home aggregate.\\
In turbulent disks, aerodynamic drift can be countered by turbulent diffusion. Using the single-particle random walk formalism from \cite{Ciesla2010, Ciesla2011}, we calculate the new radial and vertical positions of the home aggregate and monomer after timestep $\Delta t$ as:
\begin{align}
\label{eq:position}
    r(t+\Delta t) = r(t) + v_{r\text{,eff}}\Delta t + R_1\left(\frac{2}{\xi}D_\text{d}\Delta t\right)^{1/2},\\
    z(t+\Delta t) = z(t) + v_{z\text{,eff}}\Delta t + R_2\left(\frac{2}{\xi}D_\text{d}\Delta t\right)^{1/2}.
\end{align}
Here, $R_1$ and $R_2$ are uniformly drawn random numbers $R_1, R_2 \in [-1,1]$. $\xi$ denotes the variance of the distribution from which $R_1$ and $R_2$ are drawn, which means in our case that $\xi=1/3$ \citep[][]{Visser1997, Ciesla2010}. $v_{r\text{,eff}}$ and $v_{z\text{,eff}}$ denote the effective radial and vertical velocities, respectively, and are given by
\begin{align}
    v_{r\text{,eff}}=v_\text{r}+\parder{D_\text{d}}{r}+\frac{D_\text{d}}{\rho_\text{g}} \parder{\rho_\text{g}}{r},\\
    v_{z\text{,eff}}=v_\text{z}+\parder{D_\text{d}}{z}+\frac{D_\text{d}}{\rho_\text{g}}\parder{\rho_\text{g}}{z}.
\end{align}
$D_\text{d}$ denotes the dust diffusivity and is calculated from the gas diffusivity $D_\text{g}$ \citep[][]{Youdin&Lithwick2007}
\begin{align}
    D_\text{d}=\frac{D_\text{g}}{1+\text{St}^2}.
\end{align}
We here estimate the gas diffusivity $D_\text{g}$ with the turbulent viscosity, $D_\text{g}=\nu_\text{turb}=\alpha c_\text{s}H_\text{g}$ \citep[][]{Shakura&Sunyaev1973}.
\begin{figure}
    \centering
    \includegraphics[width=.49\textwidth]{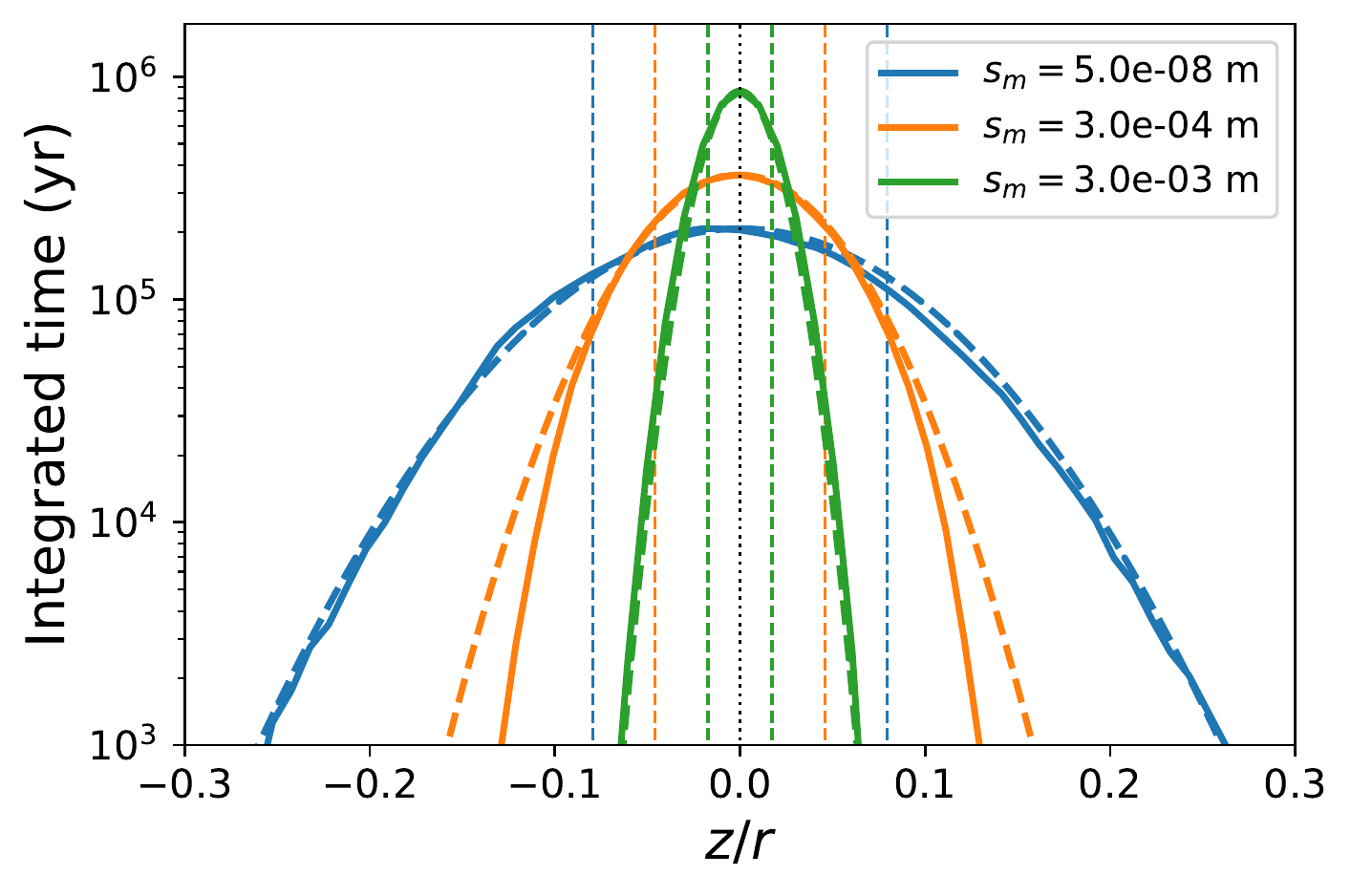}
    \caption{The integrated residence time for monomers of ${s_\text{m}=5\e{-8}}$ m (blue), $3\e{-4}$ m (orange), and $3\e{-3}$ m (green) as a function of height $z$ at $r=10$ AU. The solid lines denote the histogram obtained from the trajectories of $10^3$ monomers combined, whereas the dashed lines are the theoretical vertical density profiles expected from Eq. \eqref{eq:DubrulleSettling}. The vertical dashed lines depict the locations where $z=H_\text{d}$ for each monomer size.}
    \label{fig:22GaussTest}
\end{figure}

As a test for our dynamical model, we consider the vertical motion of $10^3$ monomers at fixed radius $r=10$ AU in Fig. \ref{fig:22GaussTest}. Our testing approach is similar to \cite{Ciesla2010} and \cite{Krijt&Ciesla2016}. We perform this test for monomers of three different sizes moving in the \texttt{const}-background disk model: A monomer of radius $s_\text{m}=5\e{-8}$ m, ${s_\text{m}=3\e{-4}}$ m, and $3\e{-3}$ m. At $r=10$ AU and $z/r=0$, the midplane Stokes numbers associated with these monomer sizes are ${\text{St} \approx 1.9\e{-7}}$, $1.2\e{-3}$, and $1.2\e{-2}$, respectively. These three monomers correspond to a fully coupled case (${\text{St} \ll \alpha}$), a partially coupled case (${\text{St} \sim \alpha}$), and a more decoupled case (${\text{St} >\alpha}$). The integration time for each monomer was chosen to be $1.1\e{5}$ yr, with a constant timestep of $1$ yr. This timestep is at least a factor 100 shorter than the turbulent stirring timescale and the settling timescale associated with the largest particles. Since the turbulent mixing timescale at $r=$10 AU is roughly $10^4$ yr, we omitted the first $10^4$ yr of each monomer trajectory from Fig. \ref{fig:22GaussTest} (see Sect. \ref{sec:2.5} and Sect. \ref{sec:3.1}).\\
The total time that these monomers spend at each height $z/r$ should be proportional to the background density profile of dust particles with the same Stokes number, which is a Gaussian profile with its standard deviation given by the dust scale height (Eq. \eqref{eq:DubrulleSettling}). Fig. \ref{fig:22GaussTest} shows that this holds for the smallest monomers (${5\e{-8}}$ m) and largest monomers (${3\e{-3}}$ m), whereas Eq. \eqref{eq:DubrulleSettling} appears to overestimate the width of the density profile of intermediate-sized monomers ($3\e{-4}$ m). This deviation was also found by \cite{Krijt&Ciesla2016}, and is a consequence of Eq. \eqref{eq:DubrulleSettling} not taking into account variations in the Stokes number as a function of height $z/r$ \citep[][]{Ciesla2010}, whereas it is in our model.

\subsection{Collisions}
\label{sec:2.3}
With the collision model, we aim to track the evolution of the home aggregate size $s_m$ and monomer depth $z_m$ as the home aggregate is altered by coagulation, fragmentation, and erosion. We here treat collisions in a similar fashion as \cite{Krijt&Ciesla2016}, i.e. collisions are allowed to occur randomly, while we track the effect of each collisional interaction on the home aggregate size $s_m$. In addition, we use a simple probabilistic model to determine the new monomer depth $z_m$ after a collision has occurred. Thus, during every global time step $\Delta t$, we first determine probabilistically whether a collision has occurred, and if so, what the effects are on the size of the home aggregate $s_\text{a}$ and on the depth at which the monomer is located $z_\text{m}$.\\
In order to determine the probability of a collision occurring during time step $\Delta t$, we are required to evaluate the collision rates for a given home aggregate of size $s_\text{a}$ with any other particle size bin present in our background dust size distribution. The collision rate for the home aggregate with a particle of size bin $i$ is
\begin{align}
\label{eq:CollisionRate}
    C_i=n_i\,v_\text{rel}\,\sigma_\text{col}.
\end{align}
Here, $n_i$ denotes the local number density of particles of size bin $i$, $\sigma_\text{col}=\pi(s_\text{a}+s_i)^2$ and $v_\text{rel}$, the relative velocity between the home aggregate and the collision partner. We relate $n_i$ to the partial disk surface density associated with particles from size bin $i$, $\Sigma_{\text{d},i}$
\begin{align}
    n_i=\frac{\Sigma_{\text{d},i}}{H_{\text{d},i}\sqrt{2\pi}}\exp{\left(-\frac{z^2}{2H_{\text{d},i}^2}\right)}
\end{align}
Here, $H_{\text{d},i}$ denotes the dust scale height associated with dust particles from size bin $i$. We can obtain the total dust surface density $\Sigma$ from $\Sigma_{\text{d},i}$ via
\begin{align}
    \Sigma_\text{d} = \sum\limits_i^{N_\text{bins}}\Sigma_{\text{d},i}.
\end{align}
The total relative velocity $v_\text{rel}$ associated with a collision between the home aggregate and a given collision partner constitutes contributions from Brownian motion ($v_\text{BM}$), differential aerodynamic drag consisting of radial drift ($v_r$), vertical settling ($v_z$), azimuthal drift ($v_\phi$), and turbulent motion ($v_\text{turb}$) \citep{Birnstiel+2010, Krijt+2016}.
\begin{align}
    v_\text{rel}=\sqrt{v_\text{BM}^2+v_r^2+v_z^2+v_\phi^2+v_\text{turb}^2}.
\end{align}
Following the approach of \cite{Krijt+2016}, we calculate the relative velocity due to Brownian motion as
\begin{align}
    v_\text{BM}=\sqrt{\frac{8k_\text{B}T_\text{d}(m_\text{a}+m_i)}{\pi m_\text{a}m_i}}.
\end{align}
Here $m_\text{a}=\frac{4}{3}\rho_\text{a}\pi s_\text{a}^3$ denotes the mass of the home aggregate (with $\rho_\text{a}$ denoting the mass density of the home aggregate), and $m_i$ denotes the mass of a collision partner from size bin $i$. All dust grains, whether part of a size bin or a monomer have equal material density.\\
Differential aeorodynamic drag constitutes a radial ($v_r$), vertical ($v_z$), and azimuthal ($v_\phi$) component, given by:
\begin{align}
    v_r&=|v_{r,\text{a}}-v_{r,i}|\\
    v_z&=|v_{z,\text{a}}-v_{z,i}|\\
    v_\phi&=\bigg|\frac{v_{r,\text{a}}}{2\text{St}_\text{a}}-\frac{v_{r,i}}{2\text{St}_i}\bigg|
\end{align}
Here, $v_{r,n}$ and $v_{z,n}$ ($n\in(a,i)$) are calculated from Eq. \eqref{eq:vr} and Eq. \eqref{eq:vz}, respectively. St$_\text{a}$ denotes the Stokes number associated with the home aggregate, while St$_i$ denotes the Stokes number of the collision partner.\\
For the contribution of turbulence to the total relative velocity, we use the approach derived by \cite{Ormel&Cuzzi2007}:
\begin{align}
    v_\text{turb}=\delta v \cdot Q.
\end{align}
Here, $\delta v=c_\text{s}\sqrt{\alpha}$ denotes the mean random velocity of the largest turbulent eddies \citep{Krijt+2016}. The value of $Q$ depends on how the stopping time $t_\text{s}=\text{St}/\Omega$ compares to the orbital period $1/\Omega$ and the turnover timescale of the smallest scale eddies, $t_\eta=1/(\Omega \sqrt{\text{Re}})$:
\begin{align}
\label{eq:turbulentQ}
    Q=\begin{cases}
    \text{Re}^\frac{1}{4}|\text{St}_1-\text{St}_2|&\qquad \text{if } t_{\text{s}1}\leq t_\eta \\
    \sqrt{\frac{1}{1+\text{St}_1}+\frac{1}{1+\text{St}_2}} &\qquad \text{if } t_{\text{s}1}\geq \Omega^{-1}\\
    1.55\sqrt{\text{St}_1}&\qquad \text{ otherwise.}
    \end{cases}
\end{align}
Here, we associate the Stokes number St$_1$ with the largest particle, either being the home aggregate or the collision partner, and St$_2$ with the smallest of the two. Re denotes the turbulent Reynolds number, which we calculate as the ratio between the turbulent viscosity $\nu_\text{turb}$ and molecular viscosity $\nu_\text{mol}$:
\begin{align}
    \text{Re} = \frac{\nu_\text{turb}}{\nu_\text{mol}}=\alpha c_\text{s}H_\text{g}\left(\sqrt{\frac{8}{\pi}}\frac{c_\text{s}\lambda_\text{mfp}}{2}\right)^{-1}.
\end{align}
The above calculations are performed for every size bin $i$, to find the associated $C_i$ via Eq. \eqref{eq:CollisionRate}. Subsequently, the total collision rate $C_\text{tot}$ can be obtained via
\begin{align}
    C_\text{tot}=\sum\limits_i^{N_\text{bins}}C_i,
\end{align}
such that the probability of a single collision event happening during timestep $\Delta t$ is
\begin{align}
    P_\text{col}=1-\exp(-C_\text{tot}\Delta t).
\end{align}
Because this expression does not incorporate the possibility of multiple collisions, we are required to set $\Delta t$ shorter than the timescale for a single collision, $\Delta t \lesssim 1/C_\text{tot}$. This can become a problem if our home aggregate is large and located near the midplane. In this situation, the home aggregate will collide frequently with (sub)miron-sized grains, resulting in $C_\text{tot}$ becoming very large \citep{Krijt&Ciesla2016}.\\
To prevent the usage of prohibitively small values of $\Delta t$, we group collisions with collision partners which have a mass $m_i$ which is smaller than a fraction $f_\text{c}=10^{-1}$ of the home aggregate mass $m_\text{a}$ \citep{Ormel&Spaans2008, Krijt+2015, Krijt&Ciesla2016}. Instead of treating each collision between the home aggregate and small collision partners separately, we only allow the home aggregate to collide with a group of small collision partners at once. The effective collision rate for such a group collision will be (much) smaller than the collision rate between the home aggregate and the individual collision partners which constitute the group. We express the number of particles per group $N_\text{col}$ as 
\begin{align}
    N_\text{col}=\frac{1}{f_\text{c}}\frac{m_i}{m_\text{a}},
\end{align}
such that the modified collision rates become
\begin{align}
\Tilde{C}_i=
    \begin{cases}
    C_i&\quad \text{if } m_i/m_\text{a}>f_\text{c},\\
    \dfrac{1}{f_\text{c}}\dfrac{m_i}{m_\text{a}}\cdot C_i&\quad\text{if } m_i/m_\text{a}\leq f_\text{c}.
    \end{cases}
\end{align}
We then calculate the modified total collision rate as
\begin{align}
\label{eq:ModifiedCollisionRate}
    \Tilde{C}_\text{tot}=\sum\limits_{i}^{N_\text{bins}}\Tilde{C}_i,
\end{align}
such that the probability of a collision occurring during a timestep $\Delta t$ is
\begin{align}
    \Tilde{P}_\text{col}=1-\exp(-\Tilde{C}_\text{tot}\Delta t).
\end{align}
To determine whether a collision event occurs during a timestep, we generate a random number $R_3\in [0,1]$, such that if $R_3\leq \Tilde{P}_\text{col}$, the monomer has undergone a collision event.\\
\\
In order to determine the effect of a collision on the monomer and home aggregate, we first need to determine the mass of the collision partner, which we determine by generating another random number $R_4\in[0,\Tilde{C}_\text{tot}]$, and sum over the various collision size bins up to bin $n$, which is the first bin which satisfies
\begin{align}
    \sum_i^n\Tilde{C}_i\geq R_4.
\end{align}
We then select a particle from size bin $n$ as the collision partner. Note that this automatically means that size bins which have a higher associated collision rate are more likely to be selected.\\
The outcome of a collision is fully determined by the mass ratio of the collision partner and the home aggregate, and their sizes. The first step of determining the collision outcome is whether the collision is constructive (net dust growth) or destructive (net dust fragmentation). We assume that the probability of a collision event resulting in fragmentation $P_\text{frag}$ can be fully expressed in terms of the relative velocity $v_\text{rel}$ \citep{Birnstiel+2011}: 
\begin{align}
\label{eq:Pfrag}
    P_\text{frag}=
    \begin{cases}
    0&\qquad\text{if } v_\text{rel}<v_\text{frag}-\delta v_\text{frag},\\
    1&\qquad\text{if } v_\text{rel}\geq v_\text{frag},\\
    1-\dfrac{v_\text{frag}-v_\text{rel}}{\delta v_\text{frag}}&\qquad \text{otherwise.}
    \end{cases}
\end{align}
We here defined $\delta v_\text{frag}=v_\text{frag}/5$, where we follow \cite{Birnstiel+2011}. After the collision partner and $P_\text{frag}$ have been determined, we generate a random number $R_5\in[0,1]$. A destructive collision will occur if $R_5\leq P_\text{frag}$.\\
Regardless of collision outcome, we need to re-calculate the home aggregate mass and size, $m_\text{a}$ and $s_\text{a}$, after each collision event. In addition, different collision events have an effect on the depth at which our monomer is embedded, $z_\text{m}$. We distinguish between the following collision outcomes:
\begin{enumerate}
    \item \textit{Coagulation}.
    In this case, $R_5>P_\text{frag}$ and the collision event results in the coagulation of the home aggregate and the collision partner(s) of mass $m_\text{c}$. The new home aggregate mass $m_{\text{a,new}}$ then follows from
    \begin{align}
    \label{eq:MassChange}
        m_{\text{a,new}}=m_\text{a,old}+m_\text{c}N_\text{col}.
    \end{align}
    Note that if $m_\text{c}/m_\text{a,old}>f_\text{c}$, the group size is $N_\text{col}=1$ (c.f. Eq. \eqref{eq:ModifiedCollisionRate}). The new monomer depth $z_{\text{m,new}}$ is calculated as
    \begin{align}
    \label{eq:DepthChange}
        z_\text{m,new} = z_\text{m,old}+s_\text{a,new}-s_\text{a,old},
    \end{align}
    where $s_{a,\text{old}}$ and $s_{a,\text{new}}$ denote the old and new home aggregate size, respectively. Effectively we thus assume that the mass of the collision partner(s) is added to the old home aggregate as a homogeneous surface layer, such that the monomer is buried deeper inside the new home aggregate.
    
    \item \textit{Fragmentation}.
    In this case, the home aggregate is catastrophically disrupted by the collision partner, which must be similar to the home aggregate in terms of mass: 
    \begin{align}
        f_\text{c}<\frac{m_\text{c}}{m_{\text{a,old}}}<\frac{1}{f_\text{c}}.
    \end{align}
    The new home aggregate of the monomer becomes one of the fragments produced during the collision. We assume these fragments follow the distribution \citep{Birnstiel+2010}
    \begin{align}
    \label{eq:FragmentDistribution}
        n(m)
        \begin{cases}
        \propto m^{-\xi_\text{frag}}&\qquad\text{if } m_\text{m}\leq m\leq m_\text{max},\\
        =0&\qquad\text{otherwise}
        \end{cases}
    \end{align}
    where $\xi_\text{frag}=1.83$ \citep{Brauer+2008a}. In addition, $m_\text{max}$ denotes the maximum possible fragment mass. In the case of fragmentation, we set $m_\text{max}=$max$(m_\text{a},m_\text{c})$. We draw the new monomer depth $z_\text{m,new}$ from a spherically uniform distribution, such that the probability $P(z_\text{m,new})$ of the monomer being located at a certain depth after a destructive collision is proportional to the home aggregate mass density $\rho_\text{a}$. This means that the monomer is more likely to be located close to the surface after a fragmentation event since $P(z_\text{m})\propto (s_\text{a}-z_\text{m})^2$. In reality, the functional form $P(z_\text{m,new})$ is likely set by the size of the original home aggregate $s_{a,\text{old}}$, size of the collision partner $s_\text{c}$ and their relative velocity $v_\text{rel}$ \citep{Dominik&Tielens1997}.
    
    \item \textit{Erosion}. Erosion occurs when $m_\text{c}/m_{a,\text{old}}\leq f_\text{c}$, as the collision partner is too small to result in full fragmentation of the home aggregate. Instead, the collision partner exhumes and ejects mass of the home aggregate. We here follow \cite{Krijt&Ciesla2016} and assume that the mass ejected is $m_\text{ej}=2m_\text{c}$. In order to determine whether the monomer is ejected, we define the ejection probability $P_\text{ej}$
    \begin{align}
        P_\text{ej}=\frac{m_\text{ej}}{m_{a,\text{old}}},
    \end{align}
    and generate another random number $R_6\in[0,1]$, such that the monomer is ejected if $R_6\leq P_\text{ej}$.
    \begin{itemize}
        \item \textit{Ejection}. In this case the new home aggregate is one of the fragments ejected, whose mass is selected according to Eq. \eqref{eq:FragmentDistribution}. In this case, we set $m_\text{max}=m_\text{c}$, and also draw the new monomer depth $z_\text{m}$ from a spherically uniform distribution. 
        \item \textit{No ejection}.
        If the monomer remains in its old home aggregate, the new lower home aggregate mass $m_\text{a,new}$ and smaller monomer depth $z_\text{m,new}$ are calculated via
        \begin{align}
            m_{a,\text{a,new}}&=m_\text{a,old}-m_\text{c}N_\text{col}\\
            z_\text{m,new}&=z_\text{m,old}+s_\text{a,new}-s_\text{a,old}.
        \end{align}
        This means that the monomer depth decreases as mass is removed isotropically from the surface. 
    \end{itemize}
    
    \item \textit{Impact}. In this case $m_\text{c}/m_\text{a,old}\geq f_\text{c}$, which means that the home aggregate is impacting and eroding the considerably larger collision partner. We here assume that the ejected material only originates from the collision partner, such that the new home aggregate becomes the collision partner, whose mass is
    \begin{align}
        m_\text{a,new}=m_\text{c}-m_\text{a,old}.
    \end{align}
    Again, we assume that the collision results in major restructuring of the new home aggregate, which is why we again use a spherically uniform distribution to determine a new value for the monomer depth $z_\text{m}$.
\end{enumerate}

\begin{table*}[ht!]
\centering
\begin{tabular}{l|ll|ll|ll}
\hline
$v_\text{frag}$ (m/s)  & 1                         &            & 5                        &            & 10                       &\\ 
            & Freq. & \% & Freq. & \%& Freq. & \% \\\hline\hline
Coagulation  & \multicolumn{1}{l}{$8.9\e{5}$}& 26.2 & \multicolumn{1}{l}{$4.3\e{6}$}& 87.5 & \multicolumn{1}{l}{$5.0\e{6}$}& 97.2   \\ 
Fragmentation& \multicolumn{1}{l}{$9.3\e{5}$}& 27.4 & \multicolumn{1}{l}{$2.0\e{5}$}& 4.0  & \multicolumn{1}{l}{0}& 0               \\ 
Erosion      & \multicolumn{1}{l}{$2.5\e{5}$}& 7.3 & \multicolumn{1}{l}{$1.0\e{5}$}& 2.1  & \multicolumn{1}{l}{$7.4\e{3}$}& 0.1      \\ 
Ejection     & \multicolumn{1}{l}{$5.8\e{5}$}& 17.4 & \multicolumn{1}{l}{$3.0\e{5}$}& 6.0  & \multicolumn{1}{l}{$1.3\e{5}$}& 2.6    \\ 
Impact       & \multicolumn{1}{l}{$7.3\e{5}$}& 21.7 & \multicolumn{1}{l}{$2.2\e{4}$}& 0.4  & \multicolumn{1}{l}{0}& 0               \\ \hline
Total        & \multicolumn{1}{l}{$3.4\e{6}$}&    & \multicolumn{1}{l}{$4.9\e{6}$}&    & \multicolumn{1}{l}{$5.1\e{6}$} &     \\ \hline
\end{tabular}
\caption{Frequency of different collision outcomes for the collisional histories in Fig. \ref{fig:23SlopeTest}.}
\label{tab:CollisionTestTable}
\end{table*}

\begin{figure}
    \centering
    \includegraphics[width=.49\textwidth]{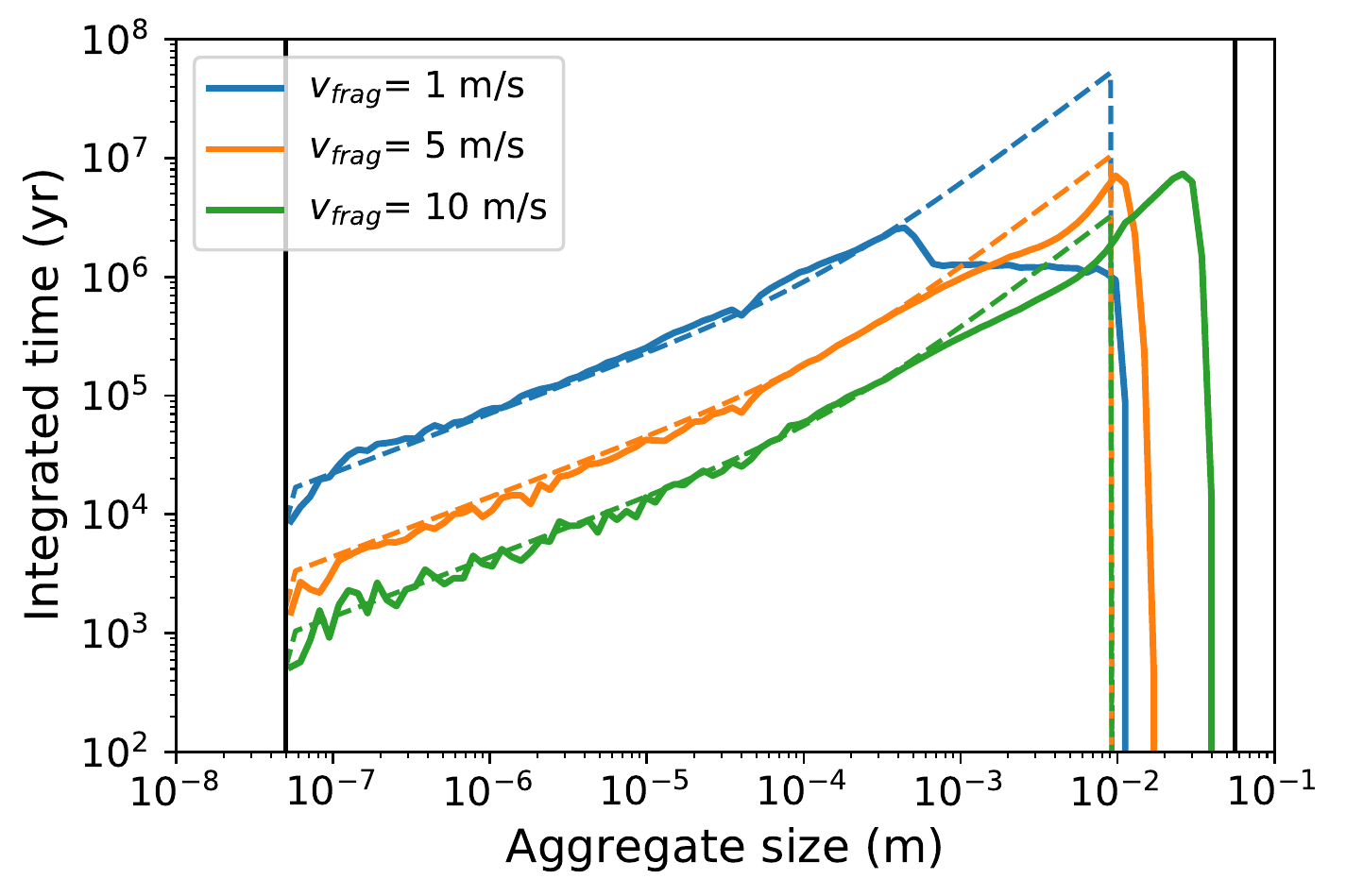}
    \caption{Integrated monomer residence time as a function of home aggregate size for different fragmentation velocities (solid coloured lines). The solid black lines on the left and right represent the minimum and maximum grain size occurring throughout the \textit{entire} \texttt{vFrag5} background disk model. The dashed lines all denote the same background dust mass distribution as a function of size, $\rho_\text{d}(a)$ at $r=10$ AU and $z=0$, shifted vertically for comparison with the integrated monomer residence time.}
    \label{fig:23SlopeTest}
\end{figure}

We consider a test similar to the one performed by \cite{Krijt&Ciesla2016}. Since tracking a monomer effectively entails tracking a unit of mass, the fraction of time which monomers spend inside aggregates of a given size $s_\text{a}$, when summed over many individual monomer histories, must be proportional to the background local dust mass density as a function of grain\footnote{Throughout this discussion, the usage "grain" signifies a dust particle which is either part of the background model or a collision partner, whereas "aggregate" refers to dust particles containing a monomer.} size $a=s_\text{a}$, $\rho_\text{d}(a,r,z)$. We compare the total integrated monomer residence time of $6000$ monomers to the \texttt{vFrag5} background model grain size distribution $\rho_\text{d}(a,r,z)$ for monomers in aggregates with fragmentation velocities of $v_\text{frag}=1,5,10$m/s. All monomer positions are kept fixed at $r=10$ AU and $z/r=0$, and initially assumed to be bare, such that the initial home aggregate size is always $s_\text{a}=s_\text{m}=5\e{-8}$ m. Furthermore, the aggregate density is assumed equal to the monomer density (i.e. $\rho_\text{a}=\rho_\text{m}$, corresponding to a filling factor $\phi=1$). Each simulation is run for 110 kyr, where we exclude the first 10 kyr of each simulation from the analysis such that we can neglect effects of our initial choice of home aggregate size. The timestep is set to $\Delta t = 10$ yr, which is several orders of magnitude below the total group collision rate $1/\Tilde{C}_\text{tot}$ at 10 AU (see Sect. \ref{sec:2.5} and Sect. \ref{sec:3.1}). \\
The integrated monomer residence times shown in Fig. \ref{fig:23SlopeTest} follow the background density distribution for all three fragmentation velocities $v_\text{frag}$ quite well for small aggregates up to $\sim 3\e{-4}$ m. In addition to the integrated monomer time in Fig. \ref{fig:23SlopeTest}, we also report the total number of occurrences of the different collision outcomes for all 6000 monomers in Table \ref{tab:CollisionTestTable}.\\
For all three fragmentation velocities, the residence time shows a small dip at ${\sim3\e{-5}}$ m. This results from the stopping time associated with this aggregate size becoming larger than the turnover timescale of the smallest scale eddies $t_\eta$ (c.f. Eq. \eqref{eq:turbulentQ}), resulting in a sudden increase in the relative velocity due to turbulence \citep{Ormel&Cuzzi2007}, which increases the collision rates associated with aggregates of $s_\text{a}\sim 3\e{-5}$ m. Therefore, these aggregates are depleted in a similar fashion as the smallest aggregates. We note that both the stopping time and $t_\eta$ depend on the local disk gas density $\rho_\text{g}$ and soundspeed $c_\text{s}$. This means that the location of this dip depends on position in the disk.\\
Above $s_a\sim 3\e{-4}$ m, the integrated residence time starts to deviate significantly from the background grain size distribution for all three values of $v_\text{frag}$. This is primarily due the fact that although our background grain size distribution incorporates settling, it does not directly account for the effects of collisions. For $v_\text{frag}=1 $m/s and $v_\text{frag}=10 $m/s, the peak in the residence time occurs respectively for aggregate sizes smaller or larger than the maximum grain size of the background grain distribution. This is expected since the maximum grain size in the background model is calculated for $v_\text{frag}=5$m/s. For $v_\text{frag}=1$m/s, aggregates larger than $\sim 3\e{-4}$ m should not be able to form (c.f. Fig. \ref{fig:21BackgroundAMAX}). This size limit is indeed the location of the peak in monomer residence time in Fig. \ref{fig:23SlopeTest}. However, the maximum grain size at 10 AU in the \texttt{vFrag5}-model is $a_\text{max}\approx 1\e{-2}$ m, which means that aggregates are still colliding with grains above the maximum aggregate size allowed by the collision model. Such a collision would occur at a relative velocity higher than $v_\text{frag}=1$m/s, although it is still possible for the monomer to end up in the larger collision partner via an impact event. Table \ref{tab:CollisionTestTable} shows that for the monomers in aggregates with $v_\text{frag}=1$m/s, impact events occur frequently. This can explain why such monomers still spend a considerable amount of time in aggregates with $s_\text{a}\gtrsim 3\e{-4}$ m. Such aggregates would have a short lifetime, as in this size regime only destructive collisions are possible. This also explains why the monomers in aggregates with $v_\text{frag}=1$m/s undergo significantly more fragmentation, erosion, and ejection events than monomers in aggregates with $v_\text{frag}=5,10$m/s (Table \ref{tab:CollisionTestTable}).\\
For monomers in aggregates with $v_\text{frag}=10$m/s, the aggregate is allowed to grow to sizes beyond the local value of $a_\text{max}\approx 1\e{-2}$ m in the background model. In this case, there are no grains of similar size present in the background model at $r=10$ AU and $z=0$, which means that the aggregate can continue to grow, primarily through the sweep-up of many groups of smaller grains. Similarly, when relative velocities start to exceed $v_\text{frag}=10$m/s, aggregates have grown to such an extent (up to $s_\text{a}\sim 4$ cm) that collisions with similarly-sized grains almost never occur. Table \ref{tab:CollisionTestTable} indeed shows no fragmentation events for monomers in aggregates with $v_\text{frag}=10$m/s, indicating that only erosion by many groups of small grains is a viable mechanism to limit further growth of the aggregate. However, Fig. \ref{fig:23SlopeTest} shows that this does not prevent monomers from spending a significant amount of time in aggregates larger than $a_\text{max}$.\\
For the case $v_\text{frag}=5$m/s, which is consistent with the value for $a_\text{max}$ calculated in the background model, the integrated residence time and $\rho_\text{d}$ are in good agreement for almost all grain sizes. However, some differences between the integrated monomer residence time and $\rho_\text{d}(a)$ still remain. The former shows a deficit with respect to $\rho_\text{d}(a)$ in the size regime below $a_\text{max}=1\e{-2}$ m, and also shows a small excess of time where monomers reside in aggregates larger than the maximum grain size in the background model $a_\text{max}$. In the collision model, the aggregate size is allowed to peak at sizes for which $v_\text{rel}>v_\text{frag}$. However, these aggregates are very short-lived, as collisions tend to result in fragmentation or erosion of such a large aggregate. This may also explain the significant number of fragmentation, ejection, and erosion events in Table \ref{tab:CollisionTestTable} for $v_\text{frag}=5$m/s. The formation of aggregates larger than $a_\text{max}$ may come at the cost of the number of aggregates between $s_a\sim 3\e{-4}$ m and $a_\text{max}$, as these are preferentially growing to aggregates slightly above $a_\text{max}$ with respect to the background model. Altogether these discrepancies appear to result from the fact that the grain size distribution in the background model is approximated with a power law distribution (Eq. \eqref{eq:dustSizeDistribution}), modified by the effects of settling. Furthermore, the grain density drops to zero in the first size bin above $a_\text{max}(r)$. However, this may be incorrect near the peak of the grain size distribution, where a dust size distribution limited by turbulence-driven fragmentation generally departs from a power law such as the one assumed in the background model \citep[][]{Birnstiel+2011, Krijt&Ciesla2016}.
\\
Altogether our collisional model is able to reproduce the background dust size distribution which is most consistent with the fragmentation velocity used in our model. As illustrated above, the usage of a background dust size distribution which is not automatically consistent with the collision model can induce errors in the collisional histories of individual monomers. This occurs when there is a difference between the fragmentation velocity in the collision model and the fragmentation velocity which determines the maximum grain size $a_\text{max}$ at a given position in the background model (Eq. \eqref{eq:amax}). The key variable affected is the home aggregate size $s_\text{a}$, specifically the average time a monomer spends in home aggregates of a given size. Furthermore, discrepancies between the value for $v_\text{frag}$ used in the collision model and the value used to calculate $a_\text{max}$ in the background model also affect the frequencies of different collision outcomes. Both of these errors propagate into the monomer depth $z_\text{m}$, which in turn can affect the ice evolution of the monomer (see also Sect. \ref{sec:4.2}). Therefore, it is key to ensure that the background model dust size distribution $\rho_\text{d}(a,r,z)$ is as consistent as possible with the assumptions in the collision model, such as the fragmentation velocity $v_\text{frag}$.

\subsection{Ice evolution}
\label{sec:2.4}
\begin{figure}
    \centering
    \includegraphics[width=.45\textwidth]{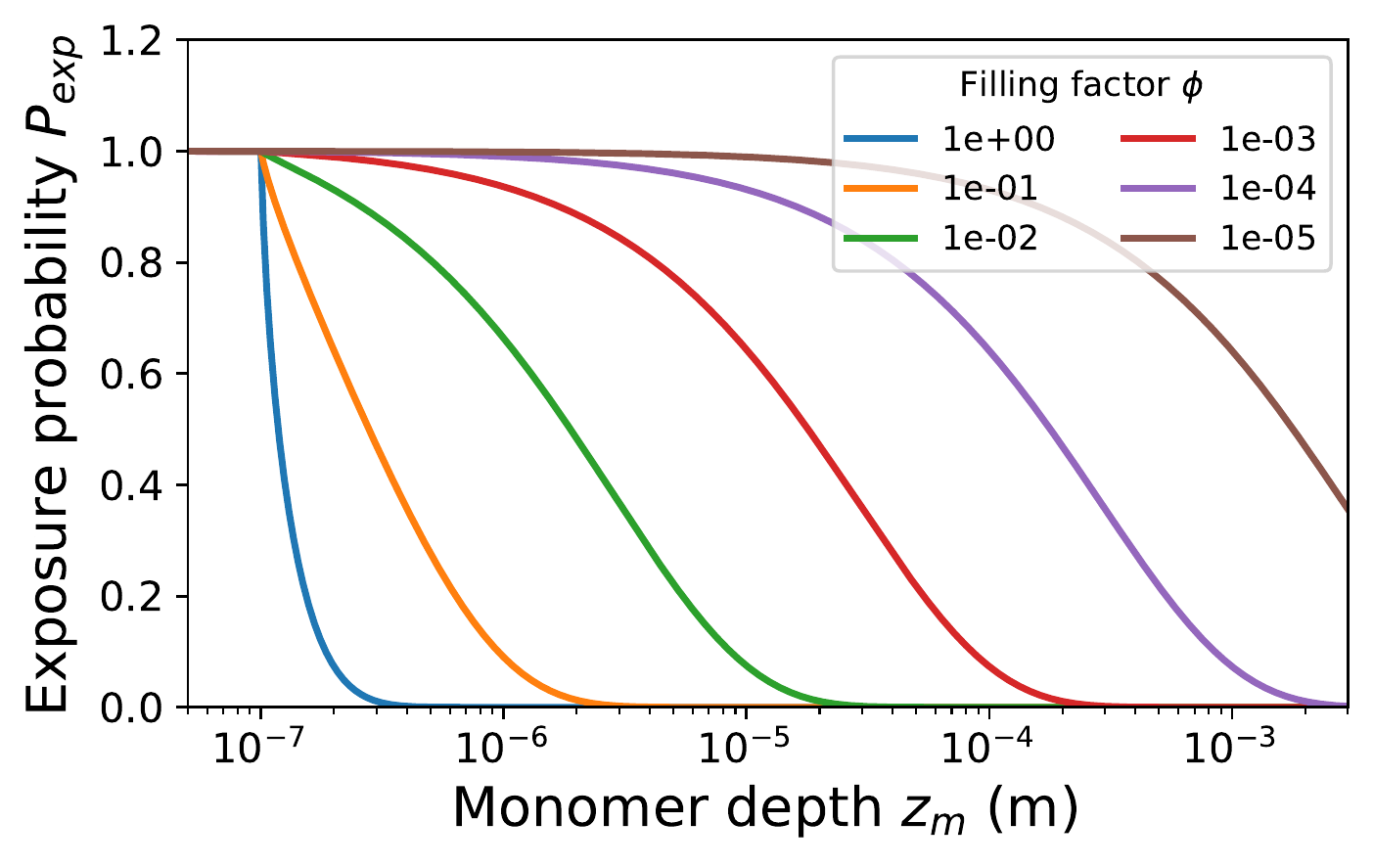}
    \caption{Exposure probability as a function of monomer depth $z_\text{m}$ for different filling factors $\phi$.}
    \label{fig:24PExp}
\end{figure}

The ice evolution model tracks the amount of ices of different species present in the ice mantle around the monomer. We here only track the ice composition of the monomer as it travels through the disk and is constantly changing home aggregate. Therefore, we aim to develop a stochastic approach for how the fluxes of volatile molecules and photons impinging onto the home aggregate affect the ices associated with the monomer. A crucial parameter in regards to both is how deep the monomer is embedded inside the home aggregate, $z_\text{m}$ (see Fig. \ref{fig:2MonomerAndHomeAggregate}). A monomer close to the surface ($z_\text{m}\sim s_\text{m}$) can more easily acquire an ice mantle from impinging molecules than a monomer buried beneath many layers of other monomers ($z_\text{m}\sim s_\text{a}\gg s_\text{m}$). However, such a buried monomer would also be more protected from photodesorption by the overlying monomers, which would absorb any incident UV radiation. If we assume that the local disk environment allows molecules to approach the home aggregate in straight lines (i.e. if the aggregate is in the Epstein regime), the monomers at $z<z_\text{m}$ can be treated as a slab of particles each of size $s_\text{m}$, absorbing individual impinging molecules and photons. From a radiative transfer viewpoint, each geometrical monomer depth $z_m$ can be associated with an optical depth $\tau$ defined as
\begin{align}
    \tau = \int\limits_0^{z_\text{m}}\sigma n_\text{m}dz.
    \label{eq:tau}   
\end{align}
Here, $\sigma$ denotes the collision cross section, and $n_m$ the monomer number density. We approximate the collision cross section with the monomer cross section $\sigma\approx \sigma_\text{m}= \pi s_\text{m}^2$, where we neglect the molecular cross section in case of impinging molecules. However, we note that this may be inappropriate for large molecular complexes which are similar in size to the monomer. Assuming constant monomer density throughout the home aggregate, we can estimate $n_\text{m}$ as
\begin{align}
\label{eq:nmonomer}
    n_\text{m}=\frac{\phi \rho_\text{m}}{m_\text{m}}=\frac{3\phi}{4\pi s_\text{m}^3},
\end{align}
where $\phi$ denotes the home aggregate mass filling factor and $m_\text{m}=\frac{4}{3}\pi s_m^3\rho_\text{m}$ denotes the monomer mass, such that Eq. \eqref{eq:tau} becomes 
\begin{align}
\label{eq:tauConst}
    \tau = \frac{3}{4}\frac{z_\text{m}}{s_\text{m}}\phi.
\end{align}
For an isotropic radiation field, the fraction of radiation which can reach down in the slab to an optical depth $\tau$ can be written as \citep[e.g.][]{Whitworth1975}
\begin{align}
\label{eq:probExp}
    E_2(\tau)&=\tau \int\limits_\tau^\infty\frac{e^{-t}}{t^2}\,dt.
\end{align}
Here, $E_2$ denotes the exponential integral, and represents the spatially averaged fraction of all impinging molecules and photons which can reach down to depth $z_\text{m}$ in the home aggregate. Note that the above equations strictly apply when $s_\text{a}\gg s_\text{m}$, and that the home aggregate has a fractal dimension $D_\text{f}=3$, such that $m_\text{a}\propto s_\text{a}^3$.\\
Although $E_2$ represents the fraction of molecules and photons which can reach down to a given optical depth $\tau$, the actual flux of molecules and photons received by an individual monomer does not have to scale with $E_2(\tau)$. Monomers may be arranged in fractal and clustered configurations inside the home aggregate as a consequence of initial hit-and-stick growth followed by compaction \citep[e.g.][]{Dominik&Tielens1997,Wurm&Blum1998, Weidling+2009}. This means that some monomers at depth $z_\text{m}$ may lie completely in the shadow of other monomers, whereas other monomers at depth $z_\text{m}$ may be fully exposed as there are no other monomers blocking incoming molecules and photons that travel inside the home aggregate effective radius $s_\text{m}$. Although full modeling of this spatial flux distribution is beyond the scope of this paper, we account for this behaviour by interpreting Eq. \eqref{eq:probExp} as an \textit{exposure probability} $P_\text{exp}$: the probability that a monomer, located at a depth $z_\text{m}$, is exposed to the outside environment. Being exposed as a monomer here means that the monomer ice mantle is able to undergo adsorption or photodesorption. Altogether we define the exposure probability as
\begin{align}
\label{eq:Pexp}
    P_\text{exp}=
    \begin{cases}
    1&\text{if } z_\text{m}<z_\text{crit}\\
    E_2(\tau); &\text{if } z_\text{m}\geq z_\text{crit},\\
    \tau=\frac{3}{4}\frac{z_\text{crit}-z_\text{m}}{s_\text{m}}\phi.
    \end{cases}
\end{align}
We here define the critical monomer depth $z_\text{crit}=2s_\text{m}=10^{-7}$ m to ensure that monomers which are closer to the surface than the diameter of a single monomer are always exposed, regardless of filling factor $\phi$. Fig. \ref{fig:24PExp} shows the value of $P_\text{exp}$ as a function of $z_\text{m}$ for different values of $\phi$. The filling factor is here varied from $\phi=1$ to $\phi=10^{-5}$. Such a low value for $\phi$ is thought to be possible for aggregates grown through efficient sticking and inefficient fragmentation, which results in very fluffy aggregates \citep[][]{Okuzumi+2012, Kataoka+2013}. For a filling factor of $10^{-5}$, it appears that molecules and photons have a significant chance to reach monomers which are embedded in the center of even the largest aggregates, amounting to more than 30 \% for $s_\text{a}= 10^{-3}$ m. For larger filling factors, monomers in the center of large aggregates are mostly shielded from gas phase molecules and photons. 

To model the evolution of the composition of the ice mantle of the tracked monomer we solve at each time step the following mass balance equation for each chemical species $x$ to find the change in ice mass of that species $M_x$:
\begin{align}
\der{M_x}{t}=
\label{eq:massbalance}
    \begin{cases}
    4\pi s_\text{m}^2\left(\frac{1}{2}\mathcal{R}_{\text{ads},x}-\mathcal{R}_{\text{tds},x}-\frac{1}{2}\mathcal{R}_{\text{pds},x}\right)&\qquad \text{if } R_7\leq P_\text{exp},\\
    &\\
    4\pi s_\text{m}^2\mathcal{R}_{\text{tds},x}&\qquad \text{otherwise.}
    \end{cases}
\end{align}
Here, $R_7\in [0,1]$ denotes a random number from a uniform distribution we generate after each collision event to determine whether the monomer is exposed. Furthermore, $\mathcal{R}_{\text{ads},x}$ denotes the specific adsorption rate (in $\unit{kg}{}\unit{m}{-2}\unit{s}{-1}$), $\mathcal{R}_{\text{tds},x}$ the specific thermal desorption rate, and $\mathcal{R}_{\text{pds},x}$ the specific photodesorption rate associated with species $x$. The factors $\frac{1}{2}$ in Eq. \eqref{eq:massbalance} associated with $\mathcal{R}_{\text{ads},x}$ and $\mathcal{R}_{\text{pds},x}$ follow from the assumption that on average, impinging molecules are only effectively adsorbed on the monomer hemisphere aimed towards the aggregate surface, whereas the same applies for photodesorption by impinging UV photons. We treat the ice transport processes as follows:
\begin{itemize}
    \item \textit{Adsorption} 
   Molecules in the gas phase can collide and stick to the monomer, such that the specific adsorption rate (in $\unit{kg}{}\unit{m}{-2}\unit{s}{-1}$) for species $x$ is given by the product of the molecule collision rate and sticking probability:
   \begin{align}
       \mathcal{R}_{\text{ads},x}=n_xm_xv_{\text{th},x}S.
   \end{align}
   Here, $n_x$ denotes the gas phase molecule number density (in $\unit{m}{-3}$), $m_x$ the molecular mass of species $x$. $v_{\text{th},x}$ denotes the mean thermal speed of species $x$ and is calculated as \citep{Woitke+2009}
   \begin{align}
       v_{\text{th},x}=\sqrt{\frac{k_\text{B}T_\text{g}}{2\pi m_x}}.
   \end{align}
   $S$ denotes the sticking probability. We calculate $S$ using the expression of \cite{He+2016}
   \begin{align}
   \label{eq:sticking}
       S=\alpha_S\left[1-\tanh\left(\beta_S\left[T_\text{d}-\gamma_S\frac{E_{\text{ads},x}}{k_\text{B}}\right]\right)\right].
   \end{align}
   Here, $\alpha_S, \beta_S$, and $\gamma_S$ are fitting parameters, which have been derived for experimental data on the sticking of H$_2$, D$_2$, N$_2$, O$_2$, CO, CH$_4$, and CO$_2$ on amorphous, nonporous water ice \citep{He+2016}. We note that $\alpha_S=0.5$ for all species to ensure $0\leq S \leq 1$, while the values of $\beta_S$ and $\gamma_S$ depend on molecule species. However, experimental data on $\beta_S$ and $\gamma_S$ for the sticking of different chemical species on different surfaces is limited. Therefore, we follow \cite{He+2016} and use ${\beta_S=0.11\,\unit{K}{-1}}$ and $\gamma_S=0.042$ for all species as an approximation.\\
   
    \item \textit{Thermal desorption} We express the thermal desorption rate $k_{\text{tds},x}$ for species $x$ as 
\begin{align}
    k_{\text{tds},x}=\nu_x\exp\left(-\frac{E_{\text{ads},x}}{k_\text{B}T_\text{d}}\right)
\end{align}
Here, $\nu_x$ denotes the ice lattice vibration frequency \citep{Tielens&Allamandola1987, Cuppen+2017}:
\begin{align}
\label{eq:latvib}
    v_x=\sqrt{\frac{2N_\text{ads}E_{\text{ads},x}}{\pi^2m_x}}.
\end{align}
$N_\text{ads}=10^{19}$ m$^{-2}$ denotes the density of adsorption sites on the monomer surface. Furthermore, $E_{\text{ads},x}$ denotes the adsorption energy associated with species $x$. The specific thermal desorption rate for an ice species $x$ can be expressed as \citep{Cuppen+2017}
\begin{figure*}[ht]
    \centering
    \includegraphics[width=.95\textwidth]{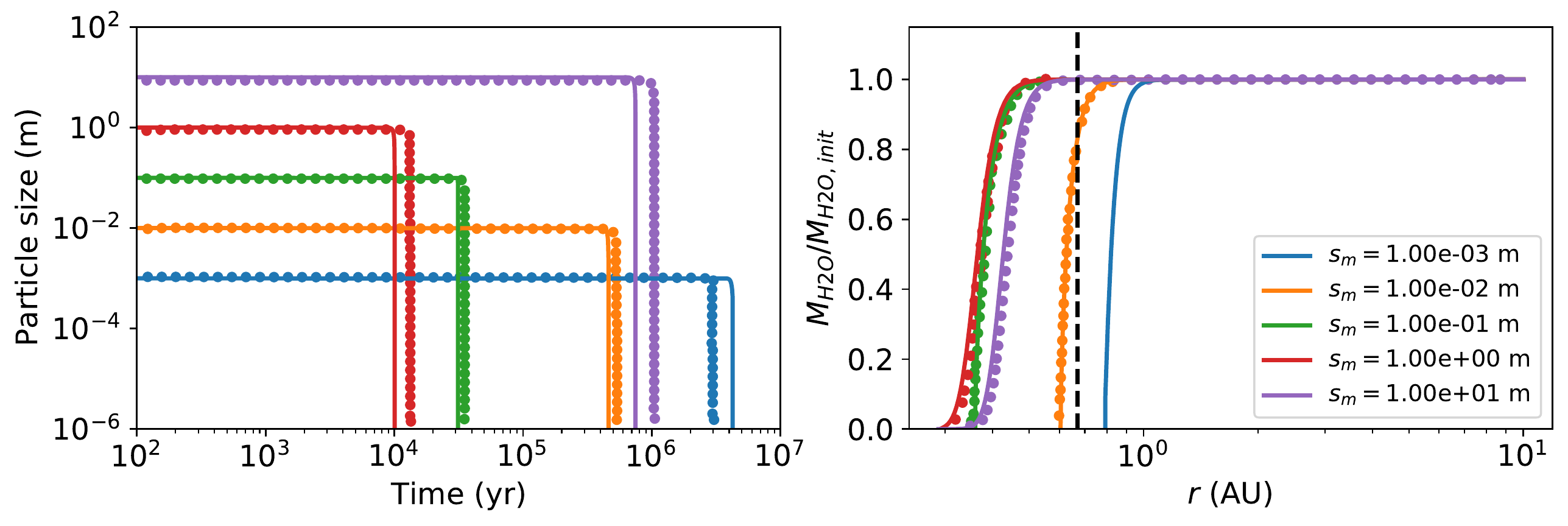}
    \caption{Particle size as a function of time (left panel) and ice mass (scaled with the initial ice mass) as a function of radial distance (right panel) for pure H$_2$O particles undergoing radial drift and thermal sublimation. Different colors denote different particles sizes. The trajectories predicted by the model of this work (solid lines) are compared with the trajectories calculated by \cite{Piso+2015} (dots). The vertical dashed line in the right panel denotes the position of the H$_2$O iceline at 0.67 AU as calculated by \cite{Piso+2015}.}
    \label{fig:24TruePisoComparisonSameEads}
\end{figure*}
\begin{align}
\label{eq:Rtds}    \mathcal{R}_{\text{tds},x}=k_{\text{tds},x}N_\text{act}f_xN_\text{ads}.
\end{align}
Here, $N_\text{act}$ is the number of actively desorbing ice molecule monolayers. Typical values for $N_\text{act}$ range from 2 to 4 \citep[][]{Cuppen+2017}, which motivated using a median value of 3 in this work. We express the ice fraction $f_x$ in terms of mass:
   \begin{align}
       f_x=\frac{M_x}{m_x\sum\limits_y M_y/m_y}.
   \end{align}
   Here, $M_x$ denotes the total mass of species $x$ present in the monomer ice mantle. \\
    \item \textit{Photodesorption} We treat photodesorption in a similar fashion as \cite{Woitke+2009}, and write the specific photodesorption rate as
       \begin{align}
       \mathcal{R}_{\text{pds},x}= Yf_x\chi F_\text{Draine}m_x.
   \end{align}
   Here $Y=1\e{-3}$ denotes the UV photon yield, and $\chi F_\text{Draine}$ denotes the local UV radiation field strength in $\unit{photons}{}\unit{m}{-2}\unit{s}{-1}$. 
\end{itemize}
In the above treatment, the escape of molecules from the home aggregate is not modelled. However, molecules liberated by thermal desorption from the ice mantles of monomers located at large $\tau$ collide with a significant number of other monomers before escaping back into the gas phase, which can result in re-adsorption if the sticking probability $S>0$. We explore the timescale associated with molecule escape and the concept of re-adsorption in Appendix \ref{sec:AB}. We find that re-adsorption may be important for particular molecular species in a comparatively small region immediately behind the iceline of the respective molecular species. Thus, our model could underestimate the amount of ice from individual species retained on monomers at large $\tau$ in these disk regions.\\
As a numerical test for our model, we compare the trajectories of pure H$_2$O monomer particles predicted by our model to the model of \cite{Piso+2015}. In our comparison, we consider particles which are allowed to undergo radial drift and thermal desorption. We here use the background disk structure and model parameters used by \cite{Piso+2015}.\\
The disk gas surface density is calculated as
\begin{align}
\Sigma_\text{P}(r) = 20 000 \left(\frac{r}{\text{AU}}\right)^{-1}\unit{kg}{}\unit{m}{-2},
\end{align}
such that the midplane gas density becomes, when assuming a Gaussian vertical density profile
\begin{align}
    \rho_\text{g,P}=\frac{\Sigma_\text{P}(r)}{H_\text{g}\sqrt{2\pi}}.
\end{align}
$H_\text{g}$ in this context is calculated as the pressure scale height ${H_\text{g}=c_\text{s}/\Omega}$. The soundspeed is obtained from the temperature profile \citep[][]{Piso+2015}
\begin{align}
    T_\text{g,P}(r)= 120\left(\frac{r}{\text{AU}}\right)^{-\frac{3}{7}}\unit{K}{}.
\end{align}
Note that since collisions, adsorption and photodesorption are not considered by \cite{Piso+2015}, the evolution of the particle is fully determined by the local gas density and temperature. Since the gas and dust temperature $T_\text{g}$ are coupled in the midplane, we assume $T_\text{g,P}=T_\text{d}$ for this test. For radial drift, the dimensionless pressure gradient $\eta$ (Eq. \eqref{eq:vr}) is estimated as
\begin{align}
    \eta_\text{P} = \frac{1}{2}\left(\frac{c_\text{s}}{r\Omega}\right)^2.
\end{align}
The sound speed $c_\text{s}$ is calculated via
\begin{align}
    c_\text{s}= \sqrt{\frac{k_\text{B}T_\text{g,P}}{\mu_\text{P} m_\text{p}}},
\end{align}
where the dimensionless mean molecular weight is set to $\mu_\text{P}=2.35$ \citep[][]{Piso+2015}.\\
We also set the value of the adsorption energy for H$_2$O, $E_{\text{ads, H}_2\text{O}}$ to $5800$ K, the same value as used by \cite{Piso+2015}. As a default value, we use the UMIST RATE12 value of $4800$ K \citep[]{McElroy+2013}. Furthermore, the particles have a density $\rho_\text{a}=2000$ kg/m$^3$, and a filling factor $\phi=1$. We calculate the ice lattice vibration frequency as \citep{Hollenbach+2009, Piso+2015}
\begin{align}
    \nu_{\H_2\O}=1.6\e{11}\sqrt{\frac{E_{\text{ads},\H_2\O}m_\text{p}}{k_\text{B}m_{\H_2\O}}}\,\unit{s}{-1},
\end{align}
while we calculate the specific thermal desorption rate $\mathcal{R}_\text{tds}$ using Eq. \eqref{eq:Rtds} with $f_{\H_2\O}=1$.\\
Fig. \ref{fig:24TruePisoComparisonSameEads} depicts the evolution of the monomer size as a function of time and H$_2$O ice mass as a function of radial distance $r$ for pure H$_2$O grains with sizes of $10^{-3}$ m up to 10 m. We here compare the evolutionary trajectories predicted by our model with the trajectories depicted in the center row of Fig. 4 of \cite{Piso+2015}, where the same particles are released in the disk midplane at 10 AU. Note that for this test, the effects of turbulent diffusion on radial migration is neglected, such that aerodynamic drag is the only process affecting the radial position of particles.\\
Both the model of this work and the model of \cite{Piso+2015} predict that regardless of initial particle size, the sublimation is almost instantaneous once a particle crosses the snowline. However, there appears to be a small difference in the time at which sublimation happens. For example, the model of \cite{Piso+2015} predicts sublimation of a particle of $10^{-3}$m after $3\e{6}$ yr, whereas the sublimation time in our model is $4\e{6}$ yr. However, a more detailed analysis revealed that the most likely cause of this deviation is inaccuracies in the digitation software used to extract the data points from Fig. 4 of \cite{Piso+2015}. This is further strengthened by the agreement of the trajectories for the ice masses as a function of radial distance shown in the right panel of Fig. \ref{fig:24TruePisoComparisonSameEads}. The authors also note the high sensitivity of the sublimation time to changes in model parameters.

\begin{table*}[ht]
    \begin{center}
    \begin{tabular}{cp{0.3\textwidth}rlc}
    \hline
         \textbf{Parameter}& \textbf{Name} &\textbf{Value} & \textbf{Unit} & \textbf{Reference} \\\hline\hline
         $s_\text{m}$ & Monomer size & $5\e{-8}$& $\unit{m}{}$ & (8) \\
         $\phi$ & Home aggregate mass filling factor & 0.1 & - & (8)\\
         $\mu$ & Mean molecular weight & $2.3$ & - & (4)\\
         $\sigma_\text{mol}$ & Mean molecular cross section & $2\e{-19}$& $\unit{m}{2}$ & (2) \\ \hline
         $N_\text{bins}$ & Number of collision partner size bins & 100 & - & (8) \\
         $f_c$ & Aggregate grouping mass fraction & $10^{-1}$ & - & (3)\\
        $v_\text{frag}$ & Fragmentation velocity & 5 & $\unit{m}{}\unit{s}{-1}$ &  (3)\\
        $\xi_\text{frag}$ & Fragment distribution power & 1.83 & - & (7)\\ \hline
        $\alpha_S$ & Sticking factor parameter & 0.5 & - & (5) \\
        $\beta_S$ & Sticking factor parameter & 0.11 & $\unit{K}{-1}$ & (5) \\
        $\gamma_S$ & Sticking factor parameter & 0.042 & - & (5) \\
        $N_\text{ads}$ & Monomer surface adsorption site density & $1\e{19}$ & $\unit{m}{-2}$ & (6) \\
        $N_\text{act}$ & Number of active surface layers & 3  & - & (8)\\
        $Y$ & UV photon yield & $1\e{-3}$ & - & (8)\\\hline
\end{tabular}
    \caption{Input parameters used for the monomer model. References: (1) \cite{Woitke+2016}, (2) \cite{Okuzumi+2012}, (3) \cite{Krijt&Ciesla2016}, (4) \cite{Krijt+2018}, (5) \cite{He+2016}, (6) \cite{Hollenbach+2009}, (7) \cite{Brauer+2008a}, (8) this work.}
    \label{tab:MonoTable}
    \end{center}
\end{table*}

\subsection{Timescales}
\label{sec:2.5}
To enable us to spatially constrain the regions where disk processes are coupled, we will compare their associated timescales later in this work (Sect. \ref{sec:3.1}). In this subsection we outline our approach to estimating these timescales.\\
For dynamical processes, we compare the radial drift, vertical settling, and turbulent mixing timescales, denoted by $\tau_\text{rm}$, $\tau_\text{zm}$, and $\tau_\text{tm}$, respectively:
\begin{align}
    \tau_\text{rm}=\frac{H_\text{g}}{v_r},\qquad \tau_\text{zm}=\frac{H_\text{g}}{v_z},\qquad 
    \tau_\text{tm}=\frac{H_\text{g}}{\alpha c_\text{s}}.
\end{align}
Here, $v_r$ and $v_z$ denote the drift velocities due to aerodynamic drag, given by Eq. \eqref{eq:vr} and Eq. \eqref{eq:vz}, respectively, while ${\nu_\text{turb}=\alpha c_\text{s} H_\text{g}}$. We here choose the gas pressure scale height $H_\text{g}$ as the length scale for the migration timescales ($\tau_\text{rm}$, $\tau_\text{zm}$) since $H_\text{g}$ is the typical length scale associated with the turbulent mixing timescale $\tau_\text{tm}$. For collision processes, the timescale over which a particle collides with a significant fraction of its own mass worth of collision partners can be estimated as 
\begin{align}
    \tau_\text{col}=\frac{1}{\Tilde{C}_\text{tot}}.
\end{align}
Here the total group collision rate $\Tilde{C}_\text{tot}$ is given by Eq. \eqref{eq:ModifiedCollisionRate}. We note that this definition differs from earlier work, where the timescale for collisional processing has been defined in terms of the time required for a particle to significantly change size as a function of the local dust-to-gas mass ratio \citep[see e.g.][]{Brauer+2008a, Birnstiel+2012}.\\
For ice formation, we consider the ice adsorption and desorption timescale for molecular species $x$ as a function of position. We here estimate the average amount of ice present on a monomer from the solid phase number density predicted by the background model, $n_{x,\text{ice}}$, as (Figs. \ref{fig:ABBackgroundModelconstAbundances}-\ref{fig:ABBackgroundModelvFrag10Abundances})
\begin{align}
\label{eq:initIce}
    M_{x,\text{ice}}=\frac{n_{x,\text{ice}}m_x}{\rho_\text{d}}m_\text{m}.
\end{align}
The adsorption and desorption timescales, $\tau_\text{ads}$ and $\tau_\text{des}$, respectively, are then given by
\begin{align}
\label{eq:tauAds}
    \tau_\text{ads}&=\frac{M_{x,\text{ice}}}{4\pi s_\text{m}^2\mathcal{R}_{\text{ads},x}}\\
    \label{eq:tauDes}
    \tau_\text{des}&=\frac{M_{x,\text{ice}}}{4\pi s_\text{m}^2\left(\mathcal{R}_{\text{tds},x}+\mathcal{R}_{\text{pds},x}\right)}.
\end{align}

    \section{Results}
    \label{sec:3}

As a first step, we use our model to constrain the disk regions where the dynamical, collisional and ice processes are highly coupled (Sect. \ref{sec:3.1}). Subsequently, we demonstrate how the coupled behaviour of these processes affects the ice mantle composition of individual monomers (Sect. \ref{sec:3.2}). For all monomer models, the parameter values denoted in Table \ref{tab:MonoTable} are used, unless noted otherwise.

\subsection{Coupling analysis}
\label{sec:3.1}
\begin{figure}
    \centering
    \includegraphics[width=.49\textwidth]{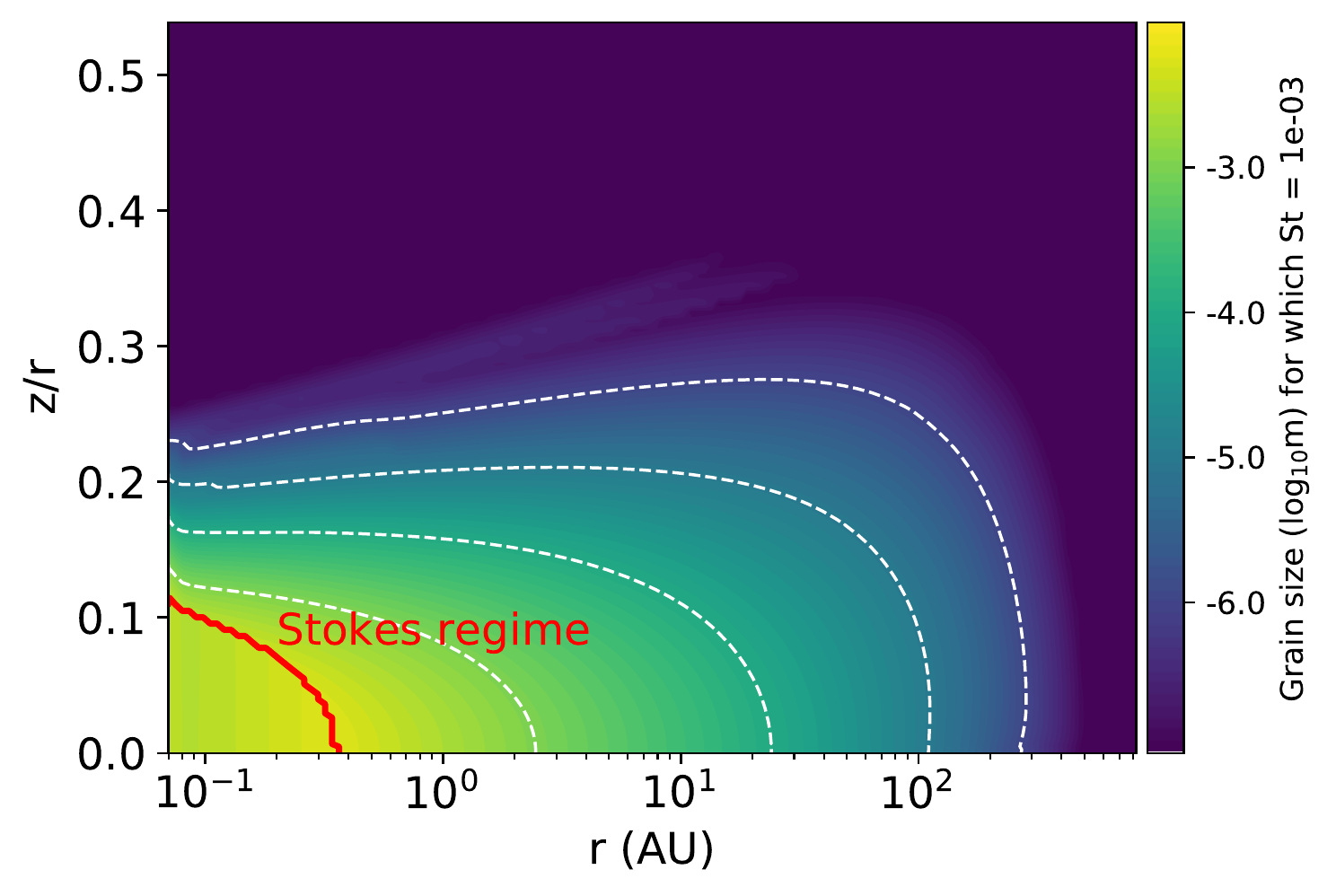}
    \caption{Grain size for which the local Stokes number is equal to $10^{-3}$, as a function of position in the \texttt{const} background disk model. The red curve encloses where particles of the size for which $\text{St} = 10^{-3}$ are in the Stokes drag regime. }
    \label{fig:31StokesNumber}
\end{figure}

As a starting point, we investigate the relative importance of the various grain transport processes. We explore the relative importance of turbulent stirring with respect to aerodynamic drift processes as a function of grain size in Fig. \ref{fig:31StokesNumber}. Throughout this section, we consider the \texttt{const}-model as the background disk model for our analysis, which means that the maximum grain size $a_\text{max}=3\e{-3}$ m remains constant as a function of $r$. Furthermore, we always assume a filling factor $\phi=1$, such that the material density is always equal to the monomer density, $\rho_\text{a}=\rho_\text{m}=2094$ kg/m$^3$. Fig. \ref{fig:31StokesNumber} highlights the physical grain size associated with $\text{St} =10^{-3}$ for the \texttt{const} background disk model discussed in Sect. \ref{sec:2.1}. We here inverted Eq. \eqref{eq:StEpstein} and Eq. \eqref{eq:StStokes} to give the grain size as a function of Stokes number. Grains with $\text{St} =10^{-3}$ would undergo significant settling (c.f Eq. \eqref{eq:DubrulleSettling}) for the turbulence strength parameter value used, $\alpha=10^{-3}$. It becomes clear that the physical grain size associated with a fixed Stokes number varies greatly throughout the disk, ranging from $\sim 10^{-2}$ m at the transition from the Epstein to the Stokes regime to $\lesssim 10^{-6}$ m in the outer disk and above $z/r\sim 0.25$. This is primarily a consequence of the background model gas density, which varies many orders of magnitude throughout the disk (Fig. \ref{fig:ABBackgroundModelconstStructure}). A lower gas density decreases the coupling of dust grains to the gas, and hence increases the Stokes number of a dust grain of given size (Eq. \eqref{eq:StEpstein} and Eq. \eqref{eq:StStokes}). We note that the grain size associated with $\text{St} =10^{-3}$ can also vary as a function of gas temperature due to the dependence of the Stokes number on the soundspeed. However, this effect is much less pronounced in Fig. \ref{fig:31StokesNumber} due to two effects. The dependence of the Stokes number on the gas temperature is weaker than its dependence on the gas density (${\text{St}\,\propto 1/c_s\propto 1/\sqrt{T_\text{g}}}$ compared to St $\propto 1/\rho_\text{g}$). Furthermore, the variation in the gas temperature is small compared to the variation in the gas density, both radially and vertically (c.f. Fig. \ref{fig:ABBackgroundModelconstStructure}). Despite the limited influence of the gas temperature, the dynamical behaviour of a dust grain of given size and its associated timescales can be expected to vary significantly as a function of position in the disk.\\
\begin{figure}
    \centering
    \includegraphics[width=.49\textwidth]{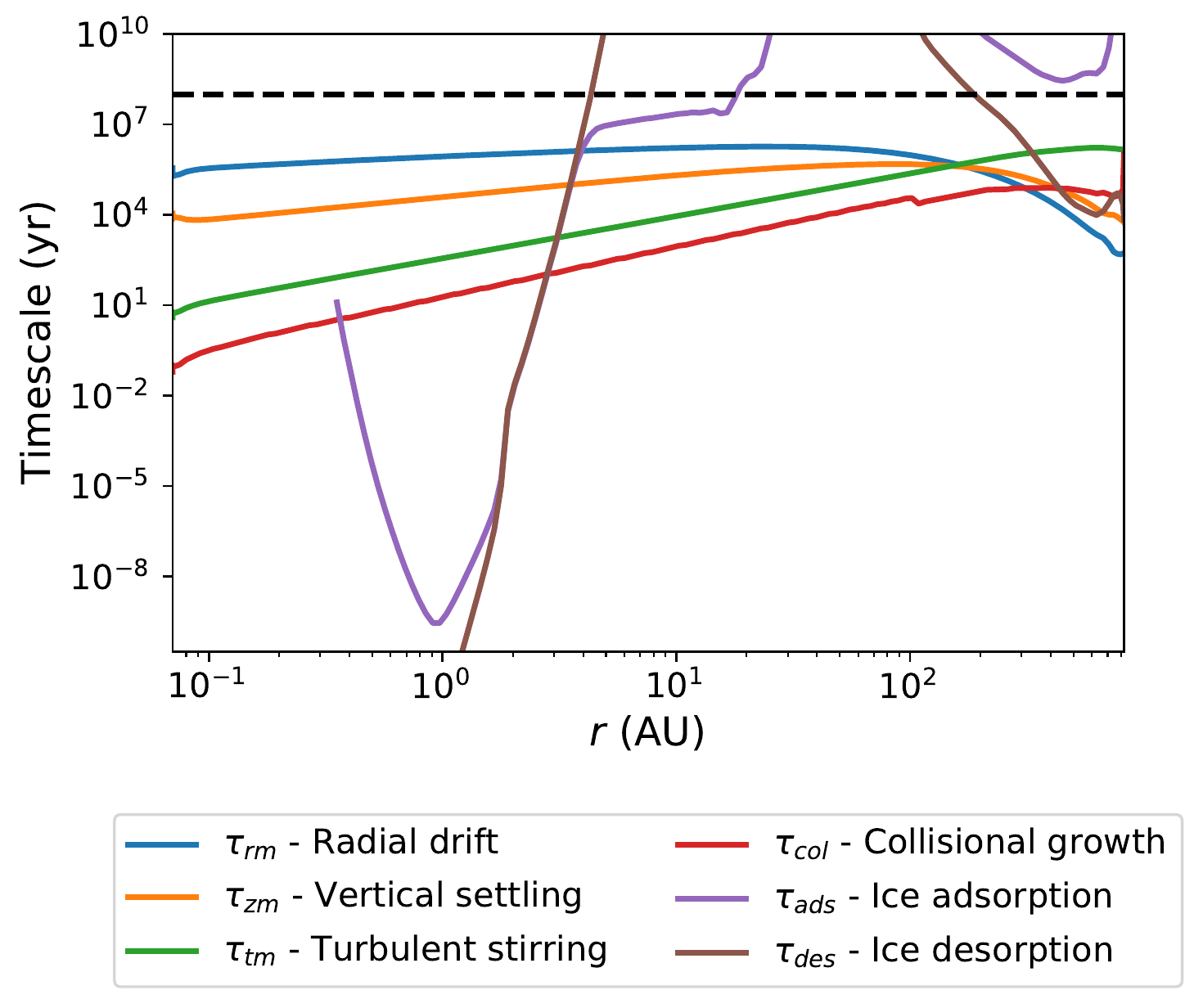}
    \caption{Behaviour of the timescales associated with various disk processes as a function of radial distance $r$. Timescales shown are for a monomer of size $s_\text{m}=10^{-5}$ m with a pure H$_2$O ice mantle, located at $z/r=0.05$. The dashed line at $t=10^8$ yr indicates the upper limit on timescales relevant for planet formation.}
    \label{fig:31TimescalesExample}
\end{figure}
\begin{figure*}[ht]
    \centering
    \includegraphics[width=.55\textwidth]{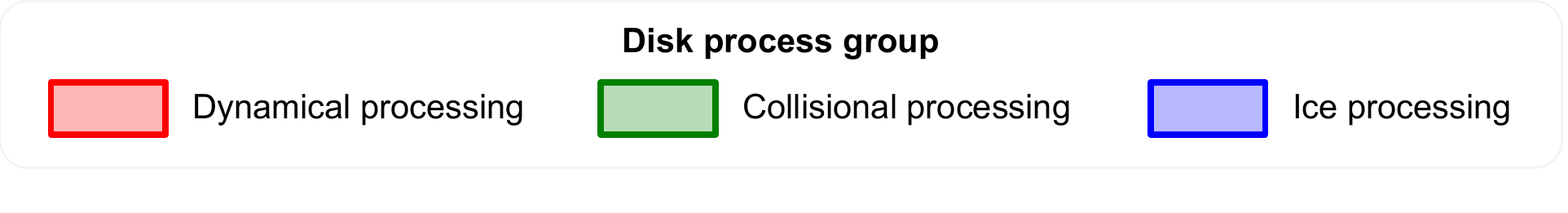}
\end{figure*}
\begin{figure*}[ht]
    \centering
    \includegraphics[width=\textwidth]{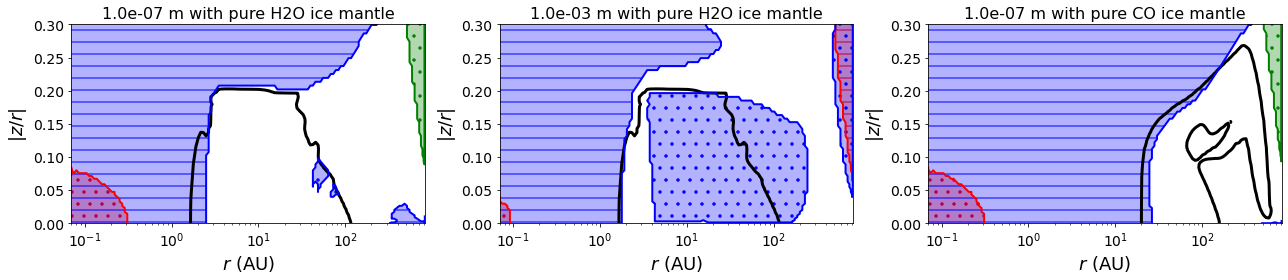}
    \caption{Process coupling behaviour for different grain sizes and ice mantle compositions. Colored areas indicate the decoupling of the different timescale groups. The filling of the coloured areas with solid lines or dots indicate the decoupling of a process category since it is much faster or slower, respectively. Black lines indicate icelines ($\tau_{ads}=\tau_{des}$).}
    \label{fig:31CouplingFigure}
\end{figure*}
\noindent As a next step, we investigate the coupling behaviour of the timescales introduced in Sect. \ref{sec:2.5}. As an example, we consider the radial behaviour of the timescales for radial drift, vertical settling, turbulent stirring, collisional growth, ice adsorption, and ice desorption in Fig. \ref{fig:31TimescalesExample}. The timescales are associated with a dust grain of $10^{-5}$ m with an H$_2$O ice mantle. The vertical height is kept fixed at $z/r=0.05$.\\
We find that in the inner disk ($r\lesssim 1$ AU), the collisional growth timescale $\tau_\text{col}$ becomes shorter than any dynamical timescale by at least an order magnitude, and drops even below $\sim 10^{-1}$ yr at $r=0.1$ AU. The shorter collision timescale for decreasing $r$ is a result of the increasing dust density closer to the star, which in turn increases collision rates (Eq. \eqref{eq:CollisionRate}).\\
The turbulent stirring timescale $\tau_\text{tm}$, although 1-3 orders of magnitude longer than $\tau_\text{col}$, is still far shorter than the aerodynamic drag timescales, $\tau_\text{rm}$ and $\tau_\text{zm}$, which are between $\sim10^4$ and $\sim10^6$ yr for $r\lesssim1$ AU. This is a consequence of the rather small particle size in combination with a high gas density in the inner disk, which together result in a small Stokes number ($\ll 10^{-3}$, c.f. Fig. \ref{fig:31StokesNumber}), and hence low drift velocities.\\
For $r\gtrsim 1$ AU, almost all timescales initially increase as a function of $r$. In case of the dynamical timescales, the increase results from the increase of gas scale height $H_\text{g}$ (Eq. \eqref{eq:GasStructure}). However, beyond 10 AU, the lower gas density results in an increasing Stokes number, and therefore increasing drift velocities result in $\tau_\text{rm}$ and $\tau_\text{zm}$ gradually decreasing around $r\gtrsim 100$ AU, overtaking turbulent stirring as the fastest dynamical processes around $r\sim150$ AU.\\
For $r \gtrsim 1$ AU, the ice formation timescales, $\tau_\text{ads}$ and $\tau_\text{des}$ both increase as a function of $r$. We define the H$_2$O iceline as the position where $\tau_\text{ads}=\tau_\text{des}$. Although Fig. \ref{fig:31TimescalesExample} indicates that $\tau_\text{ads}$ and $\tau_\text{des}$ are very similar for $r\lesssim 4$ AU, this condition puts the H$_2$O iceline at $r\approx 1.7$ AU for $z/r=0.05$. We note that both beyond the inner and outer boundary of the ice-forming region, $\tau_\text{des}$ decreases. Interior to the iceline this is due to increased thermal desorption, while in the outermost disk regions this is a consequence of higher photodesorption rates. The radial behaviour of $\tau_\text{ads}$ is primarily shaped by the local H$_2$O gas abundance, which explains its rapid decrease interior to the iceline: as H$_2$O ice sublimates interior to the iceline, the gas phase H$_2$O abundance increases significantly in the background model (c.f. Appendix \ref{sec:ABC}). The rapid increase of $\tau_\text{ads}$ for $r<1$ AU is a consequence of the decrease in the sticking probability $S$.\\
In terms of coupling behaviour, it becomes clear that in the inner disk, the various disk processes are largely decoupled since $\tau_\text{ads},\tau_\text{des}\ll\tau_\text{col}\ll\tau_\text{tm}\ll\tau_\text{rm},\tau_\text{zm}$, whereas dust dynamics, collisions, and ice processing become more coupled for increasing $r$ up to $r\approx 20$ AU. Beyond this distance, both $\tau_\text{ads}$ and $\tau_\text{des}$ become much larger than the other timescales, indicating that ice processing is very slow compared to collisional and dynamical processing in the outer disk regions.\\
It should be emphasized that this result only applies to a dust grain of size $s_\text{m}=10^{-5}$ m with a pure H$_2$O mantle. Since the collisional growth timescale is much smaller than the disk lifetime throughout most of the disk, it is expected that these dust grains will likely be incorporated rapidly into larger dust aggregates. Therefore, it is necessary to explore the coupling and decoupling of the various processes as a function of grain size, ice species, and also for the full extent of height $z/r$ above the disk midplane.\\
Fig. \ref{fig:31CouplingFigure} explores the coupling of the various timescales as a function of both $r$ and $z/r$ in case of a dust grain of $s_\text{m}=10^{-7}$ m (left panel) and $s_\text{m}=10^{-3}$ m (center panel). For $s_\text{m}=10^{-7}$ m, we also explore a case where we only allow a pure CO ice mantle to form instead of a pure H$_2$O ice mantle (right panel). We highlight the disk regions where the various timescales obey various decoupling conditions. We here categorized the timescales in a dynamical ($\tau_\text{rm}$,$\tau_\text{zm}$ and $\tau_\text{tm}$), collisional ($\tau_\text{col}$), and ice processing $\tau_\text{ads}$,$\tau_\text{des}$ group. The process with associated timescales $\tau_i$ then decouples from the other processes with associated timescales $\tau_j$ if
\begin{align}
\label{eq:muchFaster}
    f_\text{dec}\cdot\text{max}(\tau_i)<\text{min}(\tau_j),
\end{align}
or
\begin{align}
\label{eq:muchSlower}
    \text{min}(\tau_i)>f_\text{dec}\cdot\text{max}(\tau_j).
\end{align}
Here we set $f_\text{dec}=10^3$ to encapsulate that we require a particular process to act much faster (Eq. \eqref{eq:muchFaster}) or much slower (Eq. \eqref{eq:muchSlower}) than the other disk processes. For ice processing, we only consider $\tau_\text{ads}$ interior to the iceline, and $\tau_\text{des}$ exterior to the iceline in this calculation.\\
\begin{figure*}[ht!]
    \centering
    \includegraphics[width=.99\textwidth]{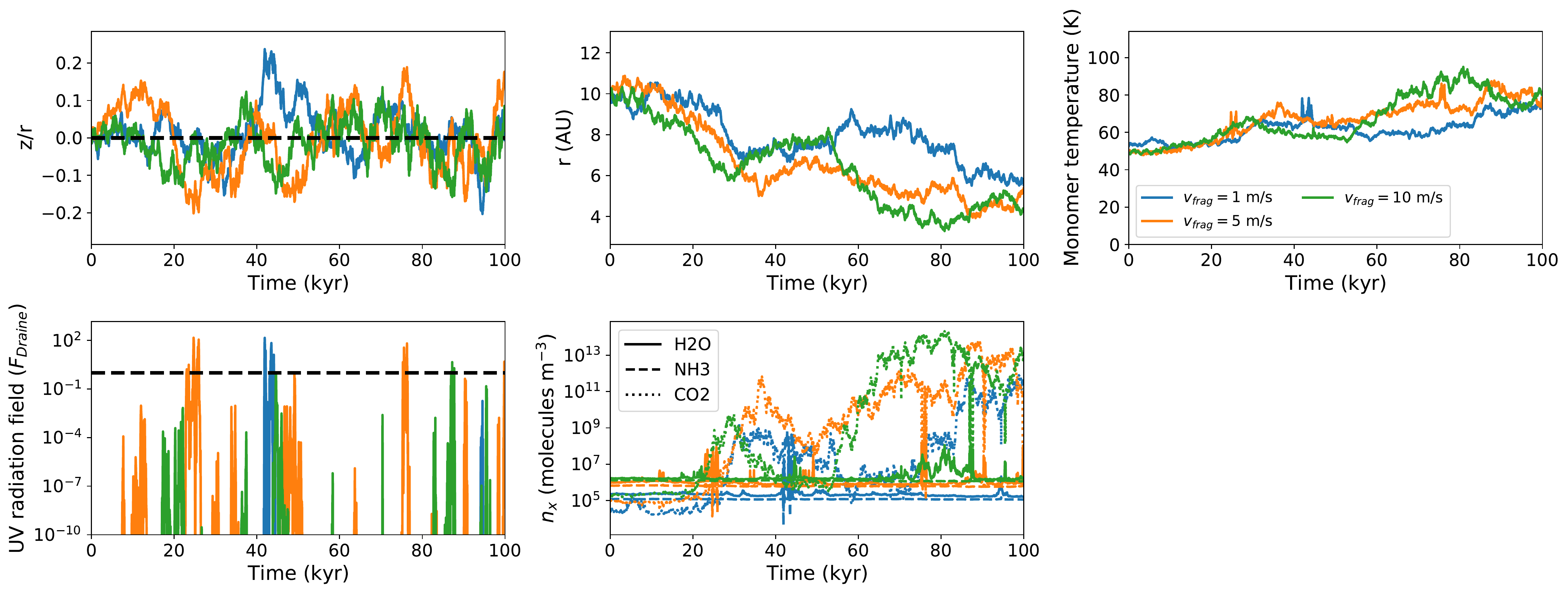}
        \caption{Evolution of the vertical positions (upper left panel), radial positions (upper center panel), temperatures (upper right panel), local UV fluxes (lower left), and local molecular number density in the gas phase (lower center) for monomers in home aggregates with different fragmentation velocity $v_\text{frag}$, released from $r=10$ AU and $z/r=0$. The black dashed line in the lower left panel indicates a UV radiation field strength of $1F_\text{Draine}$. Linestyles in the lower center panel indicate number densities of different gas phase species.}
    \label{fig:32MonomerDemoDynamics}
\end{figure*}
\noindent In all three cases, ice desorption decouples from the other processes in any disk region radially interior to the ice forming region. This decoupling originates from the short desorption timescale, since thermal desorption and photodesorption are very efficient outside the ice forming region. We also find in the cases of H$_2$O, ice formation can decouple from the other disk processes due to the low adsorption rate in the ice forming region (most clearly visible in the center panel of Fig. \ref{fig:31CouplingFigure}). This is primarily a result of the low gas phase abundance of H$_2$O inside the ice forming region (c.f. Appendix \ref{sec:ABC}). The center panel of Fig. \ref{fig:31CouplingFigure} shows that this effect is more pronounced for a large dust grain since more ice mass must be accumulated onto the grain surface to double the initial ice mantle mass (Eq. \eqref{eq:tauAds}).\\
In all three panels of Fig. \ref{fig:31CouplingFigure} we discern a region towards the innermost disk midplane where the three groups of processes fully decouple. In this region, dynamical processes become much slower than collisional growth, which is due to the high collision rates in the inner disk (c.f. Fig. \ref{fig:31TimescalesExample}). The center panel of Fig. \ref{fig:31CouplingFigure} also demonstrates that this region of full decoupling in the inner disk is smaller for larger dust grains. The shrinking of this region can be attributed to the fact that more collisions are required to result in a significant change in mass for a massive dust grain. Therefore, the effective collision rates become smaller, and hence $\tau_\text{col}$ increases.\\
The most important features of Fig. \ref{fig:31CouplingFigure} are the white regions in all three panels around the ice forming regions. For various species, all the three process categories couple to one another. However, the location of this fully coupled region is dependent on the grain size and position of the iceline. Since different grain sizes are connected via coagulation and fragmentation, and the position of the iceline depends on molecular species, there is no ubiquitously decoupled regime in the planet-forming region. In order to understand the ice composition of planetesimals, we thus require a modeling approach such as SHAMPOO which treats dynamical, collisional, and ice processing in a fully coupled fashion.

\subsection{Model demonstration and effects of different fragmentation velocity}
\label{sec:3.2}
As a first demonstration of the full SHAMPOO code, we consider the evolution of three monomers, initially not embedded in a home aggregate. The monomers are allowed to be incorporated in aggregates which have a fragmentation velocity of $v_\text{frag}=1,5$, or $10$ m/s. For each of these three scenarios, the appropriate background disk model is used (\texttt{vFrag1}, \texttt{vFrag5}, and \texttt{vFrag10}, respectively), while the monomers are released from $z=0$ at $r=10$ AU. The initial ice budget of the monomers is informed from the respective background disk model via Eq. \eqref{eq:initIce}. This means that the monomers start with a total amount of ice of $m_\text{ice}=0.0174m_\text{m}, 0.0176m_\text{m}$, and $0.0113m_\text{m}$ in the \texttt{vFrag1}-, \texttt{vFrag5}-, and \texttt{vFrag10}-model, respectively. The three monomers are allowed to be processed for $10^5$ yr. All home aggregates are assumed to have a filling factor of $\phi=0.1$ for all three scenarios. However, in practice the filling factor can differ for each home aggregate. We therefore explore the effects of different filling factors in Sect. \ref{sec:3.3}. The random number generator seed is kept fixed between all three scenarios, such that the fragmentation velocity and the associated background disk model are the only differences between the three simulations.

\subsubsection{Dynamical and environmental evolution}
Fig. \ref{fig:32MonomerDemoDynamics} shows the evolution of the vertical and radial position of the monomer, the associated monomer temperature and local UV radiation field. The monomer temperature is assumed to be equal to the local dust temperature. Despite the same initial monomer properties and model seed, the three monomers follow clearly distinct dynamical trajectories within the first 10 kyr. It becomes clear that all monomers are in the dynamical regime dominated by turbulent diffusion (c.f. Fig. \ref{fig:31StokesNumber}). In the vertical dimension, Fig. \ref{fig:32MonomerDemoDynamics} shows that in all three cases the monomer makes several vertical excursions to $|z/r|\sim 0.2$, whereas radially, all three monomers diffuse significantly inwards, with all monomers reaching inside of $r=6$ AU in 100 kyr. As the monomers gradually diffuse inwards, all three monomers experience increasing temperatures. However, the monomer in the \texttt{vFrag1}-model remains cooler than the other monomers between 60 and 80 kyr as it remains around $r=8$ AU in this period, while the monomers in the \texttt{vFrag5}- and \texttt{vFrag10}-model have already diffused interior to $r=6$ AU after 60 kyr. We also discern small peaks in the temperature evolutions of the monomers, which can be associated in time with the short-lived vertical excursions. \\
Fig. \ref{fig:32MonomerDemoDynamics} also reveals that the local UV radiation field strength is usually below $1F_\text{Draine}$ for all three monomers, although exceptions are short peaks in field strength, which can be again attributed to vertical excursions higher in the disk atmosphere. The monomer in the \texttt{vFrag1}-model here experiences significantly less peaks in field strength compared to the \texttt{vFrag5}- and \texttt{vFrag10}-model. These latter background models have a thinner dark midplane region (c.f. Fig. \ref{fig:ABBackgroundModelvFrag1Structure}, Fig. \ref{fig:ABBackgroundModelvFrag5Structure}, and Fig. \ref{fig:ABBackgroundModelvFrag10Structure}), and thus smaller vertical excursions are necessary for a monomer to reach the irradiated surface disk layers. Altogether monomers in the \texttt{vFrag5}- and \texttt{vFrag10}-model are more often in a regime where photodesorption can remove ice. The highest peaks in UV radiation field strength are found between 25 and 30 kyr for the monomer in the \texttt{vFrag5}-model and 40 and 25 kyr for the monomer in the \texttt{vFrag1}-model, where the UV radiation field very briefly peaks at $10^2F_\text{Draine}$ on both occasions. Although there exist many other peaks in UV radiation field strength, these peaks are usually well below $1F_\text{Draine}$ and of short duration. Altogether we thus expect photodesorption to affect the ice evolution of monomers only for short periods of time if a monomer is exposed during a peak in the background UV radiation field.

\begin{figure*}
    \centering
    \includegraphics[width=.8\textwidth]{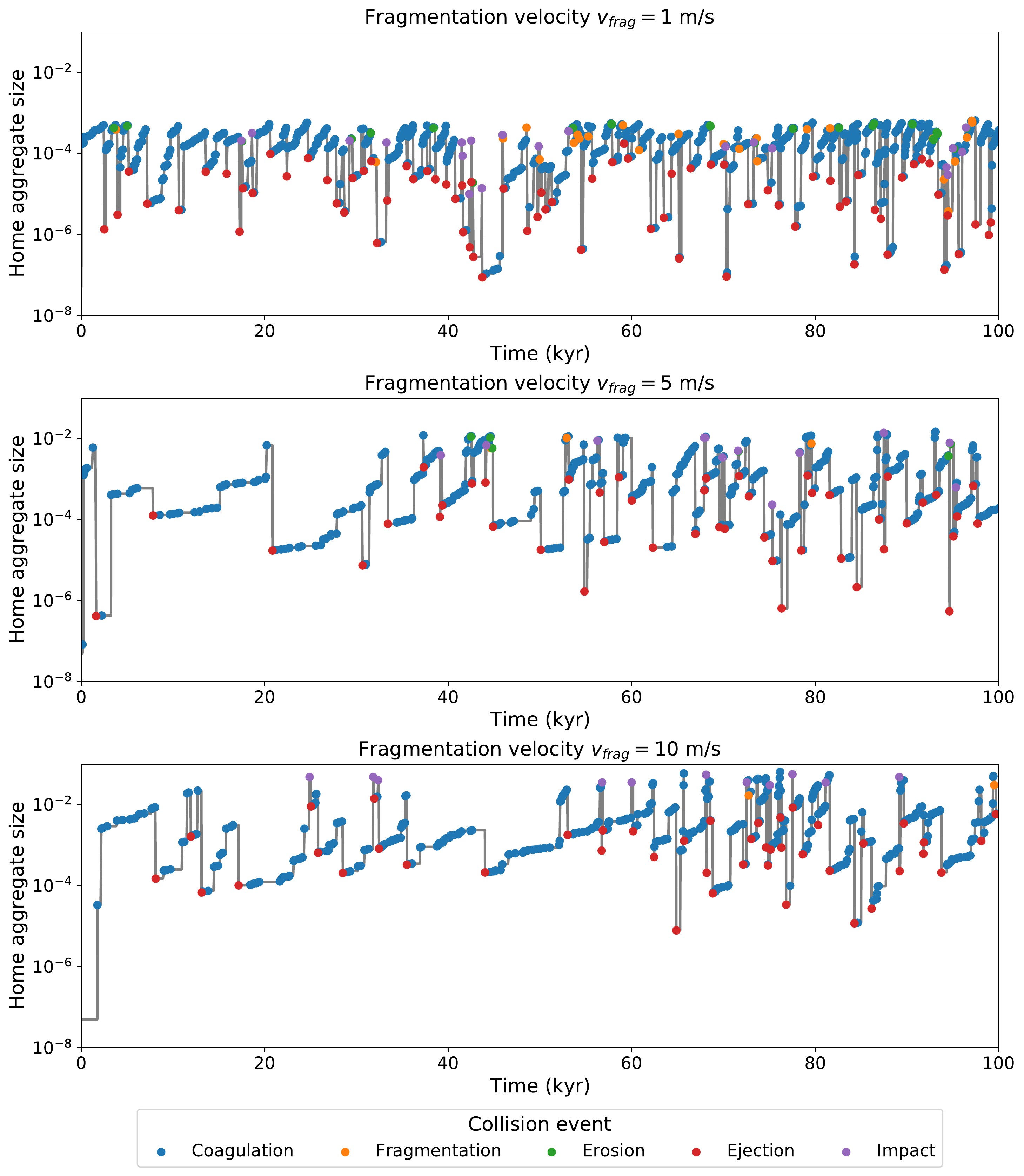}
        \caption{Collision histories and resulting home aggregate sizes $s_\text{a}$ as a function of time for monomers in aggregates of different fragmentation velocities $v_\text{frag}$. Note that the vertical position of each collision event coincides with the home aggregate size $s_\text{a}$ \textit{after} the collision has occurred.}
    \label{fig:32MonomerDemoCollisions}
\end{figure*}
\begin{figure*}
    \centering
    \includegraphics[width=.99\textwidth]{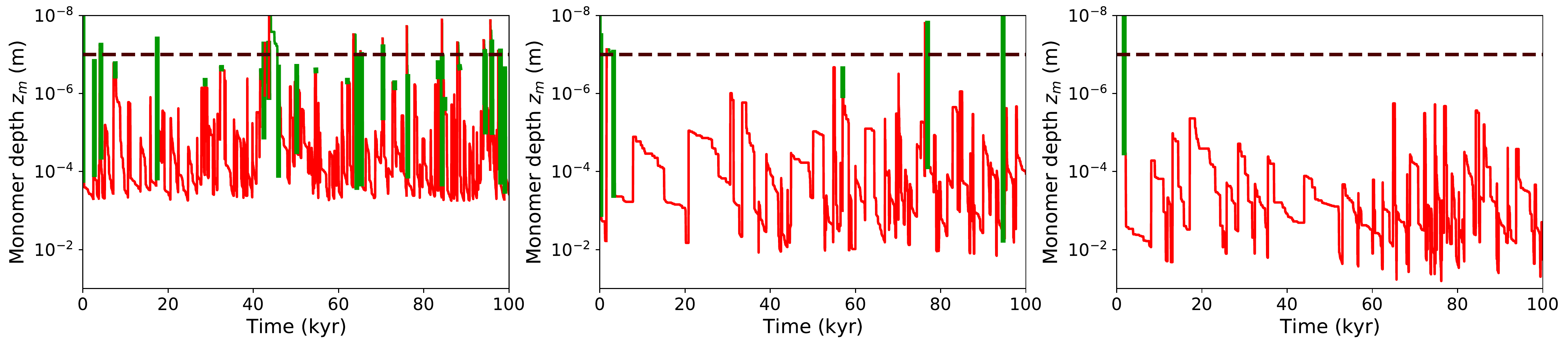}
    \caption{Evolution of monomer depth $z_\text{m}$ for monomers embedded in home aggregates with $v_\text{frag}=1$ m/s (left), $5$ m/s (center), and $10$ m/s (right). Green areas denote the times at which the monomer is exposed to gas phase molecules and UV photons, while the monomer is shielded from impinging molecules and UV photons otherwise. In each panel, the dashed line indicates $z_\text{m}=z_\text{crit}$.}
     \label{fig:32MonomerDemoDepth}
\end{figure*}
\begin{figure*}
    \centering
    \includegraphics[width=\textwidth]{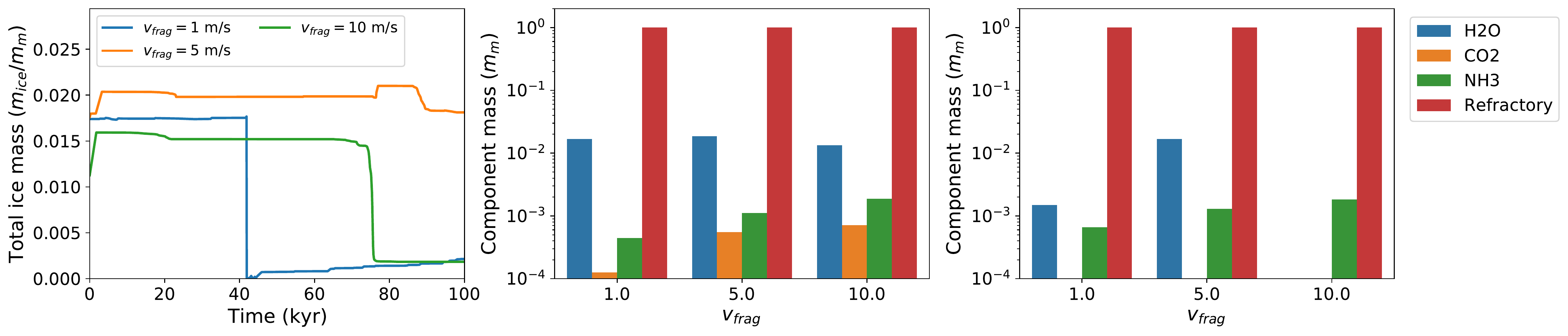}
        \caption{Evolution of the total ice mass $m_\text{ice}$ of monomers in aggregates with different fragmentation velocity $v_\text{frag}$ (left panel), together with snapshots of the monomer ice composition at 10 kyr (center panel), and 100 kyr (right panel).}
    \label{fig:32MonomerDemoIceEvolution}
\end{figure*}

\subsubsection{Collisional history}
\begin{table}[]
\centering
\begin{tabular}{l|ll|ll|ll}
\hline
$v_\text{frag}$ (m/s)  & 1                         &            & 5                        &            & 10                       &   \\
& Freq. & \% & Freq. & \%& Freq. & \% \\ \hline\hline
Coagulation           & \multicolumn{1}{l}{1018} & 87.4 & \multicolumn{1}{l}{551} & 89.0 & \multicolumn{1}{l}{473} & 89.2 \\ 
Fragmentation         & \multicolumn{1}{l}{25}  & 2.1  & \multicolumn{1}{l}{3}   & 0.4    & \multicolumn{1}{l}{2}   & 0.3          \\ 
Erosion               & \multicolumn{1}{l}{26}   & 2.2  & \multicolumn{1}{l}{7}   & 1.1  & \multicolumn{1}{l}{0}   & 0          \\ 
Ejection              & \multicolumn{1}{l}{76}  & 6.5 & \multicolumn{1}{l}{46}  & 7.4  & \multicolumn{1}{l}{44}  & 8.3 \\ 
Impact                & \multicolumn{1}{l}{20}  & 1.7 & \multicolumn{1}{l}{12}   & 1.9  & \multicolumn{1}{l}{11}   & 2.1          \\ \hline
Total                 & \multicolumn{1}{l}{1165} &         & \multicolumn{1}{l}{619}  &          & \multicolumn{1}{l}{530} &         \\ \hline
\end{tabular}
\caption{Frequency of different collision outcomes for the collisional histories in Fig. \ref{fig:32MonomerDemoCollisions}.}
\label{tab:CollisionTable}
\end{table}

The collisional histories of the three monomers are shown in Fig. \ref{fig:32MonomerDemoCollisions}, where we consider the evolution of the home aggregate size. Fig. \ref{fig:32MonomerDemoCollisions} also shows the type of interactions which result in a change in home aggregate sizes. We also count the number of collision outcomes per monomer history, giving rise to Table \ref{tab:CollisionTable}.\\
Fig. \ref{fig:32MonomerDemoCollisions} and Table \ref{tab:CollisionTable} show that fewer collisions occur over 100 kyr for increasing $v_\text{frag}$. The monomer in the \texttt{vFrag10}-model undergoes less than half as many collisions as the monomer in the \texttt{vFrag1}-model (530 collisions compared to 1165). This is a consequence of the total dust surface mass density $\Sigma_\text{d}$ in the background disk model. In the background disk models with higher $v_\text{frag}$, larger dust aggregates can form (see also Fig. \ref{fig:21BackgroundAMAX}), and thus the same amount of dust mass becomes concentrated in fewer dust particles, which in turn lowers collision rates. This is also confirmed by the typical home aggregate size seen in Fig. \ref{fig:32MonomerDemoCollisions}: $s_a$ can be larger than 1 cm in the \texttt{vFrag10}-model, while it is smaller than 1 mm in the \texttt{vFrag1}-model. We note that these home aggregate sizes are expected based on the maximum grain sizes allowed between $r=4$ AU and $r=10$ AU according to Fig. \ref{fig:21BackgroundAMAX}. Furthermore, for all three models, the frequency of collisions increases later in the evolution, which is a consequence of the radial diffusion inwards towards higher density disk regions (Fig \ref{fig:32MonomerDemoDynamics}) for all three monomers.\\
Although there are thus significant differences between the collision histories of the three monomers, coagulation is the most frequent collision outcome for all three fragmentation velocities, amounting to almost 90 \% of the collision outcomes in all three cases. Aggregate fragmentation and erosion (comprising erosion, ejection, and impact events) appear to be rather rare compared to coagulation, in particular for the \texttt{vFrag5}- and \texttt{vFrag10}-model. However, we note that coagulation is the only event allowed for collisions with $v_\text{rel}<v_\text{frag}$, which is true for most collisions in the monomer history. Once the home aggregate has grown to a size at which collisions with $v_\text{rel}>v_\text{frag}$ become commonplace, destructive collisions usually result in a smaller home aggregate, placing it back in the regime where, $v_\text{rel}<v_\text{frag}$.\\
Fig. \ref{fig:32MonomerDemoCollisions} shows that the largest decreases in home aggregate size are associated with monomer ejection, undoing the work of many coagulation events. This is expected since in these events, the monomer we track ends up in a fragment produced by erosion of the previous home aggregate. These fragments are guaranteed to be significantly smaller than the previous home aggregate (Sect. \ref{sec:2.3}).\\
Although coagulation can produce large sudden increases in home aggregate sizes (e.g. the first 5 kyr in the \texttt{vFrag10}-model), we also discern many episodes in all three models where the home aggregate grows through the sweep-up of many smaller aggregates. We note that the behaviour exhibited in Fig. \ref{fig:32MonomerDemoCollisions} is similar to the aggregate size history shown as an example in \cite{Krijt&Ciesla2016}. 

\begin{figure*} 
\centering 
\includegraphics[width=.99\textwidth]{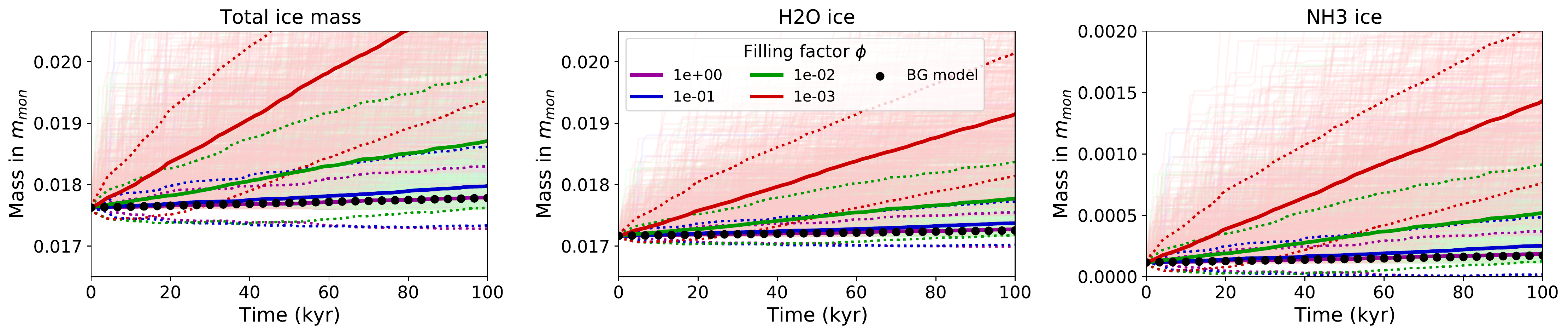} 
\caption{Evolution of the total amount of ice (left panel), H$_2$O ice (center), and NH$_3$ (right) averaged over 500 monomers for different home aggregate filling factors. Solid lines indicate the mean of the 500 monomers, whereas dotted lines indicate the standard deviation. The black dots indicate the ice evolution behaviour predicted by ProDiMo for the \texttt{vFrag5}-model. Trajectories of individual monomers are also shown in the background.} 
\label{fig:33PhiComparisonPopulation} 
\end{figure*}

\subsubsection{Ice evolution}
The evolution of the amount of ice on monomers depends on the local abundances of gas phase molecules at a given time (lower center panel of Fig. \ref{fig:32MonomerDemoDynamics}). Furthermore, the evolution of the monomer ice mantle crucially depends on the depth at which the monomer is embedded in its home aggregate $z_\text{m}$, which is depicted for the three scenarios in Fig. \ref{fig:32MonomerDemoDepth}. The monomer depth, which is a product of the collisional history, in combination with dynamical processes gives rise to the ice evolution depicted in Fig. \ref{fig:32MonomerDemoIceEvolution}. Since aggregates with lower fragmentation velocity remain smaller, it is more likely that the monomer is located close to the surface. Therefore, we find in Fig. \ref{fig:32MonomerDemoDepth} that the monomer in the \texttt{vFrag1}-model has the lowest average monomer depth compared to the \texttt{vFrag5}- and \texttt{vFrag10}-models. As a consequence, the monomer in this model is also exposed to the gas phase most often. In the \texttt{vFrag10}-model, the monomer only experiences one short initial period of exposure, which is mainly a result of our initial condition for the monomer to start without a home aggregate. For most of its evolution, the monomer in the \texttt{vFrag10}-model is located below $z_\text{m}=10^{-4}$ m. The \texttt{vFrag5}-model represents the intermediate case where the episodes of exposure after the initial phase occur less frequently than for the monomer in the \texttt{vFrag1}-model.\\
Fig. \ref{fig:32MonomerDemoIceEvolution} shows the evolution of the total amount of ice on the three monomers, along with snapshots of the monomer composition at 10 kyr and 100 kyr. Although the monomer in the \texttt{vFrag1}-model is exposed to gas phase molecules most often, the left panel of Fig. \ref{fig:32MonomerDemoIceEvolution} shows that after 100 kyr, the monomer has a similar total amount of ice ($2.1\e{-3}m_\text{m}$) as the monomer in the \texttt{vFrag10}-model ($1.8\e{-3}m_\text{m}$). Both models lost significant amounts of ice. In the \texttt{vFrag1}-model, this is the consequence of the monomer being exposed around 40 kyr, where the monomer makes a significant vertical excursion (c.f. Fig. \ref{fig:32MonomerDemoDynamics} and \ref{fig:32MonomerDemoDepth}), allowing all ice to evaporate in the strong UV radiation field via photodesorption. However, the frequent episodes of exposure do allow the monomer to regain a small amount of ice after the photodesorption around 40 kyr.\\
Although the monomer in the \texttt{vFrag10}-model is protected from photodesorption - as it is almost always unexposed, the left panel of Fig. \ref{fig:32MonomerDemoIceEvolution} shows that it still loses a significant amount of ice between 70 kyr and 80 kyr; a period where it also experiences the highest temperatures ($\sim 90$ K) of all three monomers. Moreover, the snapshot of ice composition at 100 kyr in the right panel of Fig. \ref{fig:32MonomerDemoIceEvolution} shows that the monomer in the \texttt{vFrag10}-model has lost all its H$_2$O ice, while at 10 kyr, all three monomers have similar amounts of water (center panel of Fig. \ref{fig:32MonomerDemoIceEvolution}). This is striking since the H$_2$O iceline based on the adsorption and desorption timescales is located around $r\approx 2$ AU around the disk midplane for all background models, a radial distance which is not reached by any of the monomers. Therefore, the adsorption rate is larger than the thermal desorption rate. However, since the monomer in the \texttt{vFrag10}-model is always unexposed, all the impinging molecules are adsorbed by monomers closer to the home aggregate surface. As the monomer we track is shielded from adsorption in this case, thermal desorption causes this monomer to lose all H$_2$O ice on its surface, despite the aggregate being located behind the H$_2$O ice line. However, this may be an overestimate of the amount of H$_2$O lost, as the H$_2$O lost through thermal desorption may in reality be partially replenished by the re-adsorption of H$_2$O originating from other nearby monomers (see also Appendix \ref{sec:AB}).\\
Compared to the other two monomers, the evolutionary trajectory for the monomer in the \texttt{vFrag5}-model is less dramatic, with more moderate episodes of adsorption during exposure and desorption at 20 kyr and 90 kyr. We note that this monomer is unexposed at the times of ice loss, suggesting that this loss occurs in a similar although less extreme fashion as for the \texttt{vFrag10}-model.\\
In terms of composition, the center and right panel of \mbox{Fig. \ref{fig:32MonomerDemoIceEvolution}} show that at 10 kyr, the three monomers have ice mantles of similar composition. Here, H$_2$O is the dominant ice species with smaller amounts of NH$_3$ and CO$_2$. It becomes clear that in addition to the H$_2$O ice loss for the monomers in the \texttt{vFrag1}- and \texttt{vFrag10}-model, all three monomers have lost their CO$_2$ after 100 kyr. This is a result of the low adsorption energy of CO$_2$ compared to H$_2$O and NH$_3$ ($E_{\text{ads},\C\O_2}/k_\text{B}=2990$ K compared to $E_{\text{ads},\H_2\O}/k_\text{B}=4800$ K and $E_{\text{ads},\N\H_3}/k_\text{B}=5534$ K \citep[]{McElroy+2013}). As all three monomers gradually diffuse inwards towards warmer disk regions, we expect all CO$_2$ to be lost through thermal desorption. On the contrary, the amount of NH$_3$ after 100 kyr remains fairly constant for all three monomers. For the monomer in the \texttt{vFrag5}- and \texttt{vFrag10}-model, this can be explained through the high adsorption energy of NH$_3$, which means that the role of thermal desorption is limited in comparison to CO$_2$ and H$_2$O. We note that lower values for $E_{\text{ads},\N\H_3}$ have been inferred \citep[e.g.][]{Sandford&Allamandola1993, Brown&Bolina2007}, which would result in NH$_3$ being less stable against thermal desorption in this disk region. For the \texttt{vFrag1}-model, all NH$_3$ ice at 100 kyr must have accumulated after the photodesorption event at 40 kyr, indicating that significant NH$_3$ adsorption has occurred after this catastrophic ice loss.

\subsection{Comparison to background model and effects of filling factor}
\label{sec:3.3}
Throughout this work, the home aggregates were assumed to have a filling factor of $\phi=0.1$ or $\phi=1$, which is consistent with dust aggregates which have undergone some degree of compaction. Experiments have demonstrated that typical filling factors $\phi$ associated with compact aggregates range from $\phi\eqsim 0.2$ \citep{Langkowski+2008} up to $\phi\eqsim0.4$ \citep{Weidling+2009, Weidling+2012}. Alternatively, modelling work has shown that large, fluffy aggregates with $\phi\lesssim10^{-3}$ can form if sticking is efficient and fragmentation is inefficient \citep{Okuzumi+2012, Kataoka+2013}. Since the value of $\phi$ affects how often a monomer is exposed and can accumulate ice, it is key to explore the sensitivity of ice formation in our model on $\phi$. Furthermore, ice evolution is also considered in the time-dependent chemistry of the ProDiMo background models. Therefore, we can compare the ice evolution predicted for a local monomer population by SHAMPOO to the ice evolution calculated by ProDiMo.\\
For these purposes, we consider the evolution of the average amount of ice present on 500 monomers over 100 kyr in the \texttt{vFrag5}-model in Fig. \ref{fig:33PhiComparisonPopulation}. Four different scenarios are considered in Fig \ref{fig:33PhiComparisonPopulation}, where monomers are embedded in home aggregates with a constant filling factor of $\phi=1, 10^{-1},10^{-2}$ or $10^{-3}$. As we also aim to compare the results to the fully local evolution predicted by the time-dependent chemistry in ProDiMo over the same time period, the monomer positions are fixed at $r=10$ AU in the disk midplane. This also means that the local disk environment remains the same for all simulations. Tests with dynamical processing enabled were also performed and yielded qualitatively similar results for the total amount of ice as shown in this section. However, the inclusion of dynamics results in a significant increase in the diversity of monomer trajectory and hence the standard deviation. \\
Initially, all monomers are assigned to a home aggregate of random size $s_a$ which is drawn from the background model dust mass distribution $\rho_\text{d}(a,r,z)$. Subsequently the monomer depth $z_m$ is drawn from a spherically uniform probability distribution, such that the initial values of both $s_a$ and $z_m$ are weighted with the distribution of dust mass in the background model. All monomers start with $0.0176$ times their bare monomer mass $m_\text{m}$ worth of ice, which is calculated according to Eq. \eqref{eq:initIce}. The left panel of Fig. \ref{fig:33PhiComparisonPopulation} shows that for all values of the home aggregate filling factor $\phi$, monomers gain ice over the 100 kyr evolution, primarily as a consequence of the adsorption of H$_2$O and to a lesser extent, NH$_3$ and CO$_2$. The amount of ice accumulated, however, depends on the filling factor $\phi$. Fig. \ref{fig:33PhiComparisonPopulation} shows an increase of the total amount of ice of only $0.93 \%$ for monomers in aggregates with $\phi=1$, while monomers in very porous aggregates with $\phi=10^{-3}$ on average gain $\sim 20 \%$ their initial ice mass worth of ices over 100 kyr. The intermediate cases of $\phi=10^{-2}$ and $\phi=10^{-1}$ result in an average increase in the total ice mass of $6.1\%$ and $2.0\%$. This decrease in the efficiency of ice adsorption for increasing $\phi$ is a result from the lower exposure probability $P_\text{exp}$ at given $z_\text{m}$ for higher filling factors. Fig. \ref{fig:24PExp} shows that for $\phi=10^{-3},10^{-2}$, $10^{-1}$, and $1$, gas phase molecules can respectively reach down to depths of approximately \mbox{$2\e{-4}$ m}, \mbox{$2\e{-5}$ m}, \mbox{$2\e{-6}$ m}, and \mbox{$2\e{-7}$ m}. Therefore, monomers spending time in aggregates with higher $\phi$ spend less time exposed to the gas phase. \\
The center and right panels of Fig. \ref{fig:33PhiComparisonPopulation} show that most of the ice accumulated during the 100 kyr evolution is H$_2$O and NH$_3$ for all values of $\phi$. In all cases, these two species amount to over $90\%$ of the molecules gained as ice, with the remainder being CO$_2$. This is a result from the fact that H$_2$O and NH$_3$ are the most prevalent molecules at 10 AU in the gas phase of our background model, with molecule number densities of $n_{\H_2\O}\approx9.7\e{5}\,\unit{m}{-3}$ and $n_{\N\H_3}\approx6.6\e{5}\,\unit{m}{-3}$ (c.f. Appendix \ref{sec:ABC}). Although CO and CH$_4$ are much more abundant in the gas phase ($n_{\C\O}\approx 2.4\e{14}\,\unit{m}{-3}$ and $n_{\C\H_4}\approx2.1\e{11}\,\unit{m}{-3}$), these molecules do not form any ices since temperatures at 10 AU are too high (see also Fig. \ref{fig:31CouplingFigure}). \\
Fig. \ref{fig:33PhiComparisonPopulation} also shows the average amount of ice expected on monomers based on the ice evolution predicted by the time-dependent chemistry of the \texttt{vFrag5} ProDiMo model. We linearly interpolate between the default snapshot for $n_{x,\text{ice}}$ at \mbox{$t=2\e{5}$} yr used in the background model and a new snapshot at $t=3\e{5}$ yr obtained by evolving the chemistry in the \texttt{vFrag5} ProDiMo model. The molecule number densities $n_{x,\text{ice}}$ calculated in ProDiMo are here converted to units of ice mass per monomer with Eq. \eqref{eq:initIce}, which results in an increase of the total amount of ice of only $0.89\%$ in the background model. This is roughly 10$\%$ less than the average total amount of ice expected on monomers in aggregates with $\phi=1$. We also find that in the cases of specific species, the average amount of H$_2$O and NH$_3$ on monomers in aggregates of $\phi=1$ calculated with SHAMPOO is $\sim 10\%$ above the increase calculated by ProDiMo. It is expected that monomers in aggregates with $\phi=1$ are in best agreement with ProDiMo, as ProDiMo chemically treats the refractory component of individual dust particles as rigid balls where molecules do not penetrate the aggregate surface upon adsorption. In ProDiMo, ice will only build up in a homogeneous layer on the aggregate surface, such that the rate at which ice of species $x$ is adsorbed on a single dust aggregate is given by 
\begin{align*}
    4\pi s_\text{a}^2 \mathcal{R}_{\text{ads},x}
\end{align*}
In SHAMPOO, the effective adsorption rate for the same aggregate can be estimated as
\begin{align*}
    N_\text{exp}\cdot 2\pi s_\text{m}^2 \mathcal{R}_{\text{ads},x},
\end{align*}
with $N_\text{exp}$ being the total number of exposed monomers in the aggregate. In practice, this means that a given aggregate has on average more surface area available for adsorption than a dust particle of the same size in ProDiMo, since monomers below the aggregate surface can have have nonzero exposure probability, allowing for very large $N_\text{exp}$ if the filling factor is low. Even for $\phi=1$, Fig. \ref{fig:24PExp} shows that molecules can reach a few monomer radii below the aggregate surface, although the exposure probability drops off very quickly. This nonzero exposure probability down to a few monomer radii overestimates the surface area available for adsorption with respect to ProDiMo, and thus explains $\sim 10\%$ difference.

    \section{Discussion}
    \label{sec:4}

\subsection{Applications}
\label{sec:4.1}
The SHAMPOO code provides a tool which allows the investigation of the systematic behaviour of dust monomers under the non-local disk processing of a given protoplanetary disk structure. In this section we highlight a number of possible applications of the SHAMPOO code.\\
The primary goal of SHAMPOO is to reliably predict the CHNOS abundances of dust during the first stages of planet formation, where no planetesimals have formed yet. In case of the Solar System, this would encompass the first 100-500 kyr after the formation CAIs \citep[e.g.][]{Kruijer+2014, Kruijer+2020}. In order to infer local CHNOS abundances in local dust populations, many monomer trajectories would be required. Within the framework of the background disk model, each monomer can be assigned to one spatial grid cells of the background disk model. This means that at any moment in time, every monomer is representing a unit of mass from that spatial grid cell. Predictions for the local CHNOS abundances can then be made by averaging over the individual CHNOS abundances of the monomers associated with that grid cell. A crucial requirement is that the spatial grid resolution is high enough such that the physical, thermal and chemical gradients within a grid cell are small.\\
Although this work has primarily focussed on the CHNOS-bearing molecules condensed on the monomer as ices, every monomer also consists of a refractory component of mass $m_\text{m}$. This refractory component can also contain a significant fraction of the total CHNOS elemental budget \citep[e.g.][]{Oberg&Bergin2020}. Currently, SHAMPOO does not assign a specific composition to the refractory component, although this could be incorporated in the model for example by assigning a static composition to the refractory monomer component. This refractory composition can be informed from the conditions at which the refractory component of the monomer has condensed. Altogether this would allow for a complete picture of the evolution of local CHNOS abundances. \\
We note a number of possible applications of SHAMPOO outside its main application. Specifically, SHAMPOO could be used to study the effect of dynamical and collisional processes on the transport of physically or chemically altered minerals throughout the planet-forming disk. Examples here include the possible outward diffusion of processed refractory material such as crystalline sillicates \citep[e.g.][]{Bockelee-Morvan+2002, Watson+2009, Olofsson+2009, Williams+2020}, and the transport of hydrated minerals throughout the planet-forming disk \citep[e.g.][]{Brearley+2006, DAngelo+2019}. However, such an application would involve connecting SHAMPOO to models which describe how these processes alter a single monomer given a certain disk environment. Another application lies in predicting the composition of pebble fluxes over time \citep[e.g.][]{Lambrechts+2014, Morbidelli+2015}. The growth of pebbles involves the compaction of aggregates (i.e. $\phi$ increases over time) which gradually lowers the sticking probability after a collision. Therefore, such a compact aggregate would ultimately collisionally decouple from the growth-fragmentation cycles shown in Fig. \ref{fig:32MonomerDemoCollisions} due to the bouncing barrier \citep[][]{Guttler+2010, Zsom+2010, Krijt+2018}. The application of SHAMPOO in this context would thus require the inclusion of porosity evolution. Furthermore, an application of SHAMPOO could lie in tracking the effects of non-local disk processing on local isotopic ratios such as D/H, $^{13}$C/$^{12}$C, $^{15}$N/$^{14}$N, and $^{18}$O/$^{16}$O \citep[see e.g.][]{Cleeves+2014, Visser+2018, Oberg&Bergin2020}. This would, however, require the incorporation of the isotopologues in the background model chemical network and possibly also the feedback of sublimation into the background gas phase composition.

\subsection{Model limitations}
\label{sec:4.2}
In this section we discuss the fundamental limitations of the SHAMPOO code, and highlight how we expect that these limitations are likely to affect predictions made with the SHAMPOO code both in this work and future work.

\subsubsection{Feedback of monomer evolution on the disk}
The code presented in this work offers a comprehensive description of how disk processes affect the volatile CHNOS abundances of monomer particles. However, the local conditions which affect the monomers are completely set by the background model, and are independent of the evolutionary trajectories of monomers. We highlight a number of effects this assumption may have on the evolutionary trajectories predicted by SHAMPOO.\\
In Fig. \ref{fig:23SlopeTest} we highlighted the possibility of inconsistencies between the local dust size distribution $\rho_\text{d}(a,r,z)$ in the background disk model (Eq. \ref{eq:dustSizeDistribution}) and the in-situ dust size distribution inferred from the residence time of the monomer in home aggregates of given size. However, these two distributions are not independent since the background dust size distribution determines the collision rates (Eq. \ref{eq:CollisionRate}). In Fig. \ref{fig:23SlopeTest} we demonstrated that given enough time, the time spent by the monomer in a home aggregate of given size approximately follows the residence time one would expect from the background distribution. However, we also noted significant departures, in particular when the dust size distribution of the background model is not consistent with the steady-state dust size distribution associated with a given fragmentation velocity $v_\text{frag}$ in the collision model (Sect. \ref{sec:2.3}). In the background model dust size distribution, the maximum size is specified directly by $a_\text{max}$, whereas in the collision model, the maximum home aggregate size also depends on the fragmentation velocity $v_\text{frag}$. In reality, however, $a_\text{max}$ is a function of the fragmentation velocity. In this work, we accounted for inconsistencies between the maximum aggregate size in the collision model and $a_\text{max}$ by using the parametrization from \cite{Birnstiel+2012} for a local, steady-state distribution if turbulence is the main source of relative velocity between dust particles. However, $a_\text{max}$ does remain a quantity which is not calculated self-consistently and has to be estimated separately from the SHAMPOO-code in the background model.\\
As a consequence of this discrepancy between the imposed background distribution and the steady-state distribution expected from the collision model, we found in Fig. \ref{fig:32MonomerDemoCollisions} that monomers in aggregates with too low fragmentation velocity in SHAMPOO (i.e. the background model contains too many large grains) are frequently forced into aggregates which are larger than possible given the value of $v_\text{frag}$. In these larger aggregates, monomers are more likely to be buried at larger monomer depth $z_\text{m}$, which results in a lower exposure probability. Similarly, a too high fragmentation velocity in SHAMPOO (i.e. the background model does not contain enough large grains) results in monomers spending much time in large aggregates which are not present as grains in the background model. This would thus also result in an underestimation of the exposure probability. For the volatile CHNOS budgets, this means that inconsistencies between the background model and the collision model will likely result in our model underestimating the effects of adsorption and photodesorption on the ice budgets of monomers. The results of Sect. \ref{sec:3.2} show that in particular photodesorption can have extreme effects on the evolution of monomers which are exposed frequently. Inconsistencies between the collision model and the background model would thus underestimate these effects, which highlights the importance of keeping the grain size distribution implied by the collision model and the grain size distribution of the background disk model as consistent as possible.\\
In our background disk model, we allow for the condensation and sublimation of molecules onto monomers. Since the position of the monomer between the time of condensation and sublimation of a particular molecular changes, the gas phase abundance of the molecule can become depleted in disk regions of systematic condensation onto monomers, and enriched in regions of systematic sublimation. SHAMPOO predicts the effects of a given chemical disk structure on the evolution of individual dust particles. However, it does not incorporate the effects of systematic behaviour of monomers on the disk chemical structure. It is expected that this systematic behaviour does affect the disk chemical structure on $\sim 1$ Myr timescales \citep[e.g.][]{Zhang+2020, Krijt+2020}, which may in turn affect monomers which are processed in the same region at later times. In future work, we aim to quantify the initial net gas phase depletion- and enrichment fluxes at a certain disk region using the systematic behaviour of many monomers that visit that region. This approach would enable us to constrain the timescales where the condensation or sublimation of volatile molecules have large effects on local gas phase abundances. Altogether it becomes clear that predictions of monomer trajectories made with SHAMPOO are not reliable on timescales comparable to $\sim 1$ Myr due to the expected chemical feedback on the gas phase. However, other evolutionary processes such as disk evolution and planet formation are also expected to undermine the assumption of a static background disk environment on these timescales \citep[e.g.][]{Andrews2020, Raymond&Morbidelli2020}. As a precaution, we have therefore chosen a timescale (100 kyr) significantly shorter than 1 Myr as the upper bound for the simulation time for individual monomers throughout this work. 

\subsubsection{Optical depth formalism}
In Sect. \ref{sec:2.4} we introduced a new stochastic approach for estimating ice adsorption and photodesorption for a monomer embedded at a certain depth $z_\text{m}$ inside its home aggregate. In this approach, we assumed the monomer to be buried at a depth $z_\text{m}$ below a homogeneous slab of monomers associated with an "optical depth" $\tau$. However, the expression for $\tau$ derived in Eq. \eqref{eq:tauConst} strictly applies to aggregates of constant density (i.e. $n_\text{m}$ does not vary as a function of time or position in the aggregate). In reality, however, $n_\text{m}$ does vary as a function of position in the home aggregate and evolves over time due to collisional processing.\\ 
The most notable effect here is compaction \citep[][]{Weidling+2009, Guttler+2010, Zsom+2010}. This is a process which is not explicitly modeled in this work, although the assumed filling factor of $\phi=0.1$ is consistent with aggregates which have undergone some degree of compaction \citep[e.g.][]{Langkowski+2008, Weidling+2009}. The fact that $n_\text{m}$ depends on position in the aggregate and changes over time due to collisions has important effects on the exposure probability $P_\text{exp}$ and the new monomer depth $z_\text{m}$ after a collision. The results in Sect. \ref{sec:3.3} suggest that accumulation of new ice is more efficient in porous aggregates, and thus gradual compaction implies that ice formation is a process that becomes more and more confined towards the aggregate surface at later times. \\
Another simplification made is that the optical depth $\tau$ associated with a certain monomer depth $z_\text{m}$ is assumed to be the same for impinging gas phase molecules and UV photons. In practice however, UV photons will likely be able to penetrate deeper into the aggregate since the minimum wavelength of UV photons ($\gtrsim 10^{-8}$ m) is either comparable or larger than the monomer size ($s_\text{m}=5\e{-8}$). This would imply that the exposure probability for photodesorption at a given depth would be larger, and hence more monomers would be able to lose ice via photodesorption. For $r\eqsim10$ AU, the results from Sect. \ref{sec:3.2} suggest that the effects of photodesorption are significant for individual monomers in smaller, more fragile aggregates ($v_\text{frag}=1$ m/s) in that disk region, even with no distinction between the optical depth for photons and molecules. For these monomers, our current model may thus underestimate the average amount of ice lost due to photodesorption. Similarly, a lower filling factor also reduces the optical depth, and thus is more likely to leave monomers exposed to impinging UV photons. As a function of position, Figs. \ref{fig:ABBackgroundModelconstStructure}-\ref{fig:ABBackgroundModelvFrag10Structure} show that for larger radial distances ($r\gtrsim 100$ AU), UV photons can reach deeper in the disk. Therefore, our model may also underestimate the amount of ice lost via photodesorption in small, fragile aggregates at large radial distances. However, the examples in Sect. \ref{sec:3.2} do suggest that the effect of photodesorption is less important for monomers residing in aggregates with higher $v_\text{frag}$, as the monomer is on average located at considerably larger $z_\text{m}$ in these aggregates (c.f. Fig. \ref{fig:32MonomerDemoDepth}). Therefore, we expect that the effect of a distinction between the optical depth for molecules and UV photons will remain limited for monomers in aggregates with higher $v_\text{frag}$.

    \section{Conclusions}
    \label{sec:5}
In this work we introduced and benchmarked the SHAMPOO code, which models the effects of dynamical, collisional and ice processes on the volatile CHNOS abundances of individual dust particles. In addition, we also motivated the need for this model, and used the SHAMPOO code to investigate the effects of different fragmentation velocities and aggregate filling factor on ice evolution at 10 AU. We summarize our conclusions as follows:
\begin{itemize}
    \item The dynamical behaviour of a dust grain of given size can vary greatly as a function of position throughout the disk. This is a consequence of the fact that the grain size associated with the Stokes number for which aerodynamic drag starts to significantly affect the grain dynamics (St=$10^{-3}$) varies greatly throughout the disk (Fig. \ref{fig:31StokesNumber}). This is primarily a result from the large variation of the gas density throughout the disk, and to a lesser extent also from the gas temperature.
    \item The disk regions where the timescales associated with dynamical, collisional, and ice processing couple varies, depending on grain size and ice species (Fig. \ref{fig:31CouplingFigure}). However, different grain sizes are connected through collisional processes, and the ice mantle of a dust particle can consist of multiple ice species. Therefore, we require a fully coupled modeling approach such as SHAMPOO to make physical estimates of the volatile CHNOS during the first stages of planetesimal formation.   
    \item For a fragmentation velocity of $v_\text{frag}=1$ m/s, the monomer spends more time in smaller aggregates, and is therefore on average located closer to the aggregate surface (Fig. \ref{fig:32MonomerDemoDynamics}-\ref{fig:32MonomerDemoIceEvolution}). A lower monomer depth increases the exposure probability $P_\text{exp}$, resulting in more periods of monomer exposure to the gas phase compared to monomers in aggregates of higher $v_\text{frag}$. On the one hand this means that monomers in fragile aggregates can more frequently accumulate ice through adsorption. On the other hand, the exposure to the gas phase can also result in enhanced ice loss through photodesorption if the period of exposure coincides with a large vertical excursion of the monomer to the disk surface. Altogether, our results suggest that the interiors of fragile aggregates are more likely to have undergone extensive ice processing than aggregates with a higher fragmentation velocity \mbox{($v_\text{frag}=5,10$ m/s)}.  
    \item In the absence of dynamical transport, the average ice budgets of aggregates with $\phi=1$ are in good agreement with the ice abundance predicted by the time-dependent chemistry of the ProDiMo background disk model. Furthermore, very porous aggregates ($\phi=10^{-3}$) have a higher exposure probability $P_\text{exp}$ at a given depth. Therefore, gas phase molecules can reach deeper into these aggregates, and these aggregates accumulate more ice (Fig. \ref{fig:33PhiComparisonPopulation}). As a consequence, the ice mantles of monomers inside these aggregates are likely more processed compared to monomers inside more compact aggregates.
\end{itemize} 
Altogether SHAMPOO provides a useful tool for probing the dust processing on a microscopic level. In future work, we aim to use a large set of its individual dust grain simulations to study the influence of non-local disk processing on local CHNOS abundances. Although the model has been developed to study the processing of CHNOS, we also noted applications of SHAMPOO outside its main purpose. Examples included the formation and transport of physically or chemically processed material, such as crystalized sillicates or hydrated minerals, and the formation and evolution of pebbles.

    \section*{Acknowledgements}
    The authors would like to thank Rens Waters and Kaustubh Hakim for comments and insightful discussions, and an anonymous reviewer for constructive
comments that helped to improve our simulations. This work is part of the second round of the Planetary and Exoplanetary Science Network (PEPSci-2), funded by the Netherlands Organization for Scientific Research (NWO).


\bibliographystyle{aa.bst}
\bibliography{refdata.bib}

\begin{thebibliography}{94}
\expandafter\ifx\csname natexlab\endcsname\relax\def\natexlab#1{#1}\fi

\bibitem[{{Altwegg} {et~al.}(2017){Altwegg}, {Balsiger}, {Berthelier},
  {Bieler}, {Calmonte}, {De Keyser}, {Fiethe}, {Fuselier}, {Gasc}, {Gombosi},
  {Owen}, {Le Roy}, {Rubin}, {S{\'e}mon}, \& {Tzou}}]{Altwegg+2017}
{Altwegg}, K., {Balsiger}, H., {Berthelier}, J.~J., {et~al.} 2017,
  Philosophical Transactions of the Royal Society of London Series A, 375,
  20160253, \doi{10.1098/rsta.2016.0253}
  \ads{https://ui.adsabs.harvard.edu/abs/2017RSPTA.37560253A}

\bibitem[{{Andrews}(2020)}]{Andrews2020}
{Andrews}, S.~M. 2020, Annual Review of Astronomy and Astrophysics, 58, 483,
  \doi{10.1146/annurev-astro-031220-010302}
  \ads{https://ui.adsabs.harvard.edu/abs/2020ARA&A..58..483A}

\bibitem[{{Aresu} {et~al.}(2011){Aresu}, {Kamp}, {Meijerink}, {Woitke}, {Thi},
  \& {Spaans}}]{Aresu+2011}
{Aresu}, G., {Kamp}, I., {Meijerink}, R., {et~al.} 2011, Astronomy \&
  Astrophysics, 526, A163, \doi{10.1051/0004-6361/201015449}
  \ads{https://ui.adsabs.harvard.edu/abs/2011A&A...526A.163A}

\bibitem[{{Armitage}(2010)}]{Armitage2010}
{Armitage}, P.~J. 2010, {Astrophysics of Planet Formation},
  \ads{https://ui.adsabs.harvard.edu/abs/2010apf..book.....A}

\bibitem[{{Bergner} \& {Ciesla}(2021)}]{Bergner+2021}
{Bergner}, J.~B. \& {Ciesla}, F. 2021, The Astrophysical Journal, 919, 45,
  \doi{10.3847/1538-4357/ac0fd7}
  \ads{https://ui.adsabs.harvard.edu/abs/2021ApJ...919...45B}

\bibitem[{{Birnstiel} {et~al.}(2010){Birnstiel}, {Dullemond}, \&
  {Brauer}}]{Birnstiel+2010}
{Birnstiel}, T., {Dullemond}, C.~P., \& {Brauer}, F. 2010, Astronomy \&
  Astrophysics, 513, A79, \doi{10.1051/0004-6361/200913731}
  \ads{https://ui.adsabs.harvard.edu/abs/2010A&A...513A..79B}

\bibitem[{{Birnstiel} {et~al.}(2012){Birnstiel}, {Klahr}, \&
  {Ercolano}}]{Birnstiel+2012}
{Birnstiel}, T., {Klahr}, H., \& {Ercolano}, B. 2012, Astronomy \&
  Astrophysics, 539, A148, \doi{10.1051/0004-6361/201118136}
  \ads{https://ui.adsabs.harvard.edu/abs/2012A&A...539A.148B}

\bibitem[{{Birnstiel} {et~al.}(2011){Birnstiel}, {Ormel}, \&
  {Dullemond}}]{Birnstiel+2011}
{Birnstiel}, T., {Ormel}, C.~W., \& {Dullemond}, C.~P. 2011, Astronomy \&
  Astrophysics, 525, A11, \doi{10.1051/0004-6361/201015228}
  \ads{https://ui.adsabs.harvard.edu/abs/2011A&A...525A..11B}

\bibitem[{{Blum} \& {M{\"u}nch}(1993)}]{Blum&Munch1993}
{Blum}, J. \& {M{\"u}nch}, M. 1993, Icarus, 106, 151,
  \doi{10.1006/icar.1993.1163}
  \ads{https://ui.adsabs.harvard.edu/abs/1993Icar..106..151B}

\bibitem[{{Bockel{\'e}e-Morvan} {et~al.}(2002){Bockel{\'e}e-Morvan}, {Gautier},
  {Hersant}, {Hur{\'e}}, \& {Robert}}]{Bockelee-Morvan+2002}
{Bockel{\'e}e-Morvan}, D., {Gautier}, D., {Hersant}, F., {Hur{\'e}}, J.~M., \&
  {Robert}, F. 2002, Astronomy \& Astrophysics, 384, 1107,
  \doi{10.1051/0004-6361:20020086}
  \ads{https://ui.adsabs.harvard.edu/abs/2002A&A...384.1107B}

\bibitem[{{Boogert} {et~al.}(2015){Boogert}, {Gerakines}, \&
  {Whittet}}]{Boogert+2015}
{Boogert}, A.~C.~A., {Gerakines}, P.~A., \& {Whittet}, D. C.~B. 2015, Annual
  Review of Astronomy and Astrophysics, 53, 541,
  \doi{10.1146/annurev-astro-082214-122348}
  \ads{https://ui.adsabs.harvard.edu/abs/2015ARA&A..53..541B}

\bibitem[{{Booth} \& {Ilee}(2019)}]{Booth&Ilee2019}
{Booth}, R.~A. \& {Ilee}, J.~D. 2019, Monthly Notices of the Royal Astronomical
  Society, 487, 3998, \doi{10.1093/mnras/stz1488}
  \ads{https://ui.adsabs.harvard.edu/abs/2019MNRAS.487.3998B}

\bibitem[{{Bosman} {et~al.}(2018){Bosman}, {Tielens}, \& {van
  Dishoeck}}]{Bosman+2018}
{Bosman}, A.~D., {Tielens}, A. G.~G.~M., \& {van Dishoeck}, E.~F. 2018,
  Astronomy and Astrophysics, 611, A80, \doi{10.1051/0004-6361/201732056}
  \ads{https://ui.adsabs.harvard.edu/abs/2018A&A...611A..80B}

\bibitem[{{Bower} {et~al.}(2022){Bower}, {Hakim}, {Sossi}, \&
  {Sanan}}]{Bower+2022}
{Bower}, D.~J., {Hakim}, K., {Sossi}, P.~A., \& {Sanan}, P. 2022, The Planetary
  Science Journal, 3, 93, \doi{10.3847/PSJ/ac5fb1}
  \ads{https://ui.adsabs.harvard.edu/abs/2022PSJ.....3...93B}

\bibitem[{{Brauer} {et~al.}(2008){Brauer}, {Dullemond}, \&
  {Henning}}]{Brauer+2008a}
{Brauer}, F., {Dullemond}, C.~P., \& {Henning}, T. 2008, Astronomy \&
  Astrophysics, 480, 859, \doi{10.1051/0004-6361:20077759}
  \ads{https://ui.adsabs.harvard.edu/abs/2008A&A...480..859B}

\bibitem[{{Brearley}(2006)}]{Brearley+2006}
{Brearley}, A.~J. 2006, in Meteorites and the Early Solar System II, ed. D.~S.
  {Lauretta} \& H.~Y. {McSween}, 584,
  \ads{https://ui.adsabs.harvard.edu/abs/2006mess.book..584B}

\bibitem[{{Brown} \& {Bolina}(2007)}]{Brown&Bolina2007}
{Brown}, W.~A. \& {Bolina}, A.~S. 2007, Monthly Notices of the Royal
  Astronomical Society, 374, 1006, \doi{10.1111/j.1365-2966.2006.11216.x}
  \ads{https://ui.adsabs.harvard.edu/abs/2007MNRAS.374.1006B}

\bibitem[{{Caselli} \& {Ceccarelli}(2012)}]{Caselli&Ceccarelli2012}
{Caselli}, P. \& {Ceccarelli}, C. 2012, The Astronomy and Astrophysics Review,
  20, 56, \doi{10.1007/s00159-012-0056-x}
  \ads{https://ui.adsabs.harvard.edu/abs/2012A&ARv..20...56C}

\bibitem[{{Ciesla}(2010)}]{Ciesla2010}
{Ciesla}, F.~J. 2010, The Astrophysical Journal, 723, 514,
  \doi{10.1088/0004-637X/723/1/514}
  \ads{https://ui.adsabs.harvard.edu/abs/2010ApJ...723..514C}

\bibitem[{{Ciesla}(2011)}]{Ciesla2011}
{Ciesla}, F.~J. 2011, The Astrophysical Journal, 740, 9,
  \doi{10.1088/0004-637X/740/1/9}
  \ads{https://ui.adsabs.harvard.edu/abs/2011ApJ...740....9C}

\bibitem[{{Cleeves} {et~al.}(2014){Cleeves}, {Bergin}, {Alexander}, {Du},
  {Graninger}, {{\"O}berg}, \& {Harries}}]{Cleeves+2014}
{Cleeves}, L.~I., {Bergin}, E.~A., {Alexander}, C. M.~O.~D., {et~al.} 2014,
  Science, 345, 1590, \doi{10.1126/science.1258055}
  \ads{https://ui.adsabs.harvard.edu/abs/2014Sci...345.1590C}

\bibitem[{{Cuppen} {et~al.}(2017){Cuppen}, {Walsh}, {Lamberts}, {Semenov},
  {Garrod}, {Penteado}, \& {Ioppolo}}]{Cuppen+2017}
{Cuppen}, H.~M., {Walsh}, C., {Lamberts}, T., {et~al.} 2017, Space Science
  Reviews, 212, 1, \doi{10.1007/s11214-016-0319-3}
  \ads{https://ui.adsabs.harvard.edu/abs/2017SSRv..212....1C}

\bibitem[{{Cuzzi} \& {Zahnle}(2004)}]{Cuzzi+2004}
{Cuzzi}, J.~N. \& {Zahnle}, K.~J. 2004, The Astrohpysical Journal, 614, 490,
  \doi{10.1086/423611}
  \ads{https://ui.adsabs.harvard.edu/abs/2004ApJ...614..490C}

\bibitem[{{D'Angelo} {et~al.}(2019){D'Angelo}, {Cazaux}, {Kamp}, {Thi}, \&
  {Woitke}}]{DAngelo+2019}
{D'Angelo}, M., {Cazaux}, S., {Kamp}, I., {Thi}, W.~F., \& {Woitke}, P. 2019,
  Astronomy \& Astrophysics, 622, A208, \doi{10.1051/0004-6361/201833715}
  \ads{https://ui.adsabs.harvard.edu/abs/2019A&A...622A.208D}

\bibitem[{{Dasgupta} \& {Hirschmann}(2006)}]{Dasgupta&Hirschmann2006}
{Dasgupta}, R. \& {Hirschmann}, M.~M. 2006, Nature, 440, 659,
  \doi{10.1038/nature04612}
  \ads{https://ui.adsabs.harvard.edu/abs/2006Natur.440..659D}

\bibitem[{{Dionatos} {et~al.}(2019){Dionatos}, {Woitke}, {G{\"u}del},
  {Degroote}, {Liebhart}, {Anthonioz}, {Antonellini}, {Baldovin-Saavedra},
  {Carmona}, {Dominik}, {Greaves}, {Ilee}, {Kamp}, {M{\'e}nard}, {Min},
  {Pinte}, {Rab}, {Rigon}, {Thi}, \& {Waters}}]{Dionatos+2019}
{Dionatos}, O., {Woitke}, P., {G{\"u}del}, M., {et~al.} 2019, Astronomy and
  Astrophysics, 625, A66, \doi{10.1051/0004-6361/201832860}
  \ads{https://ui.adsabs.harvard.edu/abs/2019A&A...625A..66D}

\bibitem[{{Dominik} \& {Tielens}(1997)}]{Dominik&Tielens1997}
{Dominik}, C. \& {Tielens}, A.~G.~G.~M. 1997, The Astrophysical Journal, 480,
  647, \doi{10.1086/303996}
  \ads{https://ui.adsabs.harvard.edu/abs/1997ApJ...480..647D}

\bibitem[{{Drozdovskaya} {et~al.}(2019){Drozdovskaya}, {van Dishoeck}, {Rubin},
  {J{\o}rgensen}, \& {Altwegg}}]{Drozdovskaya+2019}
{Drozdovskaya}, M.~N., {van Dishoeck}, E.~F., {Rubin}, M., {J{\o}rgensen},
  J.~K., \& {Altwegg}, K. 2019, Monthly Notices of the Royal Astronomical
  Society, 490, 50, \doi{10.1093/mnras/stz2430}
  \ads{https://ui.adsabs.harvard.edu/abs/2019MNRAS.490...50D}

\bibitem[{{Dubrulle} {et~al.}(1995){Dubrulle}, {Morfill}, \&
  {Sterzik}}]{Dubrulle+1995}
{Dubrulle}, B., {Morfill}, G., \& {Sterzik}, M. 1995, Icarus, 114, 237,
  \doi{10.1006/icar.1995.1058}
  \ads{https://ui.adsabs.harvard.edu/abs/1995Icar..114..237D}

\bibitem[{{G{\"u}ttler} {et~al.}(2010){G{\"u}ttler}, {Blum}, {Zsom}, {Ormel},
  \& {Dullemond}}]{Guttler+2010}
{G{\"u}ttler}, C., {Blum}, J., {Zsom}, A., {Ormel}, C.~W., \& {Dullemond},
  C.~P. 2010, Astronomy \& Astrophysics, 513, A56,
  \doi{10.1051/0004-6361/200912852}
  \ads{https://ui.adsabs.harvard.edu/abs/2010A&A...513A..56G}

\bibitem[{{Hakim} {et~al.}(2019){Hakim}, {Spaargaren}, {Grewal}, {Rohrbach},
  {Berndt}, {Dominik}, \& {van Westrenen}}]{Hakim+2019}
{Hakim}, K., {Spaargaren}, R., {Grewal}, D.~S., {et~al.} 2019, Astrobiology,
  19, 867, \doi{10.1089/ast.2018.1930}
  \ads{https://ui.adsabs.harvard.edu/abs/2019AsBio..19..867H}

\bibitem[{{He} {et~al.}(2016){He}, {Acharyya}, \& {Vidali}}]{He+2016}
{He}, J., {Acharyya}, K., \& {Vidali}, G. 2016, The Astrophysical Journal, 823,
  56, \doi{10.3847/0004-637X/823/1/56}
  \ads{https://ui.adsabs.harvard.edu/abs/2016ApJ...823...56H}

\bibitem[{{Helling} {et~al.}(2014){Helling}, {Woitke}, {Rimmer}, {Kamp}, {Thi},
  \& {Meijerink}}]{Helling+2014}
{Helling}, C., {Woitke}, P., {Rimmer}, P.~B., {et~al.} 2014, Life, 4, 142,
  \doi{10.3390/life4020142}
  \ads{https://ui.adsabs.harvard.edu/abs/2014Life....4..142H}

\bibitem[{{Hollenbach} {et~al.}(2009){Hollenbach}, {Kaufman}, {Bergin}, \&
  {Melnick}}]{Hollenbach+2009}
{Hollenbach}, D., {Kaufman}, M.~J., {Bergin}, E.~A., \& {Melnick}, G.~J. 2009,
  The Astrophysical Journal, 690, 1497, \doi{10.1088/0004-637X/690/2/1497}
  \ads{https://ui.adsabs.harvard.edu/abs/2009ApJ...690.1497H}

\bibitem[{{Johansen} {et~al.}(2014){Johansen}, {Blum}, {Tanaka}, {Ormel},
  {Bizzarro}, \& {Rickman}}]{Johansen+2014}
{Johansen}, A., {Blum}, J., {Tanaka}, H., {et~al.} 2014, in Protostars and
  Planets VI, ed. H.~{Beuther}, R.~S. {Klessen}, C.~P. {Dullemond}, \&
  T.~{Henning}, 547, \doi{10.2458/azu_uapress_9780816531240-ch024}
  \ads{https://ui.adsabs.harvard.edu/abs/2014prpl.conf..547J}

\bibitem[{{Johansen} {et~al.}(2007){Johansen}, {Oishi}, {Mac Low}, {Klahr},
  {Henning}, \& {Youdin}}]{Johansen+2007}
{Johansen}, A., {Oishi}, J.~S., {Mac Low}, M.-M., {et~al.} 2007, Nature, 448,
  1022, \doi{10.1038/nature06086}
  \ads{https://ui.adsabs.harvard.edu/abs/2007Natur.448.1022J}

\bibitem[{{Kama} {et~al.}(2016){Kama}, {Bruderer}, {van Dishoeck},
  {Hogerheijde}, {Folsom}, {Miotello}, {Fedele}, {Belloche}, {G{\"u}sten}, \&
  {Wyrowski}}]{Kama+2016}
{Kama}, M., {Bruderer}, S., {van Dishoeck}, E.~F., {et~al.} 2016, Astronomy \&
  Astrohpysics, 592, A83, \doi{10.1051/0004-6361/201526991}
  \ads{https://ui.adsabs.harvard.edu/abs/2016A&A...592A..83K}

\bibitem[{{Kamp} {et~al.}(2017){Kamp}, {Thi}, {Woitke}, {Rab}, {Bouma}, \&
  {M{\'e}nard}}]{Kamp+2017}
{Kamp}, I., {Thi}, W.~F., {Woitke}, P., {et~al.} 2017, Astronomy and
  Astrophysics, 607, A41, \doi{10.1051/0004-6361/201730388}
  \ads{https://ui.adsabs.harvard.edu/abs/2017A&A...607A..41K}

\bibitem[{{Kamp} {et~al.}(2010){Kamp}, {Tilling}, {Woitke}, {Thi}, \&
  {Hogerheijde}}]{Kamp+2010}
{Kamp}, I., {Tilling}, I., {Woitke}, P., {Thi}, W.~F., \& {Hogerheijde}, M.
  2010, Astronomy and Astrophysics, 510, A18, \doi{10.1051/0004-6361/200913076}
  \ads{https://ui.adsabs.harvard.edu/abs/2010A&A...510A..18K}

\bibitem[{Kasting \& Catling(2003)}]{Kasting&Catling2003}
Kasting, J.~F. \& Catling, D. 2003, Annual Review of Astronomy and
  Astrophysics, 41, 429, \doi{10.1146/annurev.astro.41.071601.170049}

\bibitem[{{Kasting} {et~al.}(1993){Kasting}, {Whitmire}, \&
  {Reynolds}}]{Kasting+1993}
{Kasting}, J.~F., {Whitmire}, D.~P., \& {Reynolds}, R.~T. 1993, Icarus, 101,
  108, \doi{10.1006/icar.1993.1010}
  \ads{https://ui.adsabs.harvard.edu/abs/1993Icar..101..108K}

\bibitem[{{Kataoka} {et~al.}(2013){Kataoka}, {Tanaka}, {Okuzumi}, \&
  {Wada}}]{Kataoka+2013}
{Kataoka}, A., {Tanaka}, H., {Okuzumi}, S., \& {Wada}, K. 2013, Astronomy \&
  Astrophysics, 557, L4, \doi{10.1051/0004-6361/201322151}
  \ads{https://ui.adsabs.harvard.edu/abs/2013A&A...557L...4K}

\bibitem[{{Kopparapu} {et~al.}(2013){Kopparapu}, {Ramirez}, {Kasting}, {Eymet},
  {Robinson}, {Mahadevan}, {Terrien}, {Domagal-Goldman}, {Meadows}, \&
  {Deshpande}}]{Kopparapu+2013}
{Kopparapu}, R.~K., {Ramirez}, R., {Kasting}, J.~F., {et~al.} 2013, The
  Astrophysical Journal, 765, 131, \doi{10.1088/0004-637X/765/2/131}
  \ads{https://ui.adsabs.harvard.edu/abs/2013ApJ...765..131K}

\bibitem[{{Krijt} {et~al.}(2020){Krijt}, {Bosman}, {Zhang}, {Schwarz},
  {Ciesla}, \& {Bergin}}]{Krijt+2020}
{Krijt}, S., {Bosman}, A.~D., {Zhang}, K., {et~al.} 2020, The Astrophysical
  Journal, 899, 134, \doi{10.3847/1538-4357/aba75d}
  \ads{https://ui.adsabs.harvard.edu/abs/2020ApJ...899..134K}

\bibitem[{{Krijt} \& {Ciesla}(2016)}]{Krijt&Ciesla2016}
{Krijt}, S. \& {Ciesla}, F.~J. 2016, The Astrophysical Journal, 822, 111,
  \doi{10.3847/0004-637X/822/2/111}
  \ads{https://ui.adsabs.harvard.edu/abs/2016ApJ...822..111K}

\bibitem[{{Krijt} {et~al.}(2016{\natexlab{a}}){Krijt}, {Ciesla}, \&
  {Bergin}}]{Krijt+2016b}
{Krijt}, S., {Ciesla}, F.~J., \& {Bergin}, E.~A. 2016{\natexlab{a}}, The
  Astrophysical Journal, 833, 285, \doi{10.3847/1538-4357/833/2/285}
  \ads{https://ui.adsabs.harvard.edu/abs/2016ApJ...833..285K}

\bibitem[{{Krijt} {et~al.}(2022){Krijt}, {Kama}, {McClure}, {Teske}, {Bergin},
  {Shorttle}, {Walsh}, \& {Raymond}}]{Krijt+2022}
{Krijt}, S., {Kama}, M., {McClure}, M., {et~al.} 2022, arXiv e-prints,
  \arxiv{arXiv:2203.10056}\ads{https://ui.adsabs.harvard.edu/abs/2022arXiv220310056K}

\bibitem[{{Krijt} {et~al.}(2015){Krijt}, {Ormel}, {Dominik}, \&
  {Tielens}}]{Krijt+2015}
{Krijt}, S., {Ormel}, C.~W., {Dominik}, C., \& {Tielens}, A.~G.~G.~M. 2015,
  Astronomy \& Astrophysics, 574, A83, \doi{10.1051/0004-6361/201425222}
  \ads{https://ui.adsabs.harvard.edu/abs/2015A&A...574A..83K}

\bibitem[{{Krijt} {et~al.}(2016{\natexlab{b}}){Krijt}, {Ormel}, {Dominik}, \&
  {Tielens}}]{Krijt+2016}
{Krijt}, S., {Ormel}, C.~W., {Dominik}, C., \& {Tielens}, A.~G.~G.~M.
  2016{\natexlab{b}}, Astronomy \& Astrophysics, 586, A20,
  \doi{10.1051/0004-6361/201527533}
  \ads{https://ui.adsabs.harvard.edu/abs/2016A&A...586A..20K}

\bibitem[{{Krijt} {et~al.}(2018){Krijt}, {Schwarz}, {Bergin}, \&
  {Ciesla}}]{Krijt+2018}
{Krijt}, S., {Schwarz}, K.~R., {Bergin}, E.~A., \& {Ciesla}, F.~J. 2018, The
  Astrophysical Journal, 864, 78, \doi{10.3847/1538-4357/aad69b}
  \ads{https://ui.adsabs.harvard.edu/abs/2018ApJ...864...78K}

\bibitem[{{Kruijer} {et~al.}(2020){Kruijer}, {Kleine}, \&
  {Borg}}]{Kruijer+2020}
{Kruijer}, T.~S., {Kleine}, T., \& {Borg}, L.~E. 2020, Nature Astronomy, 4, 32,
  \doi{10.1038/s41550-019-0959-9}
  \ads{https://ui.adsabs.harvard.edu/abs/2020NatAs...4...32K}

\bibitem[{{Kruijer} {et~al.}(2014){Kruijer}, {Touboul}, {Fischer-G{\"o}dde},
  {Bermingham}, {Walker}, \& {Kleine}}]{Kruijer+2014}
{Kruijer}, T.~S., {Touboul}, M., {Fischer-G{\"o}dde}, M., {et~al.} 2014,
  Science, 344, 1150, \doi{10.1126/science.1251766}
  \ads{https://ui.adsabs.harvard.edu/abs/2014Sci...344.1150K}

\bibitem[{{Kushiro}(1969)}]{Kushiro+1969}
{Kushiro}, I. 1969, American Journal of Science, 267-A, 269,
  \url{https://earth.geology.yale.edu/~ajs/1969/ajs_267A_11.pdf/269.pdf}
  \ads{https://ui.adsabs.harvard.edu/abs/2019Tectp.760..165T}

\bibitem[{{Lambrechts} \& {Johansen}(2014)}]{Lambrechts+2014}
{Lambrechts}, M. \& {Johansen}, A. 2014, Astronomy \& Astrophysics, 572, A107,
  \doi{10.1051/0004-6361/201424343}
  \ads{https://ui.adsabs.harvard.edu/abs/2014A&A...572A.107L}

\bibitem[{{Langkowski} {et~al.}(2008){Langkowski}, {Teiser}, \&
  {Blum}}]{Langkowski+2008}
{Langkowski}, D., {Teiser}, J., \& {Blum}, J. 2008, The Astrophysical Journal,
  675, 764

\bibitem[{{McElroy} {et~al.}(2013){McElroy}, {Walsh}, {Markwick}, {Cordiner},
  {Smith}, \& {Millar}}]{McElroy+2013}
{McElroy}, D., {Walsh}, C., {Markwick}, A.~J., {et~al.} 2013, Astronomy \&
  Astrophysics, 550, A36, \doi{10.1051/0004-6361/201220465}
  \ads{https://ui.adsabs.harvard.edu/abs/2013A&A...550A..36M}

\bibitem[{{Morbidelli} {et~al.}(2015){Morbidelli}, {Lambrechts}, {Jacobson}, \&
  {Bitsch}}]{Morbidelli+2015}
{Morbidelli}, A., {Lambrechts}, M., {Jacobson}, S., \& {Bitsch}, B. 2015,
  Icarus, 258, 418, \doi{10.1016/j.icarus.2015.06.003}
  \ads{https://ui.adsabs.harvard.edu/abs/2015Icar..258..418M}

\bibitem[{{Nixon} {et~al.}(2018){Nixon}, {King}, \& {Pringle}}]{Nixon+2018}
{Nixon}, C.~J., {King}, A.~R., \& {Pringle}, J.~E. 2018, Monthly Notices of the
  Royal Astronomical Society,, 477, 3273

\bibitem[{{{\"O}berg} \& {Bergin}(2021)}]{Oberg&Bergin2020}
{{\"O}berg}, K.~I. \& {Bergin}, E.~A. 2021, Physics Reports, 893, 1,
  \doi{10.1016/j.physrep.2020.09.004}
  \ads{https://ui.adsabs.harvard.edu/abs/2021PhR...893....1O}

\bibitem[{{Oberg} {et~al.}(2022){Oberg}, {Kamp}, {Cazaux}, {Woitke}, \&
  {Thi}}]{Oberg+2022}
{Oberg}, N., {Kamp}, I., {Cazaux}, S., {Woitke}, P., \& {Thi}, W.~F. 2022,
  Astronomy \& Astrophysics, 667, A95, \doi{10.1051/0004-6361/202244092}
  \ads{https://ui.adsabs.harvard.edu/abs/2022A&A...667A..95O}

\bibitem[{{Okuzumi} {et~al.}(2012){Okuzumi}, {Tanaka}, {Kobayashi}, \&
  {Wada}}]{Okuzumi+2012}
{Okuzumi}, S., {Tanaka}, H., {Kobayashi}, H., \& {Wada}, K. 2012, The
  Astrophysical Journal, 752, 106, \doi{10.1088/0004-637X/752/2/106}
  \ads{https://ui.adsabs.harvard.edu/abs/2012ApJ...752..106O}

\bibitem[{{Olofsson} {et~al.}(2009){Olofsson}, {Augereau}, {van Dishoeck},
  {Mer{\'\i}n}, {Lahuis}, {Kessler-Silacci}, {Dullemond}, {Oliveira}, {Blake},
  {Boogert}, {Brown}, {Evans}, {Geers}, {Knez}, {Monin}, \&
  {Pontoppidan}}]{Olofsson+2009}
{Olofsson}, J., {Augereau}, J.~C., {van Dishoeck}, E.~F., {et~al.} 2009,
  Astronomy \& Astrophysics, 507, 327, \doi{10.1051/0004-6361/200912062}
  \ads{https://ui.adsabs.harvard.edu/abs/2009A&A...507..327O}

\bibitem[{{Ormel} \& {Cuzzi}(2007)}]{Ormel&Cuzzi2007}
{Ormel}, C.~W. \& {Cuzzi}, J.~N. 2007, Astronomy \& Astrophysics, 466, 413,
  \doi{10.1051/0004-6361:20066899}
  \ads{https://ui.adsabs.harvard.edu/abs/2007A&A...466..413O}

\bibitem[{{Ormel} \& {Spaans}(2008)}]{Ormel&Spaans2008}
{Ormel}, C.~W. \& {Spaans}, M. 2008, The Astrophysical Journel, 684, 1291,
  \doi{10.1086/590052}
  \ads{https://ui.adsabs.harvard.edu/abs/2008ApJ...684.1291O}

\bibitem[{{Piso} {et~al.}(2015){Piso}, {{\"O}berg}, {Birnstiel}, \&
  {Murray-Clay}}]{Piso+2015}
{Piso}, A.-M.~A., {{\"O}berg}, K.~I., {Birnstiel}, T., \& {Murray-Clay}, R.~A.
  2015, The Astrophysical Journal, 815, 109, \doi{10.1088/0004-637X/815/2/109}
  \ads{https://ui.adsabs.harvard.edu/abs/2015ApJ...815..109P}

\bibitem[{{Raymond} \& {Morbidelli}(2020)}]{Raymond&Morbidelli2020}
{Raymond}, S.~N. \& {Morbidelli}, A. 2020, arXiv e-prints, \arxiv{2002.05756}
  \ads{https://ui.adsabs.harvard.edu/abs/2020arXiv200205756R}

\bibitem[{{Ruaud} {et~al.}(2016){Ruaud}, {Wakelam}, \& {Hersant}}]{Ruaud+2016}
{Ruaud}, M., {Wakelam}, V., \& {Hersant}, F. 2016, Monthly Notices of the Royal
  Astronomical Society, 459, 3756, \doi{10.1093/mnras/stw887}
  \ads{https://ui.adsabs.harvard.edu/abs/2016MNRAS.459.3756R}

\bibitem[{{Sandford} \& {Allamandola}(1993)}]{Sandford&Allamandola1993}
{Sandford}, S.~A. \& {Allamandola}, L.~J. 1993, The Astrophysical Journal, 417,
  815,
  \doi{10.1086/173362}\ads{https://ui.adsabs.harvard.edu/abs/1993ApJ...417..815S}

\bibitem[{{Schoonenberg} {et~al.}(2018){Schoonenberg}, {Ormel}, \&
  {Krijt}}]{Schoonenberg+2018}
{Schoonenberg}, D., {Ormel}, C.~W., \& {Krijt}, S. 2018, Astronomy \&
  Astrophysics, 620, A134, \doi{10.1051/0004-6361/201834047}
  \ads{https://ui.adsabs.harvard.edu/abs/2018A&A...620A.134S}

\bibitem[{{Shakura} \& {Sunyaev}(1973)}]{Shakura&Sunyaev1973}
{Shakura}, N.~I. \& {Sunyaev}, R.~A. 1973, Astronomy \& Astrophysics, 24, 337,
  \ads{https://ui.adsabs.harvard.edu/abs/1973A&A....24..337S}

\bibitem[{{Siess} {et~al.}(2000){Siess}, {Dufour}, \& {Forestini}}]{Siess+2000}
{Siess}, L., {Dufour}, E., \& {Forestini}, M. 2000, Astronomy \& Astrophysics,
  358, 593, \arxiv{astro-ph/0003477}
  \ads{https://ui.adsabs.harvard.edu/abs/2000A&A...358..593S}

\bibitem[{{Thi}(2015)}]{Thi+2015}
{Thi}, W.-F. 2015, in European Physical Journal Web of Conferences, Vol. 102,
  European Physical Journal Web of Conferences, 00012

\bibitem[{{Thi} {et~al.}(2013){Thi}, {Kamp}, {Woitke}, {van der Plas},
  {Bertelsen}, \& {Wiesenfeld}}]{Thi+2013}
{Thi}, W.~F., {Kamp}, I., {Woitke}, P., {et~al.} 2013, Astronomy and
  Astrophysics, 551, A49, \doi{10.1051/0004-6361/201219210}
  \ads{https://ui.adsabs.harvard.edu/abs/2013A&A...551A..49T}

\bibitem[{{Thi} {et~al.}(2011){Thi}, {Woitke}, \& {Kamp}}]{Thi+2011}
{Thi}, W.~F., {Woitke}, P., \& {Kamp}, I. 2011, Monthly Notices of the Royal
  Astronomical Society, 412, 711, \doi{10.1111/j.1365-2966.2010.17741.x}
  \ads{https://ui.adsabs.harvard.edu/abs/2011MNRAS.412..711T}

\bibitem[{{Tielens} \& {Allamandola}(1987)}]{Tielens&Allamandola1987}
{Tielens}, A.~G.~G.~M. \& {Allamandola}, L.~J. 1987, in Interstellar Processes,
  ed. D.~J. {Hollenbach} \& J.~{Thronson}, Harley~A., Vol. 134, 397,
  \doi{10.1007/978-94-009-3861-8\_16}\ads{https://ui.adsabs.harvard.edu/abs/1987ASSL..134..397T}

\bibitem[{{Tr{\o}nnes} {et~al.}(2019){Tr{\o}nnes}, {Baron}, {Eigenmann},
  {Guren}, {Heyn}, {L{\o}ken}, \& {Mohn}}]{Tronnes+2019}
{Tr{\o}nnes}, R.~G., {Baron}, M.~A., {Eigenmann}, K.~R., {et~al.} 2019,
  Tectonophysics, 760, 165, \doi{10.1016/j.tecto.2018.10.021}
  \ads{https://ui.adsabs.harvard.edu/abs/2019Tectp.760..165T}

\bibitem[{{Van Clepper} {et~al.}(2022){Van Clepper}, {Bergner}, {Bosman},
  {Bergin}, \& {Ciesla}}]{VanClepper+2022}
{Van Clepper}, E., {Bergner}, J.~B., {Bosman}, A.~D., {Bergin}, E., \&
  {Ciesla}, F.~J. 2022, arXiv e-prints, \arxiv{2202.00524}
  \ads{https://ui.adsabs.harvard.edu/abs/2022arXiv220200524V}

\bibitem[{{Visser}(1997)}]{Visser1997}
{Visser}, A. 1997, Marine Ecology Progress Series, 158, 275,
  \doi{10.3354/meps158275}
  \ads{https://ui.adsabs.harvard.edu/abs/1997MEPS..158..275V}

\bibitem[{{Visser} {et~al.}(2018){Visser}, {Bruderer}, {Cazzoletti},
  {Facchini}, {Heays}, \& {van Dishoeck}}]{Visser+2018}
{Visser}, R., {Bruderer}, S., {Cazzoletti}, P., {et~al.} 2018, Astronomy \&
  Astrophysics, 615, A75, \doi{10.1051/0004-6361/201731898}
  \ads{https://ui.adsabs.harvard.edu/abs/2018A&A...615A..75V}

\bibitem[{{Visser} {et~al.}(2021){Visser}, {Dr{\k{a}}{\.z}kowska}, \&
  {Dominik}}]{Visser+2021}
{Visser}, R.~G., {Dr{\k{a}}{\.z}kowska}, J., \& {Dominik}, C. 2021, arXiv
  e-prints, \doi{10.1051/0004-6361/202039769}
  \ads{https://ui.adsabs.harvard.edu/abs/2021arXiv210109209V}

\bibitem[{{Watson} {et~al.}(2009){Watson}, {Leisenring}, {Furlan}, {Bohac},
  {Sargent}, {Forrest}, {Calvet}, {Hartmann}, {Nordhaus}, {Green}, {Kim},
  {Sloan}, {Chen}, {Keller}, {d'Alessio}, {Najita}, {Uchida}, \&
  {Houck}}]{Watson+2009}
{Watson}, D.~M., {Leisenring}, J.~M., {Furlan}, E., {et~al.} 2009, The
  Astrophysical Journal, 180, 84, \doi{10.1088/0067-0049/180/1/84}
  \ads{https://ui.adsabs.harvard.edu/abs/2009ApJS..180...84W}

\bibitem[{{Weidenschilling}(1977)}]{Weidenschilling1977}
{Weidenschilling}, S.~J. 1977, Monthly Notices of the Royal Astronomical
  Society, 180, 57, \doi{10.1093/mnras/180.2.57}
  \ads{https://ui.adsabs.harvard.edu/abs/1977MNRAS.180...57W}

\bibitem[{{Weidling} {et~al.}(2012){Weidling}, {G{\"u}ttler}, \&
  {Blum}}]{Weidling+2012}
{Weidling}, R., {G{\"u}ttler}, C., \& {Blum}, J. 2012, Icarus, 218, 688,
  \doi{10.1016/j.icarus.2011.10.002}
  \ads{https://ui.adsabs.harvard.edu/abs/2012Icar..218..688W}

\bibitem[{{Weidling} {et~al.}(2009){Weidling}, {G{\"u}ttler}, {Blum}, \&
  {Brauer}}]{Weidling+2009}
{Weidling}, R., {G{\"u}ttler}, C., {Blum}, J., \& {Brauer}, F. 2009, The
  Astrophysical Journal, 696, 2036, \doi{10.1088/0004-637X/696/2/2036}
  \ads{https://ui.adsabs.harvard.edu/abs/2009ApJ...696.2036W}

\bibitem[{{Whitworth}(1975)}]{Whitworth1975}
{Whitworth}, A.~P. 1975, Astrophysics and Space Science, 34, 155,
  \doi{10.1007/BF00646756}
  \ads{https://ui.adsabs.harvard.edu/abs/1975Ap&SS..34..155W}

\bibitem[{{Williams} {et~al.}(2020){Williams}, {Sanborn}, {Defouilloy}, {Yin},
  {Kita}, {Ebel}, {Yamakawa}, \& {Yamashita}}]{Williams+2020}
{Williams}, C.~D., {Sanborn}, M.~E., {Defouilloy}, C., {et~al.} 2020,
  Proceedings of the National Academy of Science, 117, 23426,
  \doi{10.1073/pnas.2005235117}
  \ads{https://ui.adsabs.harvard.edu/abs/2020PNAS..11723426W}

\bibitem[{{Woitke} {et~al.}(2019){Woitke}, {Kamp}, {Antonellini}, {Anthonioz},
  {Baldovin-Saveedra}, {Carmona}, {Dionatos}, {Dominik}, {Greaves},
  {G{\"u}del}, {Ilee}, {Liebhardt}, {Menard}, {Min}, {Pinte}, {Rab}, {Rigon},
  {Thi}, {Thureau}, \& {Waters}}]{Woitke+2019}
{Woitke}, P., {Kamp}, I., {Antonellini}, S., {et~al.} 2019, Publications of the
  Astronomical Society of the Pacific, 131, 064301,
  \doi{10.1088/1538-3873/aaf4e5}
  \ads{https://ui.adsabs.harvard.edu/abs/2019PASP..131f4301W}

\bibitem[{{Woitke} {et~al.}(2009){Woitke}, {Kamp}, \& {Thi}}]{Woitke+2009}
{Woitke}, P., {Kamp}, I., \& {Thi}, W.~F. 2009, Astronomy and Astrophysics,
  501, 383, \doi{10.1051/0004-6361/200911821}
  \ads{https://ui.adsabs.harvard.edu/abs/2009A&A...501..383W}

\bibitem[{{Woitke} {et~al.}(2016){Woitke}, {Min}, {Pinte}, {Thi}, {Kamp},
  {Rab}, {Anthonioz}, {Antonellini}, {Baldovin-Saavedra}, {Carmona}, {Dominik},
  {Dionatos}, {Greaves}, {G{\"u}del}, {Ilee}, {Liebhart}, {M{\'e}nard},
  {Rigon}, {Waters}, {Aresu}, {Meijerink}, \& {Spaans}}]{Woitke+2016}
{Woitke}, P., {Min}, M., {Pinte}, C., {et~al.} 2016, Astronomy and
  Astrophysics, 586, A103, \doi{10.1051/0004-6361/201526538}
  \ads{https://ui.adsabs.harvard.edu/abs/2016A&A...586A.103W}

\bibitem[{{Wurm} \& {Blum}(1998)}]{Wurm&Blum1998}
{Wurm}, G. \& {Blum}, J. 1998, Icarus, 132, 125, \doi{10.1006/icar.1998.5891}
  \ads{https://ui.adsabs.harvard.edu/abs/1998Icar..132..125W}

\bibitem[{Youdin \& Goodman(2005)}]{Youdin&Goodman2005}
Youdin, A.~N. \& Goodman, J. 2005, The Astrophysical Journal, 620, 459,
  \doi{10.1086/426895}

\bibitem[{{Youdin} \& {Lithwick}(2007)}]{Youdin&Lithwick2007}
{Youdin}, A.~N. \& {Lithwick}, Y. 2007, Icarus, 192, 588

\bibitem[{{Zhang} {et~al.}(2020){Zhang}, {Schwarz}, \& {Bergin}}]{Zhang+2020}
{Zhang}, K., {Schwarz}, K.~R., \& {Bergin}, E.~A. 2020, The Astrophysical
  Journal Letters, 891, L17, \doi{10.3847/2041-8213/ab7823}
  \ads{https://ui.adsabs.harvard.edu/abs/2020ApJ...891L..17Z}

\bibitem[{{Zsom} {et~al.}(2010){Zsom}, {Ormel}, {G{\"u}ttler}, {Blum}, \&
  {Dullemond}}]{Zsom+2010}
{Zsom}, A., {Ormel}, C.~W., {G{\"u}ttler}, C., {Blum}, J., \& {Dullemond},
  C.~P. 2010, Astronomy \& Astrophysics, 513, A57,
  \doi{10.1051/0004-6361/200912976}
  \ads{https://ui.adsabs.harvard.edu/abs/2010A&A...513A..57Z}

\end{thebibliography}

\begin{appendix}

\onecolumn
\section{Structure and composition of the background disk models}
\label{sec:AA}
The physical disk structure of the \texttt{const}, \texttt{vFrag1}, \texttt{vFrag5} and \texttt{vFrag10} background models is shown in Sect. \ref{sec:ABP} while Sect. \ref{sec:ABC} presents the chemical structures.

\subsection{Physical background disk structures}
\label{sec:ABP}
Overview of the spatial structures of gas temperature $T_g$, dust temperature $T_d$, gas mass density $\rho_g$, dust mass density $\rho_d$ and UV radiation field strength $\chi_\text{RT}$ provided by the background disk models.

\begin{figure}[ht!]
    \centering
    \includegraphics[width=.99\textwidth]{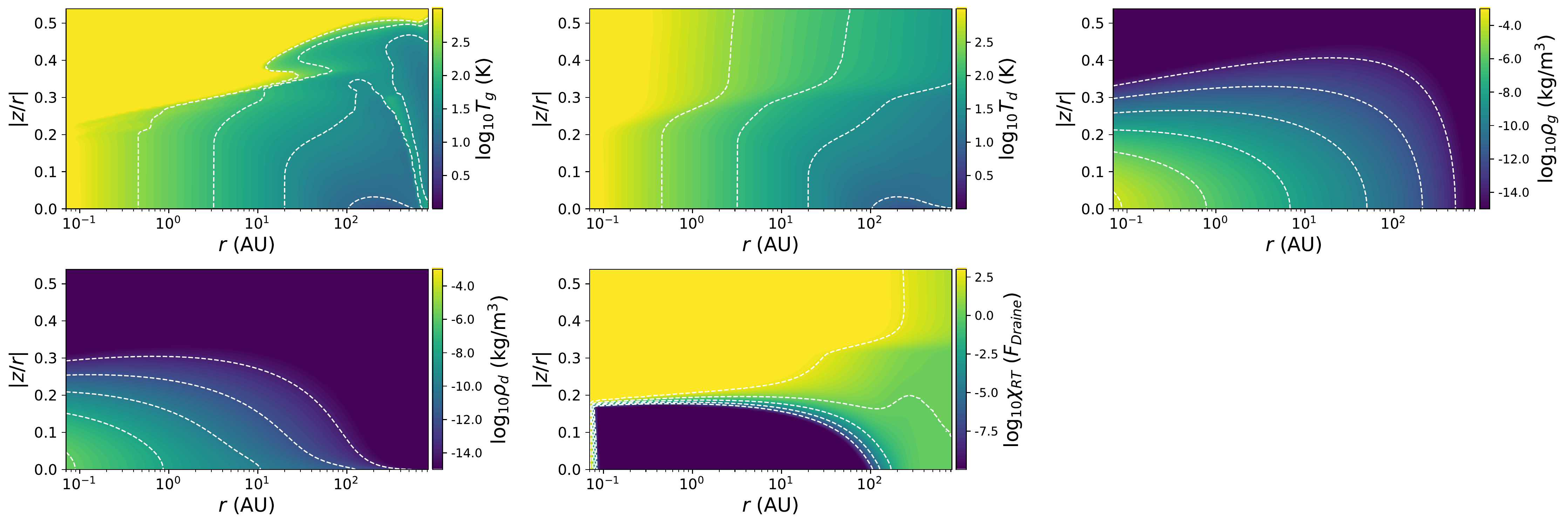}
    \caption{Physical disk structure of the \texttt{const}-model.}
    \label{fig:ABBackgroundModelconstStructure}
\end{figure}
\begin{figure}[ht!]
    \centering
    \includegraphics[width=.99\textwidth]{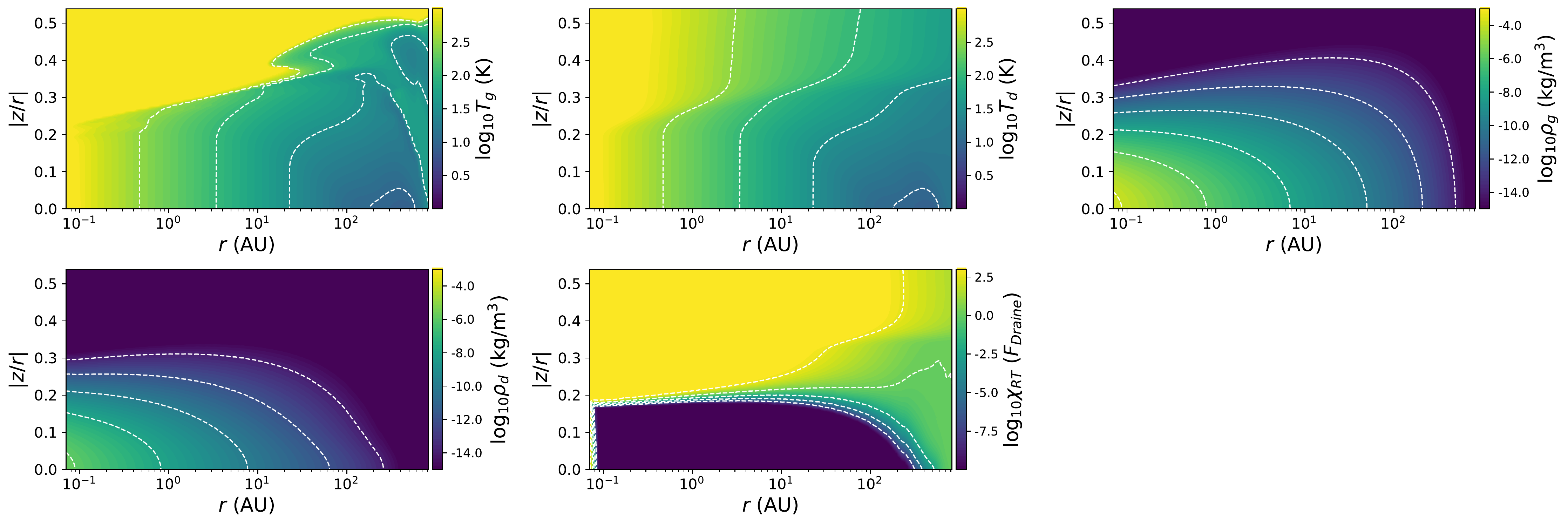}
    \caption{Physical disk structure of the \texttt{vFrag1}-model.}
    \label{fig:ABBackgroundModelvFrag1Structure}
\end{figure}
\begin{figure}[ht!]
    \centering
    \includegraphics[width=.99\textwidth]{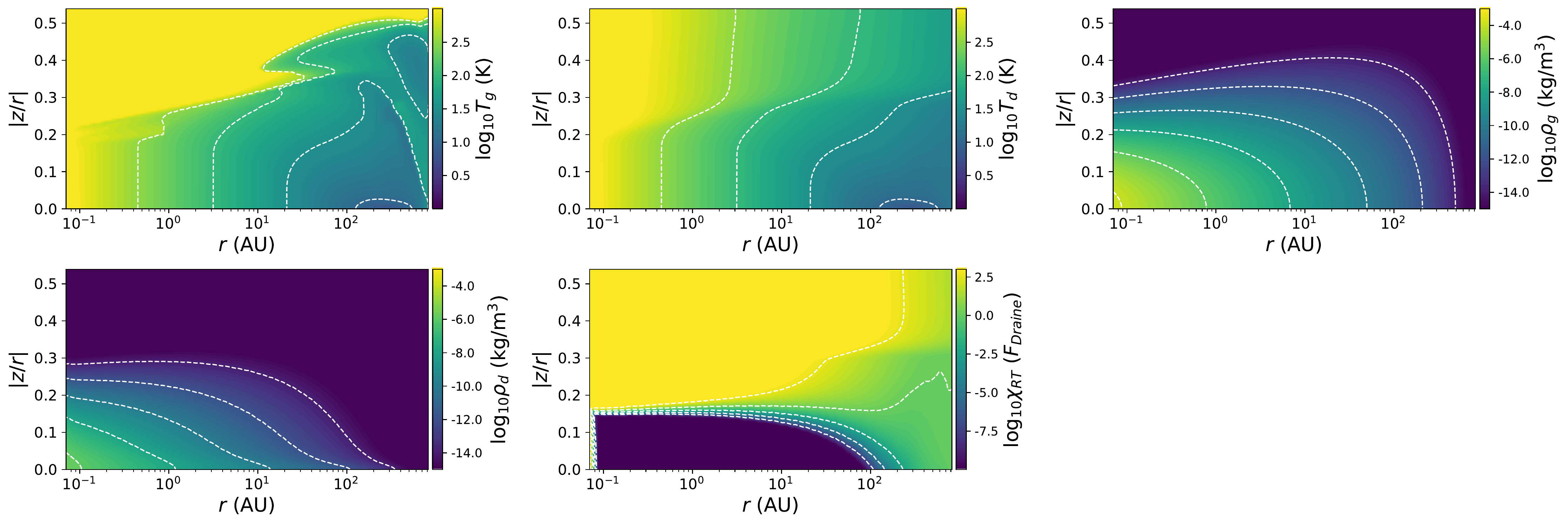}
    \caption{Physical disk structure of the \texttt{vFrag5}-model.}
    \label{fig:ABBackgroundModelvFrag5Structure}
\end{figure}
\begin{figure}[ht!]
    \centering
    \includegraphics[width=.99\textwidth]{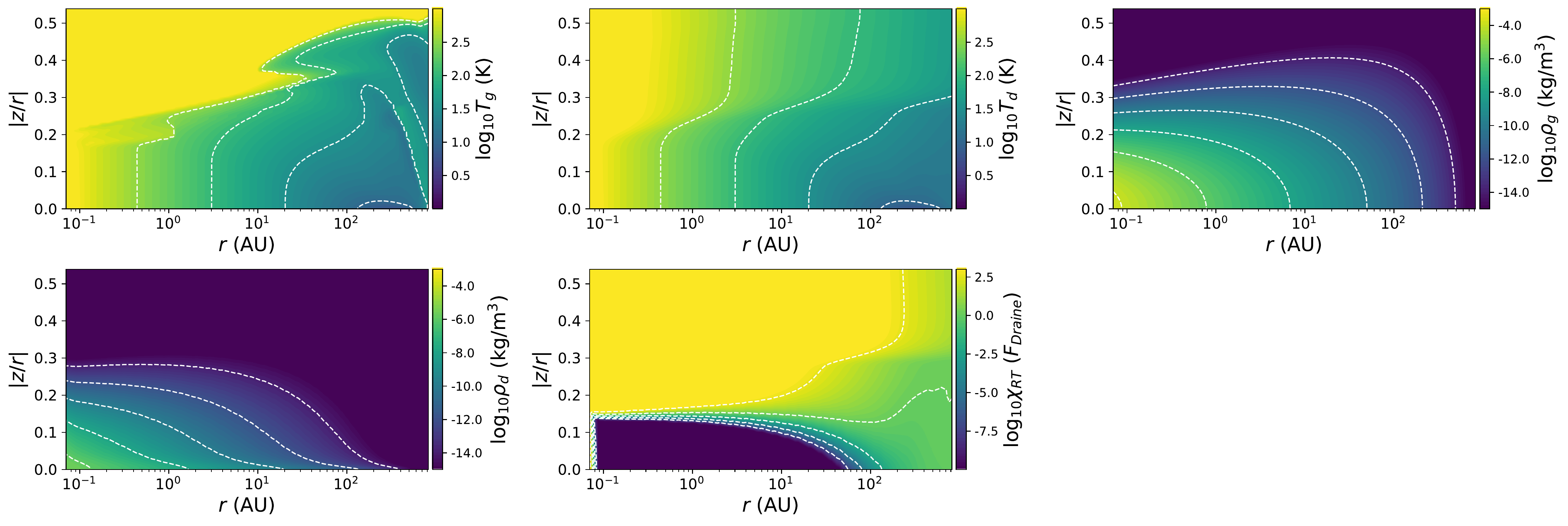}
    \caption{Physical disk structure of the \texttt{vFrag10}-model.}
    \label{fig:ABBackgroundModelvFrag10Structure}
\end{figure}
\clearpage
\subsection{Chemical background disk structure}
\label{sec:ABC}
Overview of the gas (left columns) and ice (right columns) number densities ($n_x$) of various molecules provided by the background disk models in m$^{-3}$. In each figure from top to bottom: H$_2$O, NH$_3$, CO, CO$_2$, CH$_4$ and H$_2$S.

\begin{figure}[ht!]
    \centering
    \includegraphics[width=.79\textwidth]{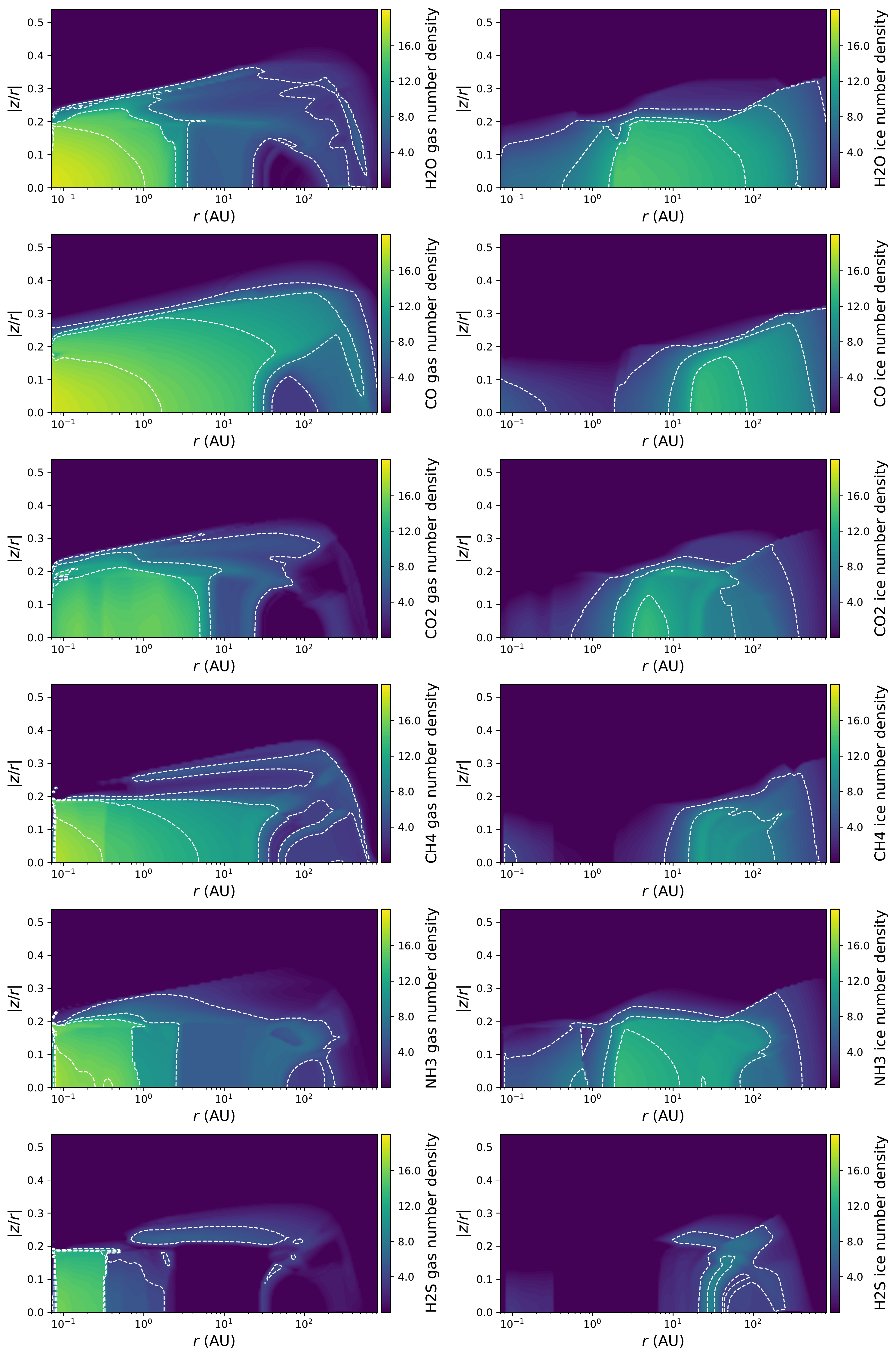}
    \caption{Chemical disk structure of the \texttt{const}-model.}
    \label{fig:ABBackgroundModelconstAbundances}
\end{figure}
\begin{figure}[ht!]
    \centering
    \includegraphics[width=.79\textwidth]{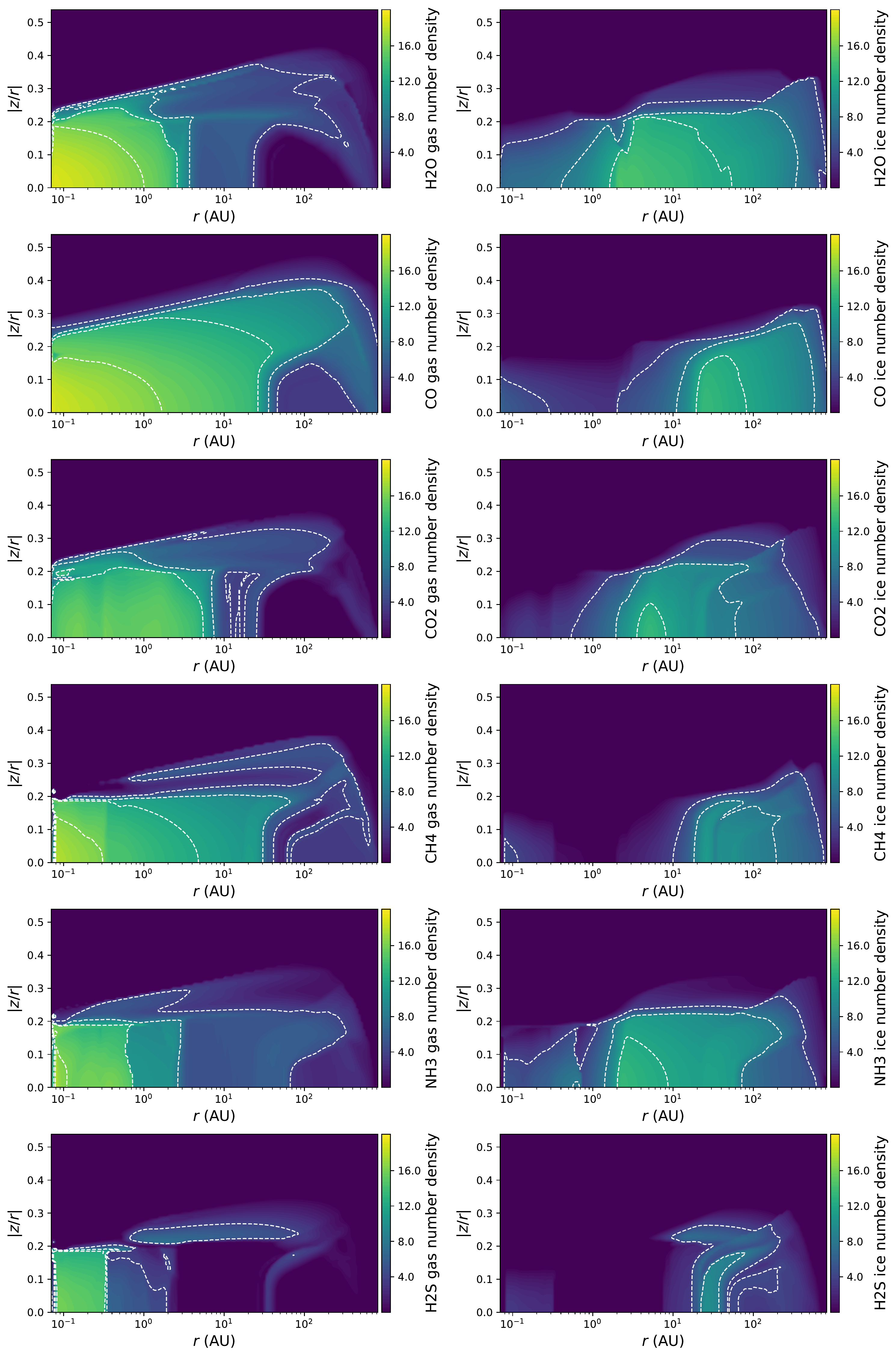}
    \caption{Chemical disk structure of the \texttt{vFrag1}-model.}
    \label{fig:ABBackgroundModelvFrag1Abundances}
\end{figure}
\begin{figure}[ht!]
    \centering
    \includegraphics[width=.79\textwidth]{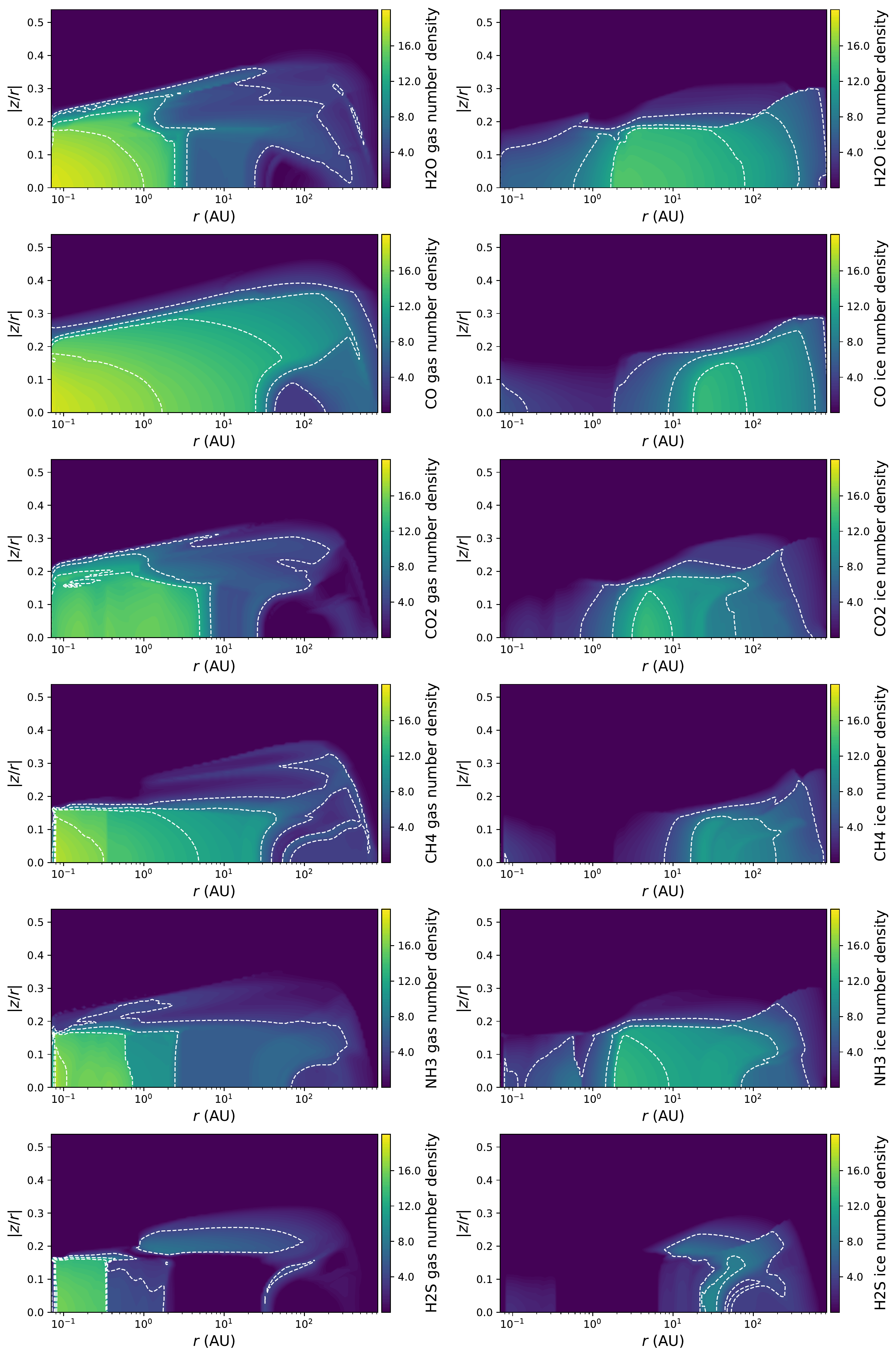}
    \caption{Chemical disk structure of the \texttt{vFrag5}-model.}
    \label{fig:ABBackgroundModelvFrag5Abundances}
\end{figure}
\begin{figure}[ht!]
    \centering
    \includegraphics[width=.79\textwidth]{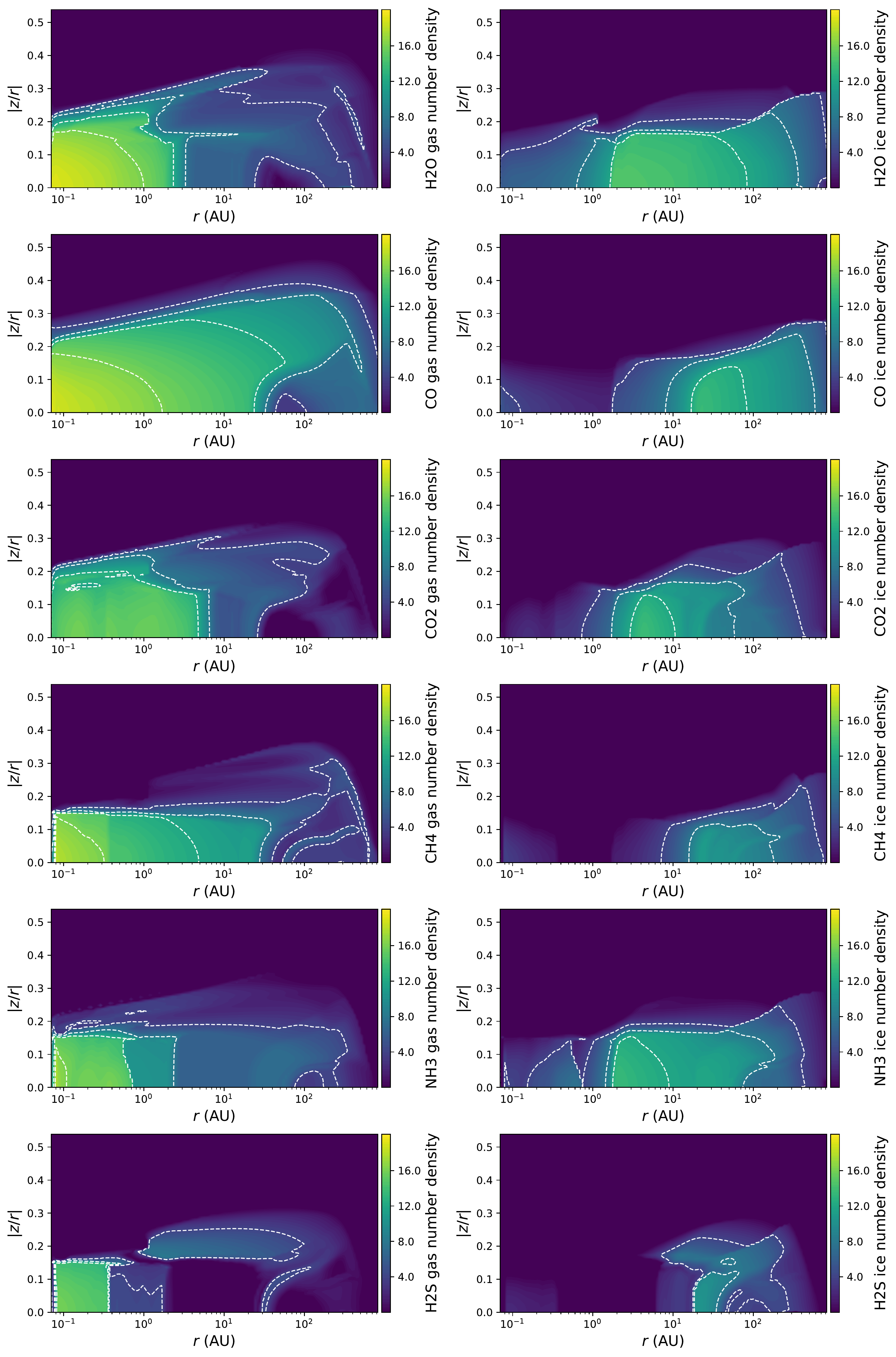}
    \caption{Chemical disk structure of the \texttt{vFrag10}-model.}
    \label{fig:ABBackgroundModelvFrag10Abundances}
\end{figure}

%\clearpage
%\section{Parameter tables}
%\subsection{Background disk model}
%\input{Tables/DiskTable.tex}

%\subsection{Monomer model}
%\input{Tables/MonoTable.tex}

%\subsection{Collision model}
%\input{Tables/ColTable.tex}

%\subsection{Ice formation model}
%\input{Tables/IceTable.tex}
\twocolumn
\section{Diffusion timescales inside the home aggregate}
\label{sec:AB}
In our treatment of desorption in Sect. \ref{sec:2.4}, molecules originating from monomers deep inside aggregates are assumed to escape the aggregate instantaneously. However, this assumption may be inappropriate for significant filling factors, as the mean free path for a molecule traveling inside an aggregate is usually shorter than $s_a$, which means that molecules liberated by desorption will undergo collisions with other monomers in an aggregate before escaping to the gas phase. This may result in re-adsorption, allowing to possibly retain more ice in the unexposed aggregate interiors than accounted for in the current model. Similarly, this mechanism could also provide a pathway for freshly accumulated ice molecules on exposed monomers to diffuse towards the unexposed monomers deeper in the aggregate. In this section, we develop a simple model to constrain the importance of this process by defining and estimating an upper bound of the timescale associated with the escape of molecules from the aggregate.\\
We assume that the liberated molecules effectively undergo a random walk as collisions with other monomers occur while escaping the aggregate. If liberated molecules only collide with monomers, the \textit{escape timescale} $\tau_\text{esc}$ associated with a molecule depends on two main factors: the number of random walk collisions between the molecule and other monomers $n_\text{c}$ before the molecule escapes the aggregate, and the fraction of these collisions which results in re-adsorption, which is given by the sticking factor $S$ (Eq. \eqref{eq:sticking}). Altogether we can thus express $\tau_\text{esc}$ as
\begin{align}
\label{eq:tauEscFirst}
    \tau_\text{esc}=n_\text{c}\left(\tau_\text{mc}+S\tau_\text{d}\right).
\end{align}
Here, $\tau_\text{mc}$ denotes the molecule collision timescale, and can be expressed as
\begin{align}
    \tau_\text{mc}=\frac{\lambda_\text{mfp}}{v}=\frac{4s_m}{3\phi v}.
\end{align}
$\lambda_\text{mfp}=1/(\pi s_m^2 n_m)$ in this case denotes the mean free path inside the aggregate. We here use the monomer density $n_m$ as given in Eq. \eqref{eq:nmonomer}. $v$ denotes the velocity of the molecule while it travels between monomers. Since we aim to obtain an upper bound on the escape timescale, we estimate $v$ with the molecule ice diffusion speed, which we estimate from the thermal hopping timescale $\tau_\text{hop}$ (i.e. the average time for a molecule to migrate from one site in the ice lattice to the next) \citep[e.g.][]{Thi+2015, Ruaud+2016, Cuppen+2017},
\begin{align}
    v = \frac{d_\text{s}}{\tau_\text{hop}}=d_\text{s}\nu_x\exp\left(-\frac{E_{\text{diff},x}}{k_\text{B}T_{d}}\right).
\end{align}
Here, $d_\text{s}\simeq 10^{-10}\,\unit{m}{}$ \citep[e.g.][]{Caselli&Ceccarelli2012} denotes the average distance between two molecule binding sites in the ice lattice. $\nu_x$ is the lattice vibration frequency as given by Eq. \eqref{eq:latvib} and $E_{\text{diff},x}$ denotes the diffusion energy, which we estimate as $E_{\text{diff},x}=\frac{1}{2}E_{\text{ads},x}$. Similarly, the desorption timescale for a single molecule $\tau_\text{d}$ (in addition to the ice mantle desorption timescale $\tau_\text{des}$ defined in Eq. \eqref{eq:tauDes}) for molecule species $x$ can be expressed as \citep[e.g.][]{Thi+2015}
\begin{align}
    \tau_\text{d}=\frac{1}{\nu_x}\exp\left(\frac{E_{\text{ads},x}}{k_\text{B}T_\text{d}}\right).
\end{align}
Altogether we write Eq. \eqref{eq:tauEscFirst} as
\begin{align}
\label{eq:tauEsc}
    \tau_\text{esc}=\frac{n_c}{\nu_x}\left[\frac{4s_m}{3\phi d_s}\exp\left(\frac{E_{\text{ads},x}}{2k_\text{B}T_\text{d}}\right)+S\exp\left(\frac{E_{\text{ads},x}}{k_\text{B}T_\text{d}}\right)\right].
\end{align}
It becomes clear from Eq. \eqref{eq:tauEsc} that $\tau_\text{esc}$ is very sensitive to $T_\text{d}$. On the one hand, this will result in extremely long $\tau_\text{esc}$ for aggregates significantly inside the ice-forming region, as thermal desorption becomes a very slow process (see also Sect. \ref{sec:3.1}), resulting in extremely long $\tau_\text{d}$. Furthermore, $S\simeq1$ in these regions, which means that even if a molecule undergoes thermal desorption, it will be re-adsorbed almost instantaneously by another nearby monomer. On the other hand in warm regions where $\tau_\text{d}$ becomes very short and $S\rightarrow 0$, we also expect that re-adsorption will not have a significant effect on monomer ice mantle composition, as $\tau_\text{esc}$ will be dominated by the molecule collision timescale $\tau_\text{mc}$. Altogether Eq. \eqref{eq:tauEsc} suggests that the effects of re-adsorption of species $x$ on monomer ice mantle composition are the largest near the iceline associated with species $x$. However, another crucial factor in determining the value of $\tau_\text{esc}$ is the number of collisions before escape $n_\text{c}$, which depends on aggregate size $s_\text{a}$ and filling factor $\phi$.\\
\begin{figure}
    \centering
    \includegraphics[width=.45\textwidth]{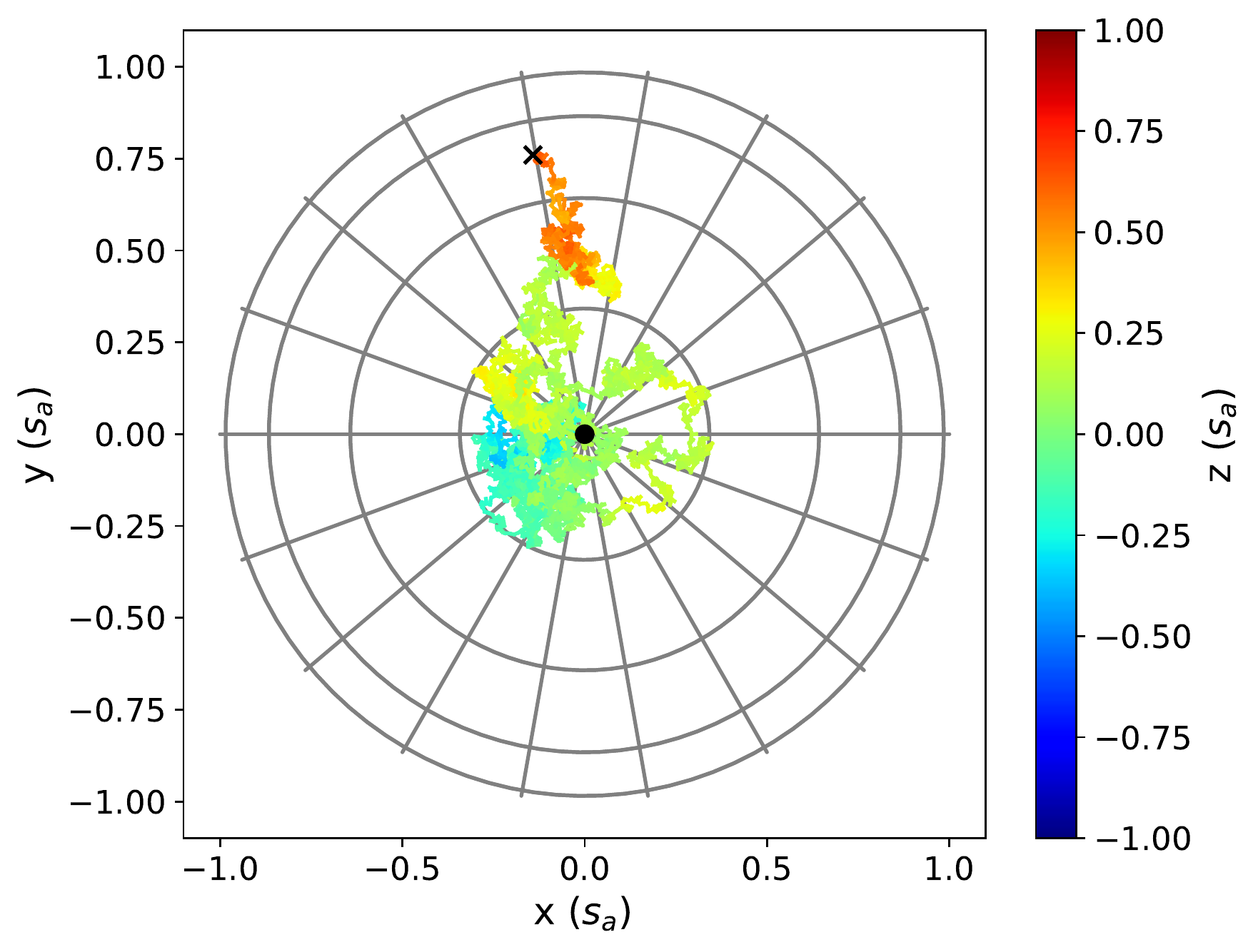}
    \caption{Example of a 3D random walk through an aggregate of radius $s_a= 10^{-5}$ m and $\phi=0.5$ (top view). The colorscale indicates the $z$-coordinate of the molecule. The black dot and cross denote the starting point and the location where the molecule crosses the aggregate surface, respectively.}
    \label{fig:ABRandomWalkExample}
\end{figure}
\begin{figure}
    \centering
    \includegraphics[width=.45\textwidth]{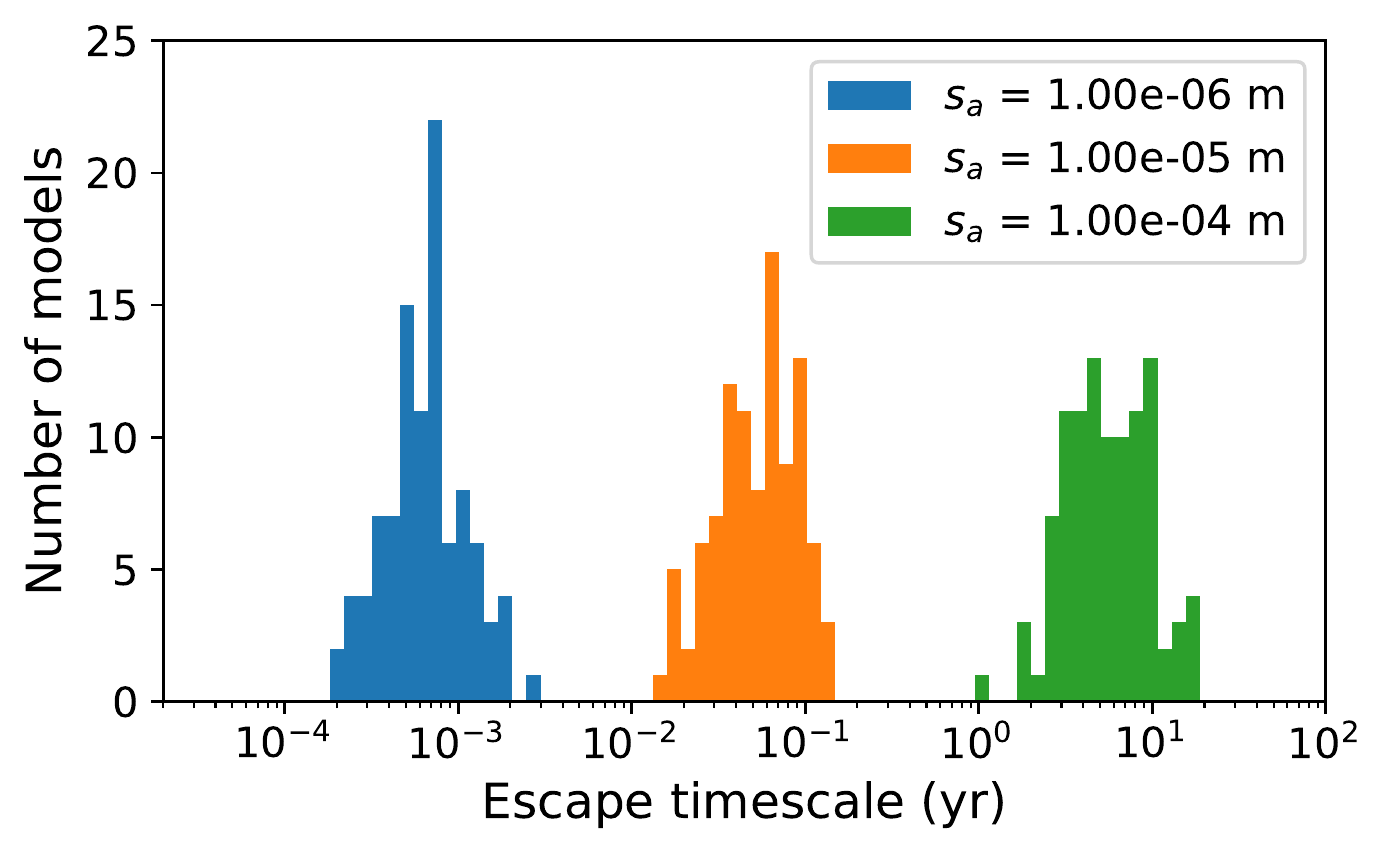}
    \caption{Distribution of escape timescales of 100 model runs in aggregates of $\phi=0.5$ and varying radius. Molecules are always released from the center of the aggregate.}
    \label{fig:ABRandomWalkStatistics}
\end{figure}
In order to constrain $n_c$ and provide an upper bound on $\tau_\text{esc}$ near the iceline, we consider a simple numerical 3D spherical random walk model. In this random walk model, individual molecules move from one position to the next in steps of size $\lambda_\text{mfp}$ in a spherically random uniform direction. The simulation ends when the molecule distance to the aggregate center is larger than the aggregate radius (i.e. $x_\text{mol}^2+y_\text{mol}^2+z_\text{mol}^2>s_a^2$). This gives rise to trajectories of individual molecules through the aggregate. An example of such a trajectory is shown in Fig. \ref{fig:ABRandomWalkExample}, where a molecule is released from the center of an aggregate with $s_a=10^{-5}$ m and $\phi=0.5$, and undergoes $n_c=7124$ collisions before escaping the aggregate. Altogether individual molecules can take significant detours through the aggregate, resulting in a considerable variation in $n_c$ and $\tau_\text{esc}$ associated with the trajectories of individual molecules. Therefore it is necessary to evaluate Eq. \eqref{eq:tauEsc} for a larger set of molecules.\\
As a next step therefore we consider the distribution of $\tau_\text{esc}$ for 100 $\H_2\O$ molecules escaping from the center of aggregates of $s_a=10^{-6},10^{-5}$, and $10^{-4}$ m at the $\H_2\O$ iceline ($T_\text{d}\approx 140$ K, c.f. Fig \ref{fig:ABBackgroundModelconstStructure} and Fig. \ref{fig:ABBackgroundModelconstAbundances}). As we expect that the effects of re-adsorption are most important for larger aggregates with a short mean free path, we consider a filling factor $\phi=0.5$. The direct dependence of $\tau_\text{esc}$ on $\phi$ in Eq. \eqref{eq:tauEsc} suggests that larger $\phi$ results in shorter $\tau_\text{esc}$. However, a shorter mean free path will also result in larger $n_\text{c}$, which effectively results in the the escape timescale being longer for higher $\phi$. The resulting escape timescales distributions $\tau_\text{esc}$ for the different aggregate sizes are shown in Fig. \ref{fig:ABRandomWalkStatistics}. Molecules escape faster from smaller aggregates due to the lower average number of collisions $n_\text{c}$ before escape. Fig. \ref{fig:ABRandomWalkStatistics} reveals that the average value of $\tau_\text{esc}$ increases orders of magnitude per order of magnitude increase in $s_m$. This suggests that for aggregates larger than $10^{-4}$ m (i.e. millimetre- and centimetre-sized aggregates), the escape timescale may become comparable to the timescales of other disk processes (c.f. Fig. \ref{fig:31TimescalesExample}). Since the largest aggregates contain most of the dust mass (c.f. Fig. \ref{fig:23SlopeTest}), Fig. \ref{fig:ABRandomWalkStatistics} implies that re-adsorption after thermal desorption in unexposed aggregate interiors may have significant effects on the average amount of ice retained in monomer ice mantles. Specifically, our model could overestimate the amount of ice lost via thermal desorption on unexposed monomers. However, an increase or decrease of $10$ K from $T_\text{d}$ was also found to result in a decrease or increase in $\tau_\text{esc}$ of almost 2 orders of magnitude, respectively. This suggests that the disk region where $\tau_\text{esc}$ couples to the other disk processes is spatially limited to a small region just behind the $\H_2\O$ iceline. However, this also means that there always is a region where $\tau_\text{esc}$ couples to the other disk processes, and thus can affect ice composition on timescales similar as the other disk processes. We expect that this result for $\H_2\O$ can be generalized to other chemical species, where the disk region where re-adsorption may affect the amount of ice of species $x$ on monomers would be shifted to just behind the iceline of the respective species. Altogether the current model may underestimate amount of ice of species $x$ retained in deeply embedded monomers just behind the iceline of \mbox{species $x$}.
\end{appendix}

\end{document}